\documentclass[12pt]{article}
\usepackage{epsf}

\renewcommand{\bar}[1]{\overline{#1}}
\newcommand{\M}{{\cal M}}
\newcommand{\VEV}[1]{\left\langle{#1}\right\rangle}
\newcommand{\etal}{{\em et al.}}
\newcommand{\ie}{{\it i.e.}}
\newcommand{\eg}{{\it e.g.}}

\newcommand{\longvec}[1]{\overrightarrow{\!\!#1}}

\newcommand{\wA}{\widetilde A}
\newcommand{\bra}[1]{\left\langle{#1}\,\right|\,}
\newcommand{\ket}[1]{\vert\,{#1}\rangle}
\newcommand{\ms}{$\overline{\mbox{MS}}$}

\newcommand{\amst}{\mbox{${\widetilde{\alpha}_{\overline{\mbox{\tiny MS}}}}$}}
\newcommand{\av}{\mbox{$\alpha_{V}$}}

\newcommand{\GeV}{{\rm GeV}}

\newcommand{\cf}{{\em c.f.}}
\newcommand{\eqcm}{\: ,}
\newcommand{\lqcd}{\Lambda_{QCD}}
\newcommand{\dpt}{\Delta_\perp}
\newcommand{\bwpsi}{\overline{\widetilde \psi}}
\newcommand{\wpsi}{\widetilde \psi}
\newcommand{\bpsi}{\overline \psi}
\newcommand{\ep}{\hbox{\hskip -.02in $\epsilon$}}
\newcommand{\D}{\hbox{\hskip -.05in $D$}}
\newcommand{\A}{\hbox{\hskip -.05in ${\widetilde A}$}}

\begin {document}
\begin{flushright}
{\small
SLAC--PUB--8315\\
December 1999\\}
\end{flushright}

\begin{center}
{{\bf\LARGE   
New Directions in Quantum Chromodynamics}\footnote{Work supported by
Department of Energy contract  DE--AC03--76SF00515.}}

\bigskip
Stanley J. Brodsky \\
{\sl Stanford Linear Accelerator Center\\
Stanford University, Stanford, California 94309}

\end{center}

\vfill

\begin{center}
{\bf\large   
Abstract }
\end{center}

I review the light-cone Fock state represention and its associated
light-cone factorization scheme as a method for encoding the
flavor, momentum, and helicity properties of hadrons in
the form of universal process-independent and frame-independent
amplitudes.  Discrete light-cone quantization (DLCQ) provides a matrix
representation of the QCD Hamiltonian and a nonperturbative method for
computing the quark and gluon bound state wavefunctions.  A number of
applications of the light-cone formalism are discussed, including an
exact light-cone Fock state representation of semi-leptonic
$B$ decay amplitudes. Hard exclusive and diffractive reactions
are shown to be sensitive to hadron distribution amplitudes, the
valence Fock state hadronic wavefuctions at small impact
separation. Semi-exclusive reactions are shown to provide new
flavor-dependent probes of distribution amplituds and new types of deep
inelastic currents.   ``Self-resolving"
diffractive processes and Coulomb dissociation are discussed as a
direct measure of the light-cone wavefunctions of hadrons. Alternatively,
one can use Coulomb dissociation to resolve nuclei in terms of their
nucleonic and mesonic degrees of freedom.  I also discuss several
theoretical tools which eliminate theoretical ambiguities in perturbative
QCD predictions.  For example, commensurate scale relations are
perturbative QCD predictions based on conformal symmetry which relate
observable to observable at fixed relative scale; such relations have no
renormalization scale or scheme ambiguity.  I also discuss the utility of
the $\alpha_V$ coupling, defined from the QCD heavy quark potential,  as
a useful physical expansion parameter for perturbative QCD and grand
unification.  New results on the  analytic fermion masses dependence  of
the $\alpha_V$ coupling at two-loop order are presented.

\vfill

\begin{center} 
{\it Invited talk presented at  \\ 
 International Summer School on Particle Production
Spanning MeV and TeV Energies (Nijmegen 99) }\\
{\it Nijmegen, The Netherlands} \\
{\it 8--20 August 1999}\\
\end{center}

\vfill\eject

\normalsize

\section{Introduction}

In quantum chromdynamics, hadrons are identified as relativistic
color-singlet bound states of confined quarks and gluons.  A primary goal
of high energy and nuclear physics is to unravel the nonperturbative
structure and dynamics of nucleons and nuclei in terms of their
fundamental quark and gluon degrees of freedom.
QCD is a relativistic quantum field theory, so that a
fundamental description of hadrons must be at the amplitude level.
Part of the complexity of hadronic physics is related
to the fact that the eigensolutions of a relativistic
theory fluctuate not only in momentum space and helicity, but also
in particle number.  For example, the heavy quark sea of the proton is
associated with higher particle number Fock states.  Thus any wavefunction
description must allow for arbitrary fluctuations in particle number.

Since the discovery of Bjorken scaling \cite{Bjorken:1966jh}
of deep inelastic lepton-proton
scattering in 1969  \cite{Breidenbach:1969kd}, high energy experiment
have provided an extraordinary amount of information on the flavor,
momentum, and helicity distributions of the quark and gluon in hadrons.
This information is generally encoded in the leading twist factorized
quark and gluon distributions
$q_{\lambda_q, \lambda_N}(x, Q), g_{\lambda_g, \lambda_N}(x, Q).$
However, since such distributions are single-par\-ticle probabilities,
they contain no information on the transverse momentum
distributions, multiparticle flavor and helicity correlations, or quantum
mechanical phases, information critical to understanding higher
twist processes or exclusive processes such as form factors, elastic
scattering, and the exclusive decays of heavy hadrons.
Although it is convenient for computational reasons to separate hard,
perturbatively calculable, and soft non-perturbative physics, the theory
has no such intrinsic division. The analysis of
QCD processes at the amplitude level is a challenging
relativistic many-body problem, mixing issues involving confinement,
chiral symmetry, non-perturbative and perturbative dynamics, and thus a
theoretical complexity far beyond traditional bound state problems.

Deep inelastic lepton-proton scattering has provided
the traditional guide to
hadron structure.  The focus in high energy physics has been on the
logarithmic DGLAP evolution of the structure functions and the associated
jet structure as a test of perturbative QCD.  However, when the photon
virtuality is small and of order of the quark intrinsic transverse
momentum, evolution from QCD radiative processes becomes quenched, and the
structure functions reveal fundamental features of the proton's
composition.  The deep inelastic scattering data in fact show that the
nonperturbative structure of nucleons is more complex than suggested by a
three-quark bound state.  For example, if the sea quarks were
generated solely by perturbative QCD evolution via gluon splitting, the
anti-quark distributions would be approximately isospin symmetric.
However, the
$\bar u(x)$ and
$\bar d(x)$ antiquark distributions of the proton at $Q^2
\sim 10$ GeV$^2$ are found to be quite different in
shape  \cite{{Nasalski:1994bh}}
and thus must reflect dynamics intrinsic to the proton's structure.
Evidence for a difference between the $\bar s(x)$ and $s(x)$ distributions
has also been claimed  \cite{Barone:1999yv}.  There have also been surprises
associated with the chirality distributions
$\Delta q = q_{\uparrow/\uparrow} - q_{\downarrow/\uparrow}$ of the valence
quarks which show that a simple valence quark
approximation to nucleon spin structure functions is far from the actual
dynamical situation  \cite{Karliner:1999fn}.

It is helpful to categorize the parton distributions as
``intrinsic"---pertain\-ing to
the long-time scale composition of the target hadron, and
``extrinsic",---reflecting
the short-time substructure of the individual quarks and gluons themselves.
Gluons carry a
significant fraction of the proton's spin as well as its momentum.  Since
gluon exchange between valence quarks contributes to the
$p-\Delta$ mass splitting, it follows that the gluon distributions
cannot be solely accounted for by gluon bremsstrahlung from
individual quarks, the
process responsible for DGLAP evolutions of the structure functions.
Similarly,  in the case of heavy quarks, $s\bar s$,
$c \bar c$, $b \bar b$, the diagrams in which the sea quarks are multiply
connected to
the valence quarks are intrinsic to the proton structure itself  \cite{IC}.
The $x$ distribution of intrinsic heavy quarks is peaked at large $x$
reflecting the fact that higher Fock state wavefunctions
containing heavy quarks are maximal when the off-shellness of
the fluctuation is minimized.  The evidence for intrinsic charm at large $x$
in deep inelastic scattering is discussed by Harris \etal  \cite{Harris:1996jx}
Thus neither gluons nor sea
quarks are solely generated by DGLAP evolution, and one cannot define a
resolution scale $Q_0$ where the sea or gluon degrees of freedom can be
neglected.

In these lectures, I shall emphasize the utility of light-cone
Hamiltonian quantization and the light-cone Fock wavefunctions for
representing hadrons in terms of their quark and gluon degrees of freedom.
The fundamental eigenvalue problem of QCD takes the form of a
Heisenberg equation:
\begin{equation}
H^{QCD}_{LC}\ket{\Psi_H} = M^2_H\ket{ \Psi_H}
\end{equation}
where the theory is quantized at fixed light-cone ``time" $\tau = t +
z/c$  \cite{PinskyPauli}.  This representation is the
extension of Schr\"odinger many-body theory to the relativistic domain.
The eigenvalues of the light-cone Hamiltonian $H^{QCD}_{LC}$ is
the square of the hadron masses $M_H$, the discrete spectrum as well as
the bound states.  Each eigenfunction can be decomposed on the complete
basis of eigensolutions $\ket{ n}$ of the free Hamiltonian $H^{0}_{LC} =
H^{QCD}_{LC}(g \to 0)$.  The light-cone Fock projections of the
eigensolution
\begin{equation}
\psi_{n/H}(x_i,\vec k_{\perp i},\lambda_i)=
\VEV{n(x_i, k_{\perp i},\lambda_i)\vert \Psi_H}, i = 1 \cdots n
\end{equation}
encode all of the information of the
hadron in terms of the flavor, helicity, and momentum content of its
quark and gluon constituents.  For example, the proton state has the Fock
expansion
\begin{eqnarray}
\ket p &=& \sum_n \VEV{n\,|\,p}\, \ket n \nonumber \\
&=& \psi^{(\Lambda)}_{3q/p} (x_i,\vec k_{\perp i},\lambda_i)\,
\ket{uud} \\[1ex]
&&+ \psi^{(\Lambda)}_{3qg/p}(x_i,\vec k_{\perp i},\lambda_i)\,
\ket{uudg} + \cdots \nonumber
\label{eq:b}
\end{eqnarray}
representing the expansion of the exact QCD eigenstate on a non-interacting
quark and gluon basis.
The probability amplitude
for each such
$n$-particle state of on-mass shell quarks and gluons in a hadron is given by a
light-cone Fock state wavefunction
$\psi_{n/H}(x_i,\vec k_{\perp i},\lambda_i)$, where the constituents have
longitudinal light-cone momentum fractions
$
x_i ={k^+_i}/{p^+} = (k^0_i+k^z_i)/(p^0+p^z)\ , \sum^n_{i=1} x_i= 1
$,
relative transverse momentum
$\vec k_{\perp i}\,, \sum^n_{i=1}\vec k_{\perp i} = \vec 0_\perp$,
and helicities $\lambda_i.$

The light-cone Fock formalism is derived in the following way:  one
first constructs the light-cone time evolution operator $P^-=P^0-P^z$
and the invariant mass operator $H_{LC}= P^- P^+-P^2_\perp $ in
light-cone gauge $A^+=0$ from the QCD Lagrangian.  The dependent field
theoretic degrees of freedom are eliminated using the QCD equations of
motion.  The total longitudinal momentum
$P^+ = P^0 + P^z$ and transverse momenta
$\vec P_\perp$ are conserved, \ie\ are independent of the interactions.
The
$P^-$ light-cone evolution operator is constructed from the independent
field theoretic degrees of freedom.  The matrix
elements of
$H_{LC}$ on the complete orthonormal basis
$\{\ket{n}\}$ of the free theory $H^0_{LC} = H_{LC}(g=0)$ can then be
constructed.  The matrix elements $\VEV{n\,|\,H_{LC}\,|\,m}$ connect
Fock states differing by 0, 1, or 2 quark or gluon quanta, and they
include the instantaneous quark and gluon contributions imposed by
eliminating dependent degrees of freedom in light-cone gauge.

The LC wavefunctions $\psi_{n/H}(x_i,\vec
k_{\perp i},\lambda_i)$ are universal, process independent, and thus
control all hadronic reactions.  For example, the quark distributions
measured in hard inclusive reactions are
\begin{eqnarray}
q_{\lambda_q/\lambda_p}(x,\Lambda)
&=& \sum_{n,q_a}
\int\prod^n_{j=1} dx_j d^2 k_{\perp j}\sum_{\lambda i}
\vert \psi^{(\Lambda)}_{n/H}(x_i,\vec k_{\perp i},\lambda_i)\vert^2
\\
&& \times \delta\left(1- \sum^n_i x_i\right) \delta^{(2)}
\left(\sum^n_i \vec k_{\perp i}\right)
\delta(x - x_q) \delta_{\lambda_a, \lambda_q}
\Theta(\Lambda^2 - {\cal M}^2_n) \nonumber
\end{eqnarray}
where the sum is over all quarks $q_a$ which match the quantum
numbers, light-cone momentum fraction $x$ and helicity of the probe
quark.  The effective lifetime of each configuration
in the laboratory frame is ${2 P_{\rm lab}/({\M}_n^2- M_p^2}) $ where
$
\M^2_n = \sum^n_{i=1}(k^2_{\perp i} + m^2_i)/x_i < \Lambda^2 $
is the off-shell invariant mass and $\Lambda$ is a global
ultraviolet regulator.
The light-cone momentum integrals are thus limited by requiring that
the invariant mass squared of the constituents of each Fock state is less
than the resolution scale $\Lambda$.  This cutoff serves to define a
factorization scheme for separating hard and soft regimes in both
exclusive and inclusive hard scattering reactions.

A crucial feature of the light-cone formalism is
the fact that the form of the
$\psi^{(\Lambda)}_{n/H}(x_i,
\vec k_{\perp i},\lambda_i)$ is invariant under longitudinal boosts; \ie,\ the
light-cone wavefunctions expressed in the relative coordinates $x_i$ and
$k_{\perp i}$ are independent of the total momentum
$P^+$,
$\vec P_\perp$ of the hadron.
The ensemble
\{$\psi_{n/H}$\} of such light-cone Fock
wavefunctions is a key concept for hadronic physics, providing a conceptual
basis for representing physical hadrons (and also nuclei) in terms of their
fundamental quark and gluon degrees of freedom.  Each Fock state interacts
distinctly; \eg, Fock states with small particle number and small impact
separation have small color dipole moments and can traverse a nucleus with
minimal interactions.  This is the basis for the predictions for ``color
transparency"  \cite{BM}.

Given the
$\psi^{(\Lambda)}_{n/H},$ one can construct any spacelike electromagnetic
or electroweak form factor from the diagonal overlap of the LC
wavefunctions  \cite{BD}.  The natural formalism for
describing the hadronic wavefunctions which enter exclusive and
diffractive amplitudes is the light-cone expansion.
Similarly, the matrix elements of the currents that define quark and gluon
structure functions can be computed from the integrated squares of the LC
wavefunctions  \cite{LB}.

Factorization theorems for hard exclusive, semi-exclusive, and
diffractive processes allow a rigorous separation of soft
non-perturbative dynamics of the bound state hadrons from the hard
dynamics of a perturbatively-calculable quark-gluon scattering
amplitude.

Roughly, the direct proofs of factorization in the light-cone scheme
proceed as follows: In hard inclusive reactions all intermediate states
are divided according to $\M^2_n < \Lambda^2 $ and $\M^2_n < \Lambda^2 $
domains.  The lower region is associated with the quark and gluon
distributions defined from the absolute squares of the LC wavefunctions in
the light cone factorization scheme.  In the high invariant mass
regime, intrinsic transverse momenta can be ignored, so that the
structure of the process at leading power has the form of hard scattering
on collinear quark and gluon constituents, as in the parton model.  The
attachment of gluons from the LC wavefunction to a propagator in the
hard subprocess is power-law suppressed in LC gauge, so that the minimal
$2 \to 2$ quark-gluon subprocesses dominate.  The
higher order loop corrections lead to the DGLAP evolution equations.

It is important to note that the
effective starting point for the PQCD evolution of the structure
functions cannot be taken as a constant
$Q^2_0$ since as
$x \to 1$ the invariant mass $\M_n$ exceeds the resolution scale
$\Lambda$.  Thus in effect, evolution is quenched at $ x \to 1$.  The
anomaly contribution to singlet helicity structure function $g_1(x,Q)$
can be explicitly identified in the LC factorization scheme as due to the
$\gamma^* g \to q
\bar q$ fusion process.  The anomaly contribution would be zero if the
gluon is on shell.  However, if the off-shellness of the state is larger
than the quark pair mass, one obtains the usual anomaly
contribution \cite{Bass:1998rn}.

In exclusive amplitudes, the LC wavefunctions are the interpolating
functions between the quark and gluon states and the hadronic states.
In an
exclusive amplitude involving a hard scale $Q^2$ all intermediate states
can be divided according to
$\M^2_n <
\Lambda^2 < Q^2 $ and $\M^2_n < \Lambda^2 $ invariant mass domains.  The
high invariant mass contributions to the amplitude has the structure of a
hard scattering process
$T_H$ in which the hadrons are replaced by their respective (collinear)
quarks and gluons.  In light-cone gauge only the minimal Fock states
contribute to the leading power-law fall-off of the exclusive amplitude.
The wavefunctions in the lower invariant mass domain can be integrated up
to the invariant mass cutoff $\Lambda$ and replaced by the gauge
invariant distribution amplitudes, $\phi_H(x_i,\Lambda)$.  Final state
and initial state corrections from gluon attachments to lines
connected to the color- singlet distribution amplitudes cancel at
leading twist.

Thus the key non-perturbative input for exclusive
processes is the gauge and frame independent hadron distribution
amplitude  \cite{LB} defined as the integral of the valence (lowest
particle number) Fock wavefunction;
\eg\ for the pion
\begin{equation}
\phi_\pi (x_i,\Lambda) \equiv \int d^2k_\perp\, \psi^{(\Lambda)}_{q\bar
q/\pi} (x_i, \vec k_{\perp i},\lambda)
\label{eq:f1}
\end{equation}
where the global cutoff $\Lambda$ is identified with the resolution $Q$.
The distribution amplitude controls leading-twist exclusive amplitudes
at high momentum transfer, and it can be related to the gauge-invariant
Bethe-Salpeter wavefunction at equal light-cone time $\tau = x^+$.  The
logarithmic evolution of hadron distribution amplitudes
$\phi_H (x_i,Q)$ can be derived from the perturbatively-computable tail
of the valence light-cone wavefunction in the high transverse momentum
regime  \cite{LB}.  Further details are provided in the following sections.

The existence of an exact formalism
provides a basis for systematic approximations and a control over neglected
terms.  For example, one can analyze exclusive semi-leptonic
$B$-decays which involve hard internal momentum transfer using a
perturbative QCD formalism \cite{BHS,Beneke:1999br}
patterned after the analysis
of form
factors at large momentum transfer  \cite{LB}.  The hard-scattering analysis
proceeds
by writing each hadronic wavefunction
as a sum of soft and hard contributions
\begin{equation}
\psi_n = \psi^{{\rm soft}}_n (\M^2_n < \Lambda^2) + \psi^{{\rm hard}}_n
(\M^2_n >\Lambda^2) ,
\end{equation}
where $\M^2_n $ is the invariant mass of the partons in the $n$-particle
Fock state and
$\Lambda$ is the separation scale.
The high internal momentum contributions to the wavefunction $\psi^{{\rm
hard}}_n $ can be calculated systematically from QCD perturbation theory
by iterating the gluon exchange kernel.  Again, the contributions from
high momentum transfer exchange to the
$B$-decay amplitude can then be written as a
convolution of a hard-scattering
quark-gluon scattering amplitude $T_H$ with the distribution
amplitudes $\phi(x_i,\Lambda)$, the valence wavefunctions obtained by
integrating the
constituent momenta up to the separation scale
${\cal M}_n < \Lambda < Q$.  This is the basis for the
perturbative hard-scattering analyses  \cite{BHS,Sz,BABR,Beneke:1999br}.
In the exact analysis, one can
identify the hard PQCD contribution as well as the soft contribution from
the convolution of the light-cone wavefunctions.
Furthermore, the hard-scattering contribution
can be systematically improved.

It is thus important to not only compute the spectrum of hadrons and
gluonic states, but also to determine the wavefunction of each QCD bound
state in terms of its fundamental quark and gluon degrees of freedom.
If we could obtain such nonperturbative solutions of QCD, then we could
compute the quark and gluon structure functions and distribution
amplitudes which control hard-scattering inclusive and exclusive
reactions as well as calculate the matrix elements of currents which
underlie electroweak form factors and the weak decay amplitudes of the
light and heavy hadrons.  The light-cone wavefunctions also determine
the multi-parton correlations which control the distribution of
particles in the proton fragmentation region as well as dynamical higher
twist effects.  Thus one can analyze not only the deep inelastic
structure functions but also the fragmentation of the spectator system.
Knowledge of hadron wavefunctions would also open a window to a deeper
understanding of the physics of QCD at the amplitude level, illuminating
exotic effects of the theory such as color transparency, intrinsic heavy
quark effects, hidden color, diffractive processes, and the QCD van der
Waals interactions.

Can we ever hope to compute the light-cone wavefunctions from first
principles in QCD?  In the Discretized Light-Cone Quantization (DLCQ)
method  \cite{DLCQ}, periodic boundary conditions are introduced in order
to render the set of light-cone momenta
$k^+_i, k_{\perp i}$ discrete.  Solving QCD then becomes reduced to
diagonalizing the mass operator of the theory.  Virtually any
$1+1$ quantum field theory, including ``reduced QCD" (which has both quark and
gluonic degrees of freedom) can be completely solved using
DLCQ  \cite{Kleb,AD}.  The method yields not only the bound-state and continuum
spectrum, but also the light-cone wavefunction for each eigensolution.  The
method is particularly elegant in the case of supersymmetric theories
 \cite{Antonuccio:1999ia}.
The
solutions for the model 1+1 theories can provide an important theoretical
laboratory for testing approximations and QCD-based models.  Recent progress in
DLCQ has been obtained for
$3+1$ theories utilizing Pauli-Villars ghost fields to provide a covariant
regularization.  Broken supersymmetry may be the key method for regulating
non-Abelian theories.  Light-cone gauge $A^+ = 0$ allows one to utilize
only the physical degrees of freedom of the gluon field.  However,
light-cone quantization in Feynman gauge has a number of attractive
features, including manifest covariance and a straightforward passage to
the Coulomb limit in the case of static quarks  \cite{Srivastava:1999gi}.

Light-cone wavefunctions thus are the natural quantities to encode hadron
properties and to bridge the gap between empirical constraints and
theoretical predictions for the bound state solutions.  We can thus
envision a program to construct the hadronic light cone Fock
wavefunctions $\psi_n(x_i, k_{\perp i},
\lambda_i)$ using not only data but constraints such as:

(1) Since the state is far off shell at large invariant mass $\M$,
one can derive rigorous limits on the
$x \to 1$, high $k_\perp$, and high
$\M^2_n$ behavior of the wavefunctions in the perturbative domain.

(2) Ladder relations connecting state of different particle number
follow from the QCD equation of motion and lead to Regge behavior of the
quark and gluon distributions at $x \to 0$.  QED provides a constraint at
$N_C \to 0.$

(3) One can obtain guides to the exact behavior of LC wavefunctions in
QCD from analytic or DLCQ solutions to toy models such as ``reduced"
$QCD(1+1).$

(4) QCD sum rules, lattice gauge theory moments, and QCD inspired models
such as the bag model, chiral theories, provide important constraints.

(5) Since the LC formalism is valid at all scales, one can utilize
empirical constraints such as the measurements of magnetic
moments, axial couplings, form factors, and distribution amplitudes.

(6) In the nonrelativistic limit, the light-cone and many-body
Schr\"odinger theory formalisms must match.

In addition to the light-cone Fock expansion, a number of other
useful theoretical tools are available to eliminate theoretical
ambiguities in QCD predictions:

(1) Conformal symmetry provides a template for
QCD predictions, leading to relations between observables which are
present even in a theory which is not scale invariant.  For example, the natural
representation of distribution amplitudes is in terms of an expansion
of orthonormal conformal functions multiplied by anomalous dimensions
determined by
QCD evolution equations  \cite{Brodsky:1980ny,Muller:1994hg}.
Thus an important guide in QCD
analyses is to identify the underlying conformal relations of QCD which are
manifest if we drop quark masses and effects due to the running of the QCD
couplings.  In fact, if QCD has an infrared fixed point (vanishing of the Gell
Mann-Low function at low momenta), the theory will closely resemble a scale-free
conformally symmetric theory in many applications.

(2) Commensurate scale relations \cite{Brodsky:1995eh} are perturbative
QCD predictions which relate observable to observable at fixed relative
scale, such as the ``generalized Crewther relation"  \cite{Brodsky:1996tb},
which
connects the Bjorken and Gross-Llewellyn Smith deep inelastic scattering sum
rules to measurements of the $e^+ e^-$ annihilation cross section.  The
relations
have no renormalization scale or scheme ambiguity.  The coefficients in
the perturbative series for commensurate scale relations are identical to
those of conformal QCD; thus no infrared renormalons are
present  \cite{Brodsky:1999gm}.  One can identify
the required conformal coefficients at any finite order by
expanding the coefficients of the usual PQCD expansion around a formal
infrared fixed point, as in the Banks-Zak method  \cite{BGGR}.  All
non-conformal
effects are absorbed by fixing the ratio of the respective momentum transfer and
energy scales.  In the case of fixed-point theories, commensurate scale
relations
relate
both the ratio of couplings and the ratio of scales as the fixed point is
approached  \cite{Brodsky:1999gm}.

(3) $\alpha_V$ Scheme.  A natural scheme for defining the QCD
coupling in exclusive and other processes is the $\alpha_V(Q^2)$ scheme defined
from the potential of static heavy quarks.  Heavy-quark lattice gauge theory
can provide highly precise values for the coupling.  All vacuum polarization
corrections due to fermion pairs are then automatically and analytically
incorporated into the Gell Mann-Low function, thus avoiding the problem of
explicitly computing and resumming quark mass corrections related to the running
of the coupling.  The use of a finite effective charge such as
$\alpha_V$ as the expansion parameter also provides a basis for regulating the
infrared nonperturbative domain of the QCD coupling.

(4) The Abelian Correspondence Principle.  One can consider QCD
predictions as analytic functions of the number of colors $N_C$ and
flavors $N_F$.  In particular, one can show at all orders of
perturbation theory that PQCD predictions reduce to those of an Abelian
theory at $N_C \to 0$ with ${\widehat \alpha} = C_F \alpha_s$ and
${\widehat N_F} = N_F/T C_F$ held fixed  \cite{Brodsky:1997jk}.  There is
thus a deep connection between QCD processes and their corresponding QED
analogs.

\section{Discretized Light-Cone Quantization}

Solving a quantum field theory such as QCD is clearly not easy.
However, highly nontrivial, one-space one-time relativistic quantum
field theories which mimic many of the features of QCD, have already
been completely solved using light-cone Hamiltonian
methods  \cite{PinskyPauli}.  Virtually any (1+1) quantum field theory can
be solved using the method of Discretized Light-Cone-Quantization
(DLCQ)  \cite{DLCQ,Brodsky:1991ir} where the matrix elements
$\VEV{n\,|\,H^{\Lambda)}_{LC}\,|\,m}$, are made discrete in momentum
space by imposing periodic or anti-periodic boundary conditions in
$x^-=x^0 - x^z$ and $\vec x_\perp$.  Upon diagonalization of $H_{LC}$,
the eigenvalues provide the invariant mass of the bound states and
eigenstates of the continuum.  In DLCQ, the Hamiltonian $H_{LC}$, which
can be constructed from the Lagrangian using light-cone time
quantization, is completely diagonalized, in analogy to Heisenberg's
solution of the eigenvalue problem in quantum mechanics.  The quantum
field theory problem is rendered discrete by imposing periodic or
anti-periodic boundary conditions.  The eigenvalues and eigensolutions
of collinear QCD then give the complete spectrum of hadrons, nuclei, and
gluonium and their respective light-cone wavefunctions.  A beautiful
example is ``collinear" QCD:  a variant of $QCD(3+1)$ defined by
dropping all of interaction terms in $H^{QCD}_{LC}$ involving transverse
momenta  \cite{Kleb}.  Even though this theory is effectively
two-dimensional, the transversely-polarized degrees of freedom of the
gluon field are retained as two scalar fields.  Antonuccio and Dalley
 \cite{AD} have used DLCQ to solve this theory.  The diagonalization of
$H_{LC}$ provides not only the complete bound and continuum spectrum of
the collinear theory, including the gluonium states, but it also yields
the complete ensemble of light-cone Fock state wavefunctions needed to
construct quark and gluon structure functions for each bound state.
Although the collinear theory is a drastic approximation to physical
$QCD(3+1)$, the phenomenology of its DLCQ solutions demonstrate general
gauge theory features, such as the peaking of the wavefunctions at
minimal invariant mass, color coherence and the helicity retention of
leading partons in the polarized structure functions at $x\rightarrow
1$.  The solutions of the quantum field theory can be obtained for
arbitrary coupling strength, flavors, and colors.

In practice it is essential to introduce an ultraviolet regulator in
order to limit the total range of $\VEV{n\,|\,H_{LC}\,|\,m}$, such as
the ``global" cutoff in the invariant mass of the free Fock state.  One
can also introduce a ``local" cutoff to limit the change in invariant
mass $|\M^2_n-\M^2_m| < \Lambda^2_{\rm local}$ which provides
spectator-independent regularization of the sub-divergences associated
with mass and coupling renormalization.  Recently, Hiller, McCartor, and
I have shown \cite{Brodsky:1998hs} that the Pauli-Villars method has
advantages for regulating light-cone quantized Hamiltonian theory.
A spectrum of Pauli-Villars ghost fields satisfying three spectral
conditions will regulate the interactions in the ultraviolet, while at
same time avoiding spectator-dependent renormalization and preserving
chiral symmetry. We have also shown that model theories in $3+ !$
dimensions can be successfully solved with such regularization.

Although gauge theories are usually quantized on the
light-cone in light-cone gauge $A^+=0$, it is also possible and
interesting to quantize the theory in Feynman gauge
\cite{Srivastava:1999gi}.  Covariant gauges are advantageous since they
preserve the rotational symmetry of the gauge interactions.

The natural renormalization scheme for the QCD coupling is
$\alpha_V(Q)$, the effective charge defined from the scattering of two
infinitely-heavy quark test charges.
This is discussed in more detail below.
The renormalization scale can then
be determined from the virtuality of the exchanged momentum, as in the
BLM and commensurate scale
methods  \cite{BLM,Brodsky:1995eh,Brodsky:1996tb,BJPR}.  Similar
effective charges have been proposed by Watson  \cite{Watson:1997fg} and
Czarnecki \etal  \cite{Czarnecki:1998sz}

In principle, we could also construct the wavefunctions of QCD(3+1)
starting with collinear QCD(1+1) solutions by systematic perturbation
theory in $\Delta H$, where $\Delta H$ contains the terms which
produce particles at non-zero $k_\perp$, including the terms linear and
quadratic in the transverse momenta $\longvec k_{\perp i}$ which are
neglected in the Hamilton $H_0$ of collinear QCD.  We can write the exact
eigensolution of the full Hamiltonian as
\[ \psi_{(3+1)} = \psi_{(1+1)} + \frac{1}{M^2-H + i \epsilon }\,
\Delta H\, \psi_{(1+1)} \ , \]
where
\[\frac{1}{M^2-H + i \epsilon }
 = {\frac{1}{M^2-H_0 + i \epsilon }} +
{\frac{1}{M^2-H+ i \epsilon }}\Delta H{\frac{1}{M^2-H_0 + i \epsilon }} \]
can be represented as the continued iteration of the Lippmann Schwinger
resolvant.
Note that the matrix
$(M^2-H_0)^{-1}$ is known to any desired precision from the DLCQ solution
of collinear QCD.

\section{Electroweak Matrix Elements and Light-Cone Wavefunctions}

\vspace{.5cm}
\begin{figure}[htb]
\begin{center}
\leavevmode
\epsfbox{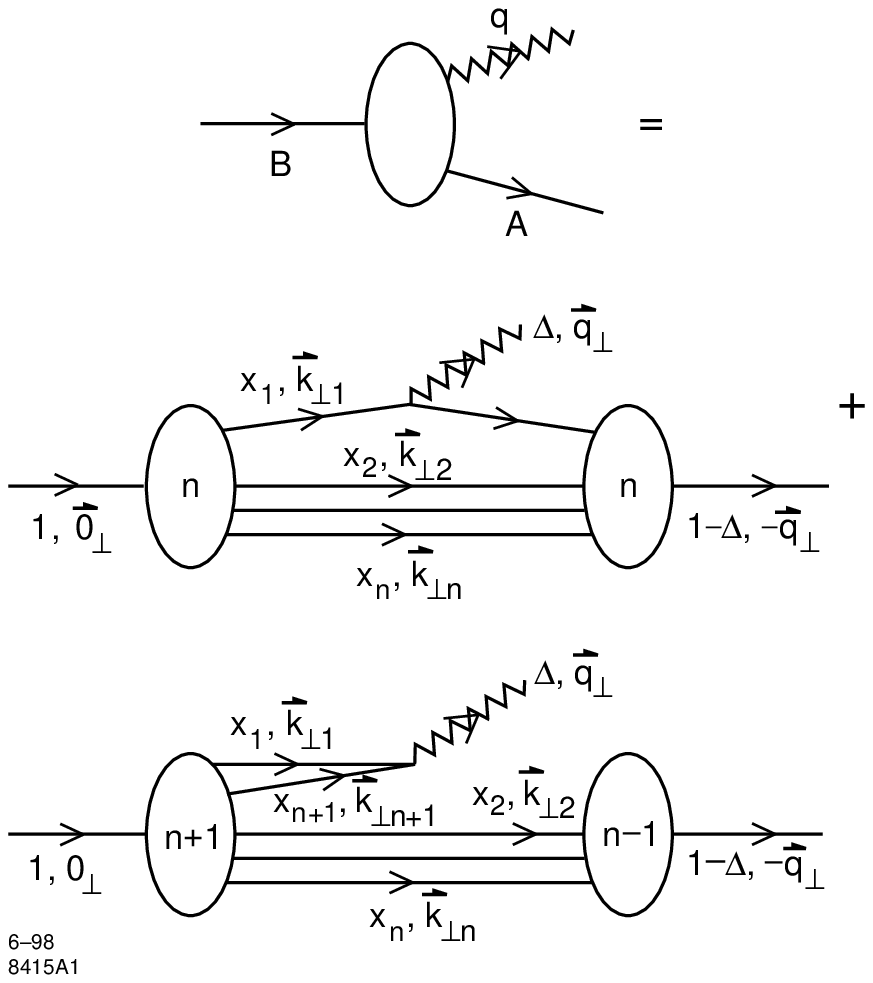}
\end{center}
\caption[*]{Exact representation of electroweak decays and time-like form
factors in the
light-cone Fock representation.
}
\label{fig1}
\end{figure}

Another remarkable advantage of the light-cone formalism is that
exclusive semileptonic
$B$-decay amplitudes such as $B\rightarrow A \ell \bar{\nu}$ can be
evaluated exactly \cite{Brodsky:1998hn}.
The timelike decay matrix elements require the computation of the
diagonal matrix element $n \rightarrow n$ where parton number is
conserved, and the off-diagonal $n+1\rightarrow n-1$ convolution where
the current operator annihilates a $q{\bar{q'}}$ pair in the initial $B$
wavefunction.  See Fig.  \ref{fig1}.  This term is a consequence of the
fact that the time-like decay $q^2 = (p_\ell + p_{\bar{\nu}} )^2 > 0$
requires a positive light-cone momentum fraction $q^+ > 0$.  Conversely
for space-like currents, one can choose $q^+=0$, as in the
Drell-Yan-West representation of the space-like electromagnetic form
factors.  However, as can be seen from the explicit analysis of the form
factor in a perturbative model, the off-diagonal convolution can yield a
nonzero $q^+/q^+$ limiting form as $q^+ \rightarrow 0$.  This extra term
appears specifically in the case of ``bad" currents such as $J^-$ in
which the coupling to $q\bar q$ fluctuations in the light-cone
wavefunctions are favored.  In effect, the $q^+ \rightarrow 0$ limit
generates $\delta(x)$ contributions as residues of the $n+1\rightarrow
n-1$ contributions.  The necessity for such ``zero mode" $\delta(x)$ terms
has been noted by Chang, Root and Yan \cite{CRY}, Burkardt \cite{BUR},
and Ji and Choi \cite{Choi:1998nf}.

The off-diagonal $n+1 \rightarrow n-1$ contributions give a new
perspective for the physics of $B$-decays.  A semileptonic decay
involves not only matrix elements where a quark changes flavor, but also
a contribution where the leptonic pair is created from the annihilation
of a $q {\bar{q'}}$ pair within the Fock states of the initial $B$
wavefunction.  The semileptonic decay thus can occur from the
annihilation of a nonvalence quark-antiquark pair in the initial hadron.
This feature will carry over to exclusive hadronic $B$-decays, such as
$B^0 \rightarrow \pi^-D^+$.  In this case the pion can be produced from
the coalescence of a $d\bar u$ pair emerging from the initial higher
particle number Fock wavefunction of the $B$.  The $D$ meson is then
formed from the remaining quarks after the internal exchange of a $W$
boson.

In principle, a precise evaluation of the hadronic matrix elements
needed for $B$-decays and other exclusive electroweak decay amplitudes
requires knowledge of all of the light-cone Fock wavefunctions of the
initial and final state hadrons.  In the case of model gauge theories
such as QCD(1+1) \cite{Horn} or collinear QCD \cite{AD} in one-space and
one-time dimensions, the complete evaluation of the light-cone
wavefunction is possible for each baryon or meson bound-state using the
DLCQ method.  It would be interesting to use such solutions as a model
for physical $B$-decays.

The existence of an exact formalism for electroweak matrix elements
gives a basis for systematic approximations and a control over
neglected terms.  For example, one can analyze exclusive semileptonic
$B$-decays which involve hard internal momentum transfer using a
perturbative QCD formalism patterned after the analysis of form factors at
large momentum transfer \cite{LB}.  The hard-scattering analysis proceeds by
writing each hadronic wavefunction as a sum of soft and hard contributions
\begin{equation}
\psi_n = \psi^{{\rm soft}}_n (\M^2_n < \Lambda^2) + \psi^{{\rm hard}}_n
(\M^2_n >\Lambda^2) ,
\end{equation}
where $\M^2_n $ is the invariant mass of the partons in the $n$-particle
Fock state and
$\Lambda$ is the separation scale.
The high internal momentum contributions to the wavefunction $\psi^{{\rm
hard}}_n $ can be calculated systematically from QCD perturbation theory
by iterating the gluon exchange kernel.  The contributions from high
momentum transfer exchange to the
$B$-decay amplitude can then be written as a convolution of a hard scattering
quark-gluon scattering amplitude $T_H$ with the distribution
amplitudes $\phi(x_i,\Lambda)$, the valence wavefunctions obtained by
integrating the
constituent momenta up to the separation scale
${\cal M}_n < \Lambda < Q$.  This is the basis for the perturbative hard
scattering analyses \cite{BHS,Sz,BALL,BABR}.
In the exact analysis, one can
identify the hard PQCD contribution as well as the soft contribution from
the convolution of the light-cone wavefunctions.  Furthermore, the hard
scattering contribution can be systematically improved.  For example, off-shell
effects can be retained in the evaluation of
$T_H$ by utilizing the exact light-cone energy denominators.

Given the solution
for the hadronic wavefunctions $\psi^{(\Lambda)}_n$ with $\M^2_n <
\Lambda^2$, one can construct the wavefunction in the hard regime with
$\M^2_n > \Lambda^2$ using projection operator techniques \cite{LB}.  The
construction can be done perturbatively in QCD since only high invariant mass,
far off-shell matrix elements are involved.  One can use this method to
derive the physical properties of the LC wavefunctions and their matrix elements
at high invariant mass.  Since $\M^2_n = \sum^n_{i=1}
\left(\frac{k^2_\perp+m^2}{x}\right)_i $, this method also allows the derivation
of the asymptotic behavior of light-cone wavefunctions at large $k_\perp$, which
in turn leads to predictions for the fall-off of form factors and other
exclusive
matrix elements at large momentum transfer, such as the quark counting rules
for predicting the nominal power-law fall-off of two-body scattering amplitudes
at fixed
$\theta_{cm}$ \cite{BrodskyLepage}.  The phenomenological successes of
these rules
can be understood within QCD if the coupling
$\alpha_V(Q)$ freezes in a range of relatively
small momentum transfer \cite{BJPR}.

\section{Other Applications of Light-Cone Quantization to QCD Phenomenology}

{\it Diffractive vector meson photoproduction.} The
light-cone Fock wavefunction representation of hadronic amplitudes
allows a simple eikonal analysis of diffractive high energy processes, such as
$\gamma^*(Q^2) p \to \rho p$, in terms of the virtual photon and the vector
meson Fock state light-cone wavefunctions convoluted with the $g p \to g p$
near-forward matrix element \cite{BGMFS}.  One can easily show that only small
transverse size $b_\perp \sim 1/Q$ of the vector meson distribution
amplitude is involved.  The hadronic interactions are minimal, and thus the
$\gamma^*(Q^2) N \to
\rho N$ reaction can occur coherently throughout a nuclear target in reactions
without absorption or shadowing.  The $\gamma^* A \to V A$ process thus
is a laboratory for testing QCD color transparency \cite{BM}.
This is discussed further in the next section.

{\it Regge behavior of structure functions.} The light-cone wavefunctions
$\psi_{n/H}$ of a hadron are not independent of each other, but rather are
coupled via the equations of motion.  Antonuccio, Dalley and I
\cite{ABD} have used the constraint of finite ``mechanical'' kinetic energy to
derive ``ladder relations" which interrelate the light-cone wavefunctions of
states differing by one or two gluons.  We then use these relations to
derive the
Regge behavior of both the polarized and unpolarized structure functions at $x
\rightarrow 0$, extending Mueller's derivation of the BFKL hard
QCD pomeron from the properties of heavy quarkonium light-cone wavefunctions at
large $N_C$ QCD \cite{Mueller}.

{\it Structure functions at large $x_{bj}$.} The behavior of structure functions
where one quark has the entire momentum requires the knowledge of LC
wavefunctions
with $x \rightarrow 1$ for the struck quark and $x \rightarrow 0$ for the
spectators.  This is a highly off-shell configuration, and thus one can
rigorously derive quark-counting and helicity-retention rules for the
power-law behavior of
the polarized and unpolarized quark and gluon distributions in the $x
\rightarrow 1$ endpoint domain.  It is interesting to note that the evolution
of structure functions is minimal in this domain because the struck quark
is highly virtual as $x\rightarrow 1$; \ie\ the starting point $Q^2_0$ for
evolution
cannot be held fixed, but must be larger than a scale of order
$(m^2 + k^2_\perp)/(1-x)$ \cite{LB,BrodskyLepage,Dmuller}.

{\it Intrinsic gluon and heavy quarks.}
The main features of the heavy sea quark-pair contributions of the Fock
state expansion of light hadrons can also be derived from perturbative QCD,
since $\M^2_n$ grows with
$m^2_Q$.  One identifies two contributions to the heavy quark sea, the
``extrinsic'' contributions which correspond to ordinary gluon splitting, and
the ``intrinsic" sea which is multi-connected via gluons to the valence quarks.
The intrinsic sea is thus sensitive to the hadronic bound state
structure \cite{IC}.  The maximal contribution of the
intrinsic heavy quark occurs at $x_Q \simeq {m_{\perp Q}/ \sum_i m_\perp}$
where $m_\perp = \sqrt{m^2+k^2_\perp}$;
\ie\ at large $x_Q$, since this minimizes the invariant mass $\M^2_n$.  The
measurements of the charm structure function by the EMC experiment are
consistent with intrinsic charm at large $x$ in the nucleon with a
probability of order $0.6 \pm 0.3 \% $ \cite{Harris:1996jx}.  Similarly, one can
distinguish intrinsic
gluons which are associated with multi-quark interactions and extrinsic gluon
contributions associated with quark substructure \cite{BS}.  One can also
use this
framework to isolate the physics of the anomaly contribution to the Ellis-Jaffe
sum rule.

{\it Materialization of far-off-shell configurations.}
In a high energy hadronic collisions, the highly-virtual states of a hadron
can be
materialized into physical hadrons simply by the soft interaction of any of the
constituents \cite{BHMT}.  Thus a proton state with intrinsic charm $\ket{ u
u d \bar c c}$ can be materialized, producing a $J/\psi$ at large $x_F$,
by the
interaction of a light-quark in the target.  The production
occurs on the front-surface of a target nucleus, implying an $A^{2/3}$
$J/\psi$ production cross section at large $x_F,$  which is consistent
with experiment, such as Fermilab experiments E772 and E866.

{\it Rearrangement mechanism in heavy quarkonium decay.}
It is usually assumed that a heavy quarkonium state such as the
$J/\psi$ always decays to light hadrons via the annihilation of its heavy quark
constituents to gluons.  However, as Karliner and I \cite{Brodsky:1997fj}
have recently
shown, the transition $J/\psi \to \rho
\pi$ can also occur by the rearrangement of the $c \bar c$ from the $J/\psi$
into the $\ket{ q \bar q c \bar c}$ intrinsic charm Fock state of the $\rho$ or
$\pi$.  On the other hand, the overlap rearrangement integral in the
decay $\psi^\prime \to \rho \pi$ will be suppressed since the intrinsic
charm Fock state radial wavefunction of the light hadrons will evidently
not have nodes in its radial wavefunction.  This observation gives a
natural explanation of the long-standing puzzle why the $J/\psi$ decays
prominently to two-body pseudoscalar-vector final states, whereas the
$\psi^\prime$ does not.

{\it Asymmetry of intrinsic heavy quark sea.}
The higher Fock state of the proton $\ket{u u d s \bar s}$ should
resemble a $\ket{ K \Lambda}$ intermediate state, since this minimizes its
invariant mass $\M$.  In such a state, the
strange quark has a higher mean momentum fraction $x$ than the $\bar
s$  \cite{Warr,Signal,BMa}.  Similarly, the helicity intrinsic strange
quark in this configuration will be anti-aligned with the helicity of the
nucleon
 \cite{Warr,BMa}.  This $Q \leftrightarrow \bar Q$ asymmetry
is a striking feature of the intrinsic heavy-quark sea.

{\it Comover phenomena.}
Light-cone wavefunctions describe not only the partons that interact in a hard
subprocess but also the associated partons freed from the projectile.  The
projectile partons which are comoving (\ie, which have similar rapidity) with
final state quarks and gluons can interact strongly producing (a) leading
particle effects, such as those seen in open charm hadroproduction; (b)
suppression of quarkonium \cite{BrodskyMueller} in favor of open heavy hadron
production, as seen in the E772 experiment; (c) changes in color configurations
and selection rules in quarkonium hadroproduction, as has been emphasized by
Hoyer and Peigne  \cite{Hoyer:1998ha}.  All of these effects violate the
usual ideas of factorization for inclusive reactions.  Further, more than
one parton from the
projectile can enter the hard subprocess, producing dynamical higher twist
contributions, as seen for example in
Drell-Yan experiments  \cite{BrodskyBerger,Brandenburg}.

{\it Jet hadronization in light-cone QCD.}
One of the goals of nonperturbative analysis in QCD is to compute jet
hadronization from first principles.  The DLCQ solutions provide a possible
method to accomplish this.  By inverting the DLCQ solutions, we can write the
``bare'' quark state of the free theory as
$\ket{q_0} = \sum \ket n \VEV{n\,|\,q_0}$
 where now $\{\ket n\}$ are the exact DLCQ eigenstates of
$H_{LC}$, and
$\VEV{n\,|\,q_0}$ are the DLCQ projections of the eigensolutions.  The expansion
in automatically infrared and ultraviolet regulated if we impose global cutoffs
on the DLCQ basis:
$\lambda^2 < \Delta\M^2_n < \Lambda^2
$
where $\Delta\M^2_n = \M^2_n-(\Sigma \M_i)^2$.  It would be
interesting to study jet hadronization at the amplitude level for
the existing DLCQ solutions to QCD (1+1) and collinear QCD.

{\it Hidden Color.}
The deuteron form factor at high $Q^2$ is sensitive to wavefunction
configurations where all six quarks overlap within an impact
separation $b_{\perp i} < {\cal O} (1/Q);$ the leading power-law
fall off predicted by QCD is $F_d(Q^2) = f(\alpha_s(Q^2))/(Q^2)^5$,
where, asymptotically, $f(\alpha_s(Q^2)) \propto
\alpha_s(Q^2)^{5+2\gamma}$  \cite{Brodsky:1976rz}.  The derivation of the
evolution equation for the deuteron distribution amplitude and its
leading anomalous dimension $\gamma$ is given in Ref.  \cite{bjl83}
In general, the six-quark wavefunction of a deuteron
is a mixture of five different color-singlet states.  The dominant
color configuration at large distances corresponds to the usual
proton-neutron bound state.  However at small impact space
separation, all five Fock color-singlet components eventually
acquire equal weight, \ie, the deuteron wavefunction evolves to
80\%\ ``hidden color.'' The relatively large normalization of the
deuteron form factor observed at large $Q^2$ points to sizable
hidden color contributions  \cite{Farrar:1991qi}.

{\it Spin-Spin Correlations in Nucleon-Nucleon
Scattering and the Charm Threshold.}
One of the most striking anomalies in elastic proton-proton
scattering is the large spin correlation $A_{NN}$ observed at large
angles  \cite{krisch92}.  At $\sqrt s \simeq 5 $ GeV, the rate for
scattering with incident proton spins parallel and normal to the
scattering plane is four times larger than that for scattering with
anti-parallel polarization.  This strong polarization correlation can
be attributed to the onset of charm production in the intermediate
state at this energy  \cite{Brodsky:1988xw}.  The intermediate state 
$\ket{u u d u u d c \bar c}$ has odd intrinsic parity and couples to
the $J=S=1$ initial state, thus strongly enhancing scattering when
the incident projectile and target protons have their spins parallel
and normal to the scattering plane.  The charm threshold can also
explain the anomalous change in color transparency observed at the
same energy in quasi-elastic $ p p$ scattering.  A crucial test is
the observation of open charm production near threshold with a
cross
section of order of $1 \mu$b.

\section{Features of Hard Exclusive Processes in QCD }

Exclusive and diffractive reactions are highly challenging to analyze in QCD
since they require knowledge of the hadron wavefunctions at the amplitude level.
There has been much progress analyzing
exclusive and diffractive reactions at large momentum transfer from first
principles
in QCD.  Rigorous statements can be made on the basis of asymptotic freedom and
factorization theorems which separate the underlying hard quark and gluon
subprocess amplitude from the nonperturbative physics incorporated into the
process-independent hadron distribution amplitudes
$\phi_H(x_i,Q)$  \cite{LB}, the valence light-cone
wavefunctions integrated over $k^2_\perp<Q^2$.

In general, hard exclusive hadronic amplitudes such as quarkonium
decay, heavy hadron decay,  and scattering amplitudes where hadrons
are scattered with large momentum transfer can be factorized at leading
power as a
convolution of distribution amplitudes and hard-scattering quark/gluon matrix
elements \cite{LB}
\begin{eqnarray}
\M_{\rm Hadron} &=& \prod_H \sum_n \int
\prod^{n}_{i=1} d^2k_\perp \prod^{n}_{i=1}dx\, \delta
\left(1-\sum^n_{i=1}x_i\right)\, \delta
\left(\sum^n_{i=1} \vec k_{\perp i}\right) \nonumber \\[2ex]
&& \times \psi^{(\lambda)}_{n/H} (x_i,\vec k_{\perp i},\lambda_i)\,
T_H^{(\Lambda)} \ .
\label{eq:e}
\end{eqnarray}
Here $T_H^{(\Lambda)}$ is the underlying quark-gluon
subprocess scattering amplitude in which the (incident and final) hadrons are
replaced by their respective quarks and gluons with momenta $x_ip^+$, $x_i\vec
p_{\perp}+\vec k_{\perp i}$ and invariant mass above the
separation scale $\M^2_n > \Lambda^2$.  At large $Q^2$ one can integrate
over the
transverse momenta.  The leading power behavior of the hard quark-gluon
scattering
amplitude $T_H(\vec k_{\perp i}=0),$ defined for the case where the quarks are
effectively collinear with their respective parent hadron's momentum,
provides the
basic scaling and helicity features of the hadronic amplitude.  The essential
part of the hadron wavefunction is the hadronic distribution amplitudes
 \cite{LB}, defined as the integral over transverse momenta of the valence
(lowest
particle number) Fock wavefunction, as defined in Eq. (\ref{eq:f1})
where the global cutoff $\Lambda$ is identified with the
resolution $Q$.  The distribution amplitude controls leading-twist exclusive
amplitudes at high momentum transfer, and it can be related to the
gauge-invariant Bethe-Salpeter wavefunction at equal light-cone time
$\tau = x^+$.

The $\log Q$ evolution of the hadron distribution amplitudes
$\phi_H (x_i,Q)$ can be derived from the
perturbatively-computable tail of the valence light-cone wavefunction in the
high transverse momentum regime.  The LC ultraviolet
regulators provide a factorization scheme for elastic and inelastic
scattering, separating the hard dynamical contributions with invariant mass
squared $\M^2 > \Lambda^2_{\rm global}$ from the soft physics with
$\M^2 \le \Lambda^2_{\rm global}$ which is incorporated in the
nonperturbative LC wavefunctions.  The DGLAP evolution of quark and gluon
distributions can also be derived in an analogous way by computing the
variation of
the Fock expansion with respect to $\Lambda^2$.
The renormalization scale ambiguities in hard-scattering
amplitudes via commensurate scale
relations \cite{Brodsky:1995eh,Brodsky:1996tb,Brodsky:1999gm} which connect the
couplings entering exclusive amplitudes to the
$\alpha_V$ coupling which controls the QCD heavy quark
potential  \cite{Brodsky:1998dh}.

The features of exclusive processes to leading power in the transferred
momenta are well known:

(1) The leading power fall-off is given by dimensional counting rules for
the hard-scattering amplitude: $T_H \sim 1/Q^{n-1}$, where $n$ is the total
number
of fields
(quarks, leptons, or gauge fields) participating in the hard
scattering  \cite{BF,Matveev:1973ra}.  Thus the reaction is dominated by
subprocesses
and Fock states involving the minimum number of interacting fields.  The
hadronic
amplitude follows this fall-off modulo logarithmic corrections from the
running of
the QCD coupling, and the evolution of the hadron distribution amplitudes.
In some
cases, such as large angle $p p \to p p $ scattering, pinch contributions from
multiple hard-scattering processes must also be
included  \cite{Landshoff:1974ew}.
The general success of dimensional counting rules implies that the
effective coupling
$\alpha_V(Q^*)$ controlling the gluon exchange propagators in
$T_H$ are frozen in the infrared, \ie, have an infrared fixed point, since the
effective momentum transfers $Q^*$ exchanged by the gluons are often a
small fraction
of the overall momentum transfer  \cite{Brodsky:1998dh}. The pinch contributions
are then suppressed by a factor decreasing faster than a fixed power  \cite{BF}.

(2) The leading power dependence is given by hard-scattering amplitudes $T_H$
which conserve quark helicity  \cite{Brodsky:1981kj,Chernyak:1999cj}.  Since the
convolution of $T_H$ with the light-cone wavefunctions projects out states with
$L_z=0$, the leading hadron amplitudes conserve hadron helicity; \ie, the
sum of
initial and final hadron helicities are conserved.  Hadron helicity conservation
thus follows from the underlying chiral structure of QCD.

(3) Since the convolution of the hard scattering amplitude $T_H$ with the
light-cone
wavefunctions projects out the valence states with small impact parameter,
the essential part of the hadron wavefunction entering a hard exclusive
amplitude has
a small color dipole moment.  This leads to the absence of initial or final
state
interactions among the scattering hadrons as well as the color transparency
of quasi-elastic interactions in a nuclear target  \cite{BM,Frankfurt:1992dx}.
Color transparency reflects the underlying gauge theoretic basis
of the strong interactions.
For example, the amplitude for diffractive vector meson photoproduction
$\gamma^*(Q^2) p \to \rho p$, can be written as convolution of the virtual
photon and
the vector meson Fock state light-cone wavefunctions the $g p \to g p$
near-forward matrix element  \cite{BGMFS}.  One can easily show that only
small transverse size $b_\perp \sim 1/Q$ of the vector meson distribution
amplitude is involved.  The sum over the interactions of the exchanged
gluons tend to
cancel reflecting its small color dipole moment.  Since the hadronic
interactions are
minimal,  the
$\gamma^*(Q^2) N \to
\rho N$ reaction at large $Q^2$ can occur coherently throughout a nuclear
target in
reactions without absorption or final state interactions.  The $\gamma^*A
\to V A$ process thus provides a natural framework for testing QCD color
transparency.  Evidence for color transparency in such reactions has been
found by Fermilab experiment E665  \cite{Adams:1997bh}.

(4) The evolution
equations for distribution amplitudes which incorporate the operator product
expansion, renormalization group invariance, and conformal symmetry
 \cite{LB,Brodsky:1980ny,Muller:1994hg,Ball:1998ff,Braun:1999te}.

(5) Hidden color
degrees of freedom in nuclear wavefunctions reflects the complex color
structure of hadron and nuclear wavefunctions  \cite{bjl83}.  The hidden color
increases the normalization of nuclear amplitudes such as the deuteron form
factor at large momentum transfer.

The
field of analyzable exclusive processes has recently been expanded to a new
range of QCD processes, such as the highly virtual diffractive processes
$\gamma^* p \to \rho p$  \cite{BGMFS,Collins:1997hv}, and semi-exclusive
processes such as
$\gamma^* p \to \pi^+ X$  \cite{acw,Brodsky:1998sr,BB} where the $\pi^+$ is
produced in isolation at large $p_T$.  An important new
application of the perturbative QCD analysis of exclusive processes is the
recent analysis of hard $B$ decays such as $B \to \pi \pi$ by Beneke, {\it et
al.} \cite{Beneke:1999br}

Exclusive hard-scattering reactions and hard diffractive reactions are now
giving
a valuable window into the structure and
dynamics of hadronic amplitudes.  Recent measurements of the
photon-to-pion transition form factor at CLEO  \cite{Gronberg:1998fj}, the
diffractive dissociation of pions into jets at Fermilab  \cite{E791},
diffractive vector meson leptoproduction at Fermilab and HERA, and the new
program
of experiments on exclusive proton and deuteron processes at Jefferson
Laboratory
are now yielding fundamental information on hadronic wavefunctions,
particularly the
distribution amplitude of mesons.  Such information is also critical for
interpreting exclusive heavy hadron decays and the matrix elements and
amplitudes
entering $CP$-violating processes at the $B$ factories.

There has been much progress analyzing
exclusive and diffractive reactions at large momentum transfer from first
principles
in QCD.  Rigorous statements can be made on the basis of asymptotic freedom and
factorization theorems which separate the underlying hard quark and gluon
subprocess amplitude from the nonperturbative physics incorporated into the
process-independent hadron distribution amplitudes
$\phi_H(x_i,Q)$  \cite{LB}.  An important new
application is the
recent analysis of hard exclusive
$B$ decays by Beneke, {\it et al.}  \cite{Beneke:1999br} Key features of such
analyses are: (a) evolution equations for distribution amplitudes which
incorporate the operator product expansion, renormalization group
invariance, and conformal symmetry
 \cite{LB,Brodsky:1980ny,Brodsky:1986ve,Muller:1994hg,Ball:1998ff,Braun:1999te};
(b) hadron helicity conservation which follows from the underlying chiral
structure of QCD  \cite{Brodsky:1981kj}; (c) color transparency, which
eliminates corrections to hard exclusive amplitudes from initial and final state
interactions at leading power and reflects the underlying gauge theoretic basis
for the strong interactions  \cite{BM,Frankfurt:1992dx}; and (d) hidden color
degrees of freedom in nuclear wavefunctions, which reflects the color
structure of hadron and nuclear wavefunctions  \cite{bjl83}.  There have
also been
recent advances eliminating renormalization scale ambiguities in hard-scattering
amplitudes via commensurate scale
relations \cite{Brodsky:1995eh,Brodsky:1996tb,Brodsky:1999gm} which connect the
couplings entering exclusive amplitudes to the
$\alpha_V$ coupling which controls the QCD heavy quark
potential  \cite{Brodsky:1998dh}.  The postulate that the QCD coupling has an
infrared fixed-point can explain the applicability of
conformal scaling and dimensional counting
rules to physical QCD processes  \cite{BF,Matveev:1973ra,Brodsky:1998dh}.  The
field of analyzable exclusive processes has recently been expanded to a new
range
of QCD processes, such as electroweak decay
amplitudes, highly virtual diffractive processes such as
$\gamma^* p \to \rho p$  \cite{BGMFS,Collins:1997fb}, and semi-exclusive
processes such as
$\gamma^* p \to \pi^+ X$  \cite{acw,Brodsky:1998sr,BB} where the $\pi^+$ is
produced in isolation at large $p_T$.

The natural
renormalization scheme
for the QCD coupling in hard exclusive processes is $\alpha_V(Q)$, the
effective charge defined from the scattering of two infinitely-heavy
quark test charges.  The renormalization scale can then be determined
from the virtuality of the exchanged momentum of the gluons, as in the BLM and
commensurate scale
methods  \cite{BLM,Brodsky:1995eh,Brodsky:1996tb,Brodsky:1999gm}.  We
will discuss these theoretical tools and methods in the later sections.

The main features of exclusive processes to leading power in the
transferred momenta are:

(1) The leading power fall-off is given by dimensional counting rules for
the hard-scattering amplitude: $T_H \sim 1/Q^{n-1}$, where $n$ is the total
number
of fields
(quarks, leptons, or gauge fields) participating in the hard
scattering  \cite{BF,Matveev:1973ra}.  Thus the reaction is dominated by
subprocesses
and Fock states involving the minimum number of interacting fields.  The
hadronic
amplitude follows this fall-off modulo logarithmic corrections from the
running of
the QCD coupling, and the evolution of the hadron distribution amplitudes.
In some
cases, such as large angle $p p \to p p $ scattering, pinch contributions from
multiple hard-scattering processes must also be
included  \cite{Landshoff:1974ew}.
The general success of dimensional counting rules implies that the
effective coupling
$\alpha_V(Q^*)$ controlling the gluon exchange propagators in
$T_H$ are frozen in the infrared, \ie, have an infrared fixed point, since the
effective momentum transfers $Q^*$ exchanged by the gluons are often a
small fraction
of the overall momentum transfer  \cite{Brodsky:1998dh}.  The pinch
contributions
are suppressed by a factor decreasing faster than a fixed power  \cite{BF}.

(2) The leading power dependence is given by hard-scattering amplitudes $T_H$
which conserve quark helicity  \cite{Brodsky:1981kj,Chernyak:1999cj}.  Since the
convolution of $T_H$ with the light-cone wavefunctions projects out states with
$L_z=0$, the leading hadron amplitudes conserve hadron helicity; \ie, the
sum of
initial and final hadron helicities are conserved.

(3) Since the convolution of the hard scattering amplitude $T_H$ with the
light-cone
wavefunctions projects out the valence states with small impact parameter,
the essential part of the hadron wavefunction entering a hard exclusive
amplitude has
a small color dipole moment.  This leads to the absence of initial or final
state
interactions among the scattering hadrons as well as the color transparency.
of quasi-elastic interactions in a nuclear target  \cite{BM,Frankfurt:1992dx}.
For example, the amplitude for diffractive vector meson photoproduction
$\gamma^*(Q^2) p \to \rho p$, can be written as convolution of the virtual
photon and
the vector meson Fock state light-cone wavefunctions the $g p \to g p$
near-forward matrix element  \cite{BGMFS}.  One can easily show that only
small transverse size $b_\perp \sim 1/Q$ of the vector meson distribution
amplitude is involved.  The sum over the interactions of the exchanged
gluons tend to
cancel reflecting its small color dipole moment.  Since the hadronic
interactions are
minimal,  the
$\gamma^*(Q^2) N \to
\rho N$ reaction at large $Q^2$ can occur coherently throughout a nuclear
target in
reactions without absorption or final state interactions.  The $\gamma^*A
\to V A$ process thus provides a natural framework for testing QCD color
transparency.  Evidence for color transparency in such reactions has been
found by Fermilab experiment E665  \cite{Adams:1997bh}.

Diffractive multi-jet production in heavy
nuclei provides a novel way to measure the shape of the LC Fock
state wavefunctions and test color transparency.  For example, consider the
reaction
 \cite{Bertsch,MillerFrankfurtStrikman,Frankfurt:1999tq}
$\pi A \rightarrow {\rm Jet}_1 + {\rm Jet}_2 + A^\prime$
at high energy where the nucleus $A^\prime$ is left intact in its ground
state.  The transverse momenta of the jets have to balance so that
$
\vec k_{\perp i} + \vec k_{\perp 2} = \vec q_\perp < {R^{-1}}_A \ ,
$
and the light-cone longitudinal momentum fractions have to add to
$x_1+x_2 \sim 1$ so that $\Delta p_L < R^{-1}_A$.  The process can
then occur coherently in the nucleus.  Because of color transparency,  \ie,
the cancelation of color interactions in a small-size color-singlet
hadron,  the valence wavefunction of the pion with small impact
separation will penetrate the nucleus with minimal interactions,
diffracting into jet pairs  \cite{Bertsch}.
The $x_1=x$, $x_2=1-x$ dependence of
the di-jet distributions will thus reflect the shape of the pion distribution
amplitude; the $\vec k_{\perp 1}- \vec k_{\perp 2}$
relative transverse momenta of the jets also gives key information on
 the underlying shape of the valence pion
wavefunction  \cite{MillerFrankfurtStrikman,Frankfurt:1999tq}. The QCD
analysis can be
confirmed by the observation that the diffractive nuclear amplitude
extrapolated to
$t = 0$ is linear in nuclear number $A$, as predicted by QCD color
transparency.  The integrated diffractive rate should scale as $A^2/R^2_A \sim
A^{4/3}$.  A diffractive dissociation experiment of this type, E791,  is now in
progress at Fermilab using 500 GeV incident pions on nuclear
targets  \cite{E791}.  The preliminary results from E791 appear to be consistent
with color transparency.  The momentum fraction distribution of the jets is
consistent with a valence light-cone wavefunction of the pion consistent with
the shape of the asymptotic distribution amplitude, $\phi^{\rm asympt}_\pi (x) =
\sqrt 3 f_\pi x(1-x)$.  As discussed below, data from
CLEO \cite{Gronberg:1998fj} for the
$\gamma
\gamma^* \rightarrow \pi^0$ transition form factor also favor a form for
the pion distribution amplitude close to the asymptotic solution \cite{LB}
to the perturbative QCD evolution
equation  \cite{Kroll,Rad,Brodsky:1998dh,Feldmann:1999wr,Schmedding:1999ap}.
It will also be interesting to study diffractive tri-jet production using proton
beams
$ p A \rightarrow {\rm Jet}_1 + {\rm Jet}_2 + {\rm Jet}_3 + A^\prime $ to
determine the fundamental shape of the 3-quark structure of the valence
light-cone wavefunction of the nucleon at small transverse
separation  \cite{MillerFrankfurtStrikman}.  One interesting possibility is
that the distribution amplitude of the
$\Delta(1232)$ for $J_z = 1/2, 3/2$ is close to the asymptotic form $x_1
x_2 x_3$,  but that the proton distribution amplitude is more complex.
This would explain why the $p \to\Delta$ transition form factor appears to
fall faster at large $Q^2$ than the elastic $p \to p$ and the other $p \to
N^*$ transition form factors  \cite{Stoler:1999nj}.
Conversely, one can use incident real and virtual photons:
$ \gamma^* A \rightarrow {\rm Jet}_1 + {\rm Jet}_2 + A^\prime $ to
confirm the shape of the calculable light-cone wavefunction for
transversely-polarized and longitudinally-polarized virtual photons.  Such
experiments will open up a direct window on the amplitude
structure of hadrons at short distances.

There are a large number of measured exclusive reactions in which the
empirical power
law fall-off predicted by dimensional counting and PQCD appears to be
accurate over a large range of momentum transfer.
These include processes such as the proton form factor, time-like meson pair
production in $e^+ e^-$ and $\gamma
\gamma$ annihilation, large-angle scattering processes such as pion
photoproduction
$\gamma p \to \pi^+ p$, and nuclear processes such as the deuteron form
factor at
large momentum transfer and deuteron photodisintegration
 \cite{Brodsky:1976rz}.  A
spectacular example is the recent measurements at CESR of the photon to pion
transition form factor in the reaction $e \gamma \to e
\pi^0$  \cite{Gronberg:1998fj}.
As predicted by leading twist QCD \cite{LB} $Q^2 F_{\gamma
\pi^0}(Q^2)$ is essentially constant for 1 GeV$^2 < Q^2 < 10$ GeV$^2.$
Further, the
normalization is consistent with QCD at NLO if one assumes that the pion
distribution
amplitude takes on the form $\phi^{\rm asympt}_\pi (x) =
\sqrt 3 f_\pi x(1-x)$ which is the asymptotic solution \cite{LB} to the
evolution equation for the pion
distribution amplitude  \cite{Kroll,Rad,Brodsky:1998dh,Schmedding:1999ap}.

If the pion distribution amplitude is close to its asymptotic form, then one can
predict the normalization of exclusive amplitudes such as the spacelike
pion form factor $Q^2 F_\pi(Q^2)$.  Next-to-leading order
predictions are now becoming available which incorporate higher order
corrections
to the pion distribution amplitude as well as the hard scattering
amplitude  \cite{Muller:1994hg,Melic:1999hg,Szczepaniak:1998sa}.
However, the
normalization of the PQCD prediction for the pion form factor depends
directly on the
value of the effective coupling
$\alpha_V(Q^*)$ at momenta $Q^{*2} \simeq Q^2/20$.  Assuming
$\alpha_V(Q^*) \simeq 0.4$, the QCD LO prediction appears to be
smaller by approximately a factor of 2 compared to the presently available data
extracted
from the original pion electroproduction experiments from
CEA  \cite{Bebek:1976ww}.  A
definitive comparison will require a careful extrapolation to the pion pole and
extraction of the longitudinally polarized photon contribution of the $e p
\to \pi^+ n$ data.

The measured deuteron form factor and the deuteron photodisintegration
cross section
appear to follow the leading-twist QCD predictions at large momentum
transfers in the
few GeV region  \cite{Holt:1990ze,Bochna:1998ca}.  The normalization of the
measured deuteron form factor is large compared to model calculations
 \cite{Farrar:1991qi} assuming that the deuteron's six-quark wavefunction can be
represented at short distances with the color structure of two color
singlet baryons.
This provides indirect evidence for the presence of hidden color components as
required by PQCD  \cite{bjl83}.

There are, however, experimental exceptions to the general success of the
leading twist PQCD approach, such as (a) the dominance of the $J/\psi \to
\rho \pi$
decay which is forbidden by hadron helicity conservation and (b) the strong
normal-normal spin asymmetry $A_{NN}$ observed in polarized elastic $p p
\to p p$
scattering and an apparent breakdown of color transparency at large CM
angles and
$E_{CM} \sim 5$ GeV.  These conflicts with leading-twist PQCD predictions can be
used to identify the presence of new physical effects.  For example,
It is usually assumed that a heavy quarkonium state such as the
$J/\psi$ always decays to light hadrons via the annihilation of its heavy
quark
constituents to gluons.  However, the transition $J/\psi \to \rho
\pi$ can also occur by the rearrangement of the $c \bar c$ from the $J/\psi$
into the $\ket{ q \bar q c \bar c}$ intrinsic charm Fock state of the $\rho$ or
$\pi$  \cite{Brodsky:1997fj}.  On the other hand, the overlap rearrangement
integral in the decay $\psi^\prime \to \rho \pi$ will be suppressed since the
intrinsic charm Fock state radial wavefunction of the light hadrons will
evidently
not have nodes in its radial wavefunction.  This observation provides a natural
explanation of the long-standing puzzle why the $J/\psi$ decays prominently to
two-body pseudoscalar-vector final states, whereas the $\psi^\prime$
does not.  The unusual effects seen in elastic proton-proton scattering at
$E_{CM}
\sim 5$ GeV and large angles could be related to the charm threshold
and the effect of a $\ket{ uud uud c \bar c }$ resonance which would appear
as in
the $J=L=S=1$ $p p $ partial wave  \cite{Brodsky:1988xw}.

Recent experiments at Jefferson laboratory utilizing a new polarization transfer
technique indicate that $G_E(Q^2)/G_M(Q^2)$ falls with increasing momentum
transfer
$-t = Q^2$ in the measured domain $1 < Q^2 < 3 $ GeV$^2$  \cite{Jones:1999rz}.
This observation
implies that the helicity-changing Pauli form factor $F_2(Q^2)$ is comparable to
the helicity conserving form factor $F_2(Q^2)$ in this domain.  If such a trend
continues to larger $Q^2$ it would be in severe conflict with the
hadron-helicity conserving principle of perturbative QCD.  If $F_2$ were
comparable to $F_1$ at large $Q^2$ in the case of timelike
processes, such as $p \bar p \to e^+ e^-$, where
$G_E= F_1 + {Q^2\over 4 M_N^2} F_2,$ one would see strong deviations from
the usual $1 + \cos^2{\theta}$ dependence of the differential cross section as
well as PQCD scaling.  This seems to be in conflict with the available
data from the $E835$
$\bar p p \to e^+ e^-$ experiment at Fermilab  \cite{Ambrogiani:1999bh}.

A debate has continued on whether processes such as the pion
and proton
form factors and elastic Compton scattering $\gamma p \to \gamma p$ might be
dominated by higher twist mechanisms until very large momentum
transfers  \cite{Isgur:1989iw,Radyushkin:1998rt,Bolz:1996sw}.  For example,
if one
assumes that the light-cone wavefunction of the pion has the form
$\psi_{\rm soft}(x,k_\perp) = A \exp (-b {k_\perp^2\over x(1-x)})$, then the
Feynman endpoint contribution to the overlap integral at small $k_\perp$ and
$x \simeq 1$ will dominate the form factor compared to the hard-scattering
contribution until
very large $Q^2$.  However, the above form of $\psi_{\rm soft}(x,k_\perp)$
has no
suppression at $k_\perp =0$ for any $x$; \ie, the
wavefunction in the hadron rest frame does not fall-off at all for $k_\perp
= 0$ and
$k_z \to - \infty$.  Thus such wavefunctions do not represent
soft QCD contributions.  Furthermore, such endpoint contributions will be
suppressed
by the QCD Sudakov form factor, reflecting the fact that a near-on-shell
quark must
radiate
if it absorbs large momentum.  If the endpoint contribution dominates
proton Compton
scattering, then both photons will interact on the same
quark line in a local fashion, and the
amplitude is predicted to be real, in strong contrast to the complex
phase structure of the PQCD predictions.  It should be noted that there is no
apparent endpoint contribution
which could explain the success of dimensional counting ($s^{-7}$ scaling
at fixed
$\theta_{cm}$) in large-angle pion photoproduction.

The perturbative QCD predictions \cite{Kronfeld:1991kp} for the
Compton amplitude phase can be tested in
virtual Compton scattering by interference with Bethe-Heitler
processes  \cite{Brodsky:1972vv}.  One can also measure the interference of
deeply
virtual Compton amplitudes with the timelike form factors by studying reactions
in $e^+ e^-$ colliders such as $e^+ e^- \to \pi^+ \pi^- \gamma$.  The asymmetry
with respect to the electron or positron beam measures the interference of the
Compton diagrams with the amplitude in which the photon is emitted from the
lepton line.

It is interesting to compare the corresponding calculations of form
factors of bound states in QED.  The soft wavefunction
is the Schr\"odinger-Coulomb solution $\psi_{1s}(\vec k) \propto (1 + {{\vec
p}^2/(\alpha m_{\rm red})^2})^{-2}$, and the full wavefunction,  which
incorporates transversely polarized photon exchange, only differs by
a factor $(1 + {\vec p}^2/m^2_{\rm red})$.  Thus the leading twist
dominance of form
factors in QED occurs at relativistic scales $Q^2 > {m^2_{\rm
red}} $  \cite{Brodsky:1989pv}.
Furthermore, there are no extra relative factors of $\alpha$ in the
hard-scattering contribution.  If the QCD coupling $\alpha_V$ has
an infrared fixed-point, then the fall-off of the valence wavefunctions of
hadrons will have analogous power-law
forms, consistent with the Abelian correspondence
principle  \cite{Brodsky:1997jk}.  If such power-law wavefunctions are
indeed applicable to the soft domain of QCD then, the transition to
leading-twist
power law behavior will occur in the nominal hard perturbative QCD domain where
$Q^2 \gg \VEV{k^2_\perp}, m_q^2$.

\section{Measurement of Light-cone Wavefunctions and Tests of Color
Transparency via Diffractive Dissociation.}

Diffractive multi-jet production in heavy
nuclei provides a novel way to measure the shape of the LC Fock
state wavefunctions and test color transparency.  For example, consider the
reaction
 \cite{Bertsch,MillerFrankfurtStrikman,Frankfurt:1999tq}
$\pi A \rightarrow {\rm Jet}_1 + {\rm Jet}_2 + A^\prime$
at high energy where the nucleus $A^\prime$ is left intact in its ground
state.  The transverse momenta of the jets have to balance so that
$
\vec k_{\perp i} + \vec k_{\perp 2} = \vec q_\perp < {R^{-1}}_A \ ,
$
and the light-cone longitudinal momentum fractions have to add to
$x_1+x_2 \sim 1$ so that $\Delta p_L < R^{-1}_A$.  The process can
then occur coherently in the nucleus.  Because of color transparency,  \ie,
the cancelation of color interactions in a small-size color-singlet
hadron,  the valence wavefunction of the pion with small impact
separation will penetrate the nucleus with minimal interactions,
diffracting into jet pairs  \cite{Bertsch}.  The two-gluon exchange process in
effect differentiates the transverse momentum dependence of the hadron's
wavefunction twice.  Thus the $x_1=x$,
$x_2=1-x$ dependence of the di-jet distributions will reflect the shape of
the pion distribution amplitude; the $\vec k_{\perp 1}- \vec k_{\perp 2}$
relative transverse momenta of the jets also gives key information on the
underlying shape of the valence pion
wavefunction  \cite{MillerFrankfurtStrikman,Frankfurt:1999tq}.  The QCD
analysis can
be confirmed by the observation that the diffractive nuclear amplitude
extrapolated to
$t = 0$ is linear in nuclear number $A$, as predicted by QCD color
transparency.  The integrated diffractive rate should scale as $A^2/R^2_A \sim
A^{4/3}$.  A diffractive dissociation experiment of this type, E791,  is now in
progress at Fermilab using 500 GeV incident pions on nuclear
targets  \cite{E791}.  The preliminary results from E791 appear to be consistent
with color transparency.  The momentum fraction distribution of the jets is
consistent with a valence light-cone wavefunction of the pion consistent with
the shape of the asymptotic distribution amplitude, $\phi^{\rm asympt}_\pi (x) =
\sqrt 3 f_\pi x(1-x)$.  Data from
CLEO \cite{Gronberg:1998fj} for the
$\gamma
\gamma^* \rightarrow \pi^0$ transition form factor also favor a form for
the pion distribution amplitude close to the asymptotic solution \cite{LB}
to the perturbative QCD evolution
equation  \cite{Kroll,Rad,Brodsky:1998dh,Feldmann:1999wr,Schmedding:1999ap}.
It is also possible that the distribution amplitude of the
$\Delta(1232)$ for $J_z = 1/2, 3/2$ is close to the asymptotic form $x_1
x_2 x_3$,  but that the proton distribution amplitude is more complex.
This would explain why the $p \to\Delta$ transition form factor appears to
fall faster at large $Q^2$ than the elastic $p \to p$ and the other $p \to
N^*$ transition form factors  \cite{Stoler:1999nj}.
It will thus be very interesting to study diffractive tri-jet production using
proton beams dissociating into three jets on a nuclear target.
$ p A \rightarrow {\rm Jet}_1 + {\rm Jet}_2 + {\rm Jet}_3 + A^\prime $ to
determine the fundamental shape of the 3-quark structure of the valence
light-cone wavefunction of the nucleon at small transverse
separation  \cite{MillerFrankfurtStrikman}.

It is also interesting to consider the Coulomb dissociation of hadrons as a
means
to resolve their light-cone wavefunctions  \cite{BHP}.  In the case of photon
exchange, the transverse momentum dependence of the light-cone wavefunction is
differentiated only once.  For example, consider the process $e p \to
e^\prime {\rm
Jet}_1 + {\rm Jet}_2 + {\rm Jet}_3$ in which the proton dissociates into
three distinct jets at large transverse momentum by scattering on an
electron.  In
the case
of an $e p$ collider such as HERA, one can require all of the hadrons to be
produced outside a forward annular exclusion zone,
$\theta_H >
\theta_{\rm min}$, thus ensuring a minimum transverse momentum of each produced
final
state particle.  The
distribution of hadron longitudinal momentum in each azimuthal sector can
be used
to determine the underlying
$x_1, x_2, x_3$ dependence of the proton's valence three-quark wavefunction.
Such a procedure will allow the proton to self-resolve its fundamental
structure.  Similarly at lower momentum scales, one can study the
dissociation of light nuclei into their nucleon and mesonic components in
diffractive high momentum reactions.

One can use incident real and virtual photons:
$ \gamma^* A \rightarrow {\rm Jet}_1 + {\rm Jet}_2 + A^\prime $ to
confirm the shape of the calculable light-cone wavefunction for
transversely-polarized and longitudinally-polarized virtual photons.  At low
transverse momentum, one expects interesting nonperturbative modifications.
Such
experiments will open up a direct window on the amplitude structure of
hadrons at
short distances.

\section{Semi-Exclusive Processes:  New Probes of Hadron Structure}

A new class of hard ``semi-exclusive''
processes of the form $A+B \to C + Y$, have been proposed as new probes of
QCD  \cite{BB,acw,Brodsky:1998sr}.  These processes are characterized
by a large momentum transfer $t= (p_A-p_C)^2$ and a large rapidity gap between
the final state particle $C$ and the inclusive system $Y$.
Here $A, B$
and $C$ can be hadrons or (real or virtual) photons.  The cross
sections for such processes factorize in terms of the distribution
amplitudes of $A$ and $C$ and the parton distributions in the target
$B$.  Because of this factorization semi-exclusive reactions provide a
novel array of generalized currents, which not only give insight into
the dynamics of hard scattering QCD processes, but also allow
experimental access to new combinations of the universal quark and
gluon distributions.

\begin{figure}[htb]
\begin{center}
 \leavevmode
 \epsfxsize=3.5in
 \epsfbox{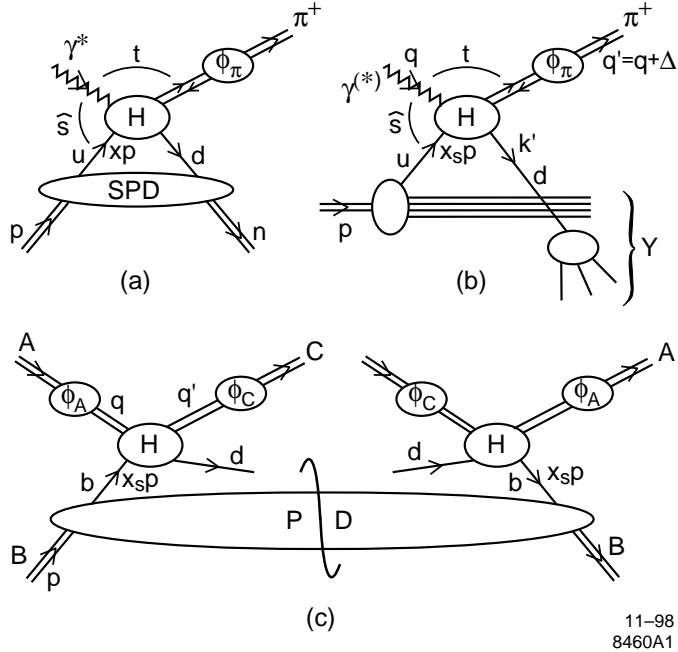}
\end{center}
\caption{{} (a): Factorization of $\gamma^* p \to \pi^+ n$ into a
 skewed parton distribution (SPD), a hard scattering $H$ and the pion
 distribution amplitude $\phi_\pi$.  (b): Semi-exclusive process
 $\gamma^{(*)} p \to \pi^+ Y$.  The $d$-quark produced in the hard
 scattering $H$ hadronizes independently of the spectator partons in
 the proton.  (c): Diagram for the cross section of a generic
 semi-exclusive process.  It involves a hard scattering $H$,
 distribution amplitudes $\phi_A$ and $\phi_C$ and a parton
 distribution (PD) in the target $B$.}
\label{one}
\end{figure}

QCD scattering amplitude for
deeply virtual exclusive processes like Compton scattering $\gamma^* p
\to \gamma p$ and meson production $\gamma^* p \to M p$ factorizes
into a hard subprocess and soft universal hadronic matrix
elements  \cite{JiRad,Collins:1997fb,BGMFS}.
For example, consider exclusive meson
electroproduction such as $e p \to e \pi^+ n$ (Fig.~\ref{one}a).  Here one
takes (as in DIS) the Bjorken limit of large photon virtuality, with
$x_B = Q^2/(2 m_p \nu)$ fixed, while the momentum transfer $t =
(p_p-p_n)^2$ remains small.  These processes involve `skewed' parton
distributions, which are generalizations of the usual parton
distributions measured in DIS.  The skewed distribution in Fig.~\ref{one}a
describes the emission of a $u$-quark from the proton target together
with the formation of the final neutron from the $d$-quark and the
proton remnants.  As the subenergy $\hat s$ of the scattering process
$\gamma^* u \to \pi^+ d$ is not fixed, the amplitude involves an
integral over the $u$-quark momentum fraction $x$.

An essential condition for the factorization of the deeply virtual
meson production amplitude of Fig.\ \ref{one}a is the existence of a large
rapidity gap between the produced meson and the neutron.  This
factorization remains valid if the
neutron is replaced with a hadronic system $Y$ of invariant mass $M_Y^2 \ll
W^2$, where $W$ is
the c.m.\ energy of the $\gamma^* p$ process.
For $M_Y^2 \gg m_p^2$ the momentum $k'$ of the $d$-quark in Fig.~\ref{one}b is
large with respect to the proton remnants, and hence it forms a jet.
This jet hadronizes independently of the other particles in the final
state if it is not in the direction of the meson, \ie, if the meson
has a large transverse momentum $q'_\perp = \Delta^{\phantom .}_\perp$
with respect to the photon direction in the $\gamma^* p$ c.m.  Then the
cross section for an inclusive system $Y$ can be calculated as in DIS,
by treating the $d$-quark as a final state particle.

The large $\dpt$ furthermore allows only transversally compact
configurations of the projectile $A$ to couple to the hard subprocess
$H$ of Fig.~\ref{one}b, as it does in exclusive processes  \cite{LB}.  Hence the
above discussion applies not only to incoming virtual photons at large
$Q^2$, but also to real photons $(Q^2=0)$ and in fact to any hadron
projectile.

Let us then consider the general process $A+B\to C+Y$, where $B$ and
$C$ are hadrons or real photons, while the projectile $A$ can also be
a virtual photon.  In the semi-exclusive kinematic limit
$\lqcd^2,\, M_B^2,\, M_C^2 \ll
M_Y^2,\, \dpt^2 \ll W^2$
we have a large rapidity gap
$|y_C - y_d| = \log \frac{W^2}{\dpt^2 + M_Y^2}$
between $C$ and the parton $d$ produced in the hard scattering (see
Fig.~\ref{one}c).  The cross section then factorizes into the form
\begin{eqnarray}
\lefteqn{ \frac{d\sigma}{dt\,dx_S}(A+B\to C+Y) }
 \hspace{4em} \nonumber \\
&=& \sum_{b} f_{b/B}(x_S,\mu^2) \frac{d\sigma}{dt} (A b \to C d)
 \eqcm \label{gencross}
\end{eqnarray}
where $t=(q-q')^2$ and $f_{b/B}(x_S,\mu^2)$ denotes the distribution
of quarks, antiquarks and gluons $b$ in the target $B$.  The momentum
fraction $x_S$ of the struck parton $b$ is fixed by kinematics to the
value
$x_S = \frac{-t}{M_Y^2-t}$
and the factorization scale $\mu^2$ is characteristic of the hard
subprocess $A b \to C d$.

It is conceptually helpful to regard the hard scattering amplitude $H$
in Fig.~\ref{one}c as a generalized current of momentum $q-q'=p_A - p_C$,
which interacts with the target parton $b$.  For $A= \gamma^*$ we
obtain a close analogy to standard DIS when particle $C$ is removed.
With $q' \to 0$ we thus find $-t \to Q^2$, $M_Y^2 \to W^2$, and see
that $x_S$ goes over into $x_B = Q^2 /(W^2 + Q^2)$.  The
possibility to control the value of $q'$ (and hence the momentum
fraction $x_S$ of the struck parton) as well as the quantum numbers of
particles $A$ and $C$ should make semi-exclusive processes a versatile
tool for studying hadron structure.  The cross section further depends
on the distribution amplitudes $\phi_A$, $\phi_C$ (\cf\ Fig.~\ref{one}c),
allowing new ways of measuring these quantities.

\section{ Conformal Symmetry as a Template}

Testing quantum chromodynamics to high precision is not easy.  Even in
processes involving high momentum transfer, perturbative QCD predictions
are complicated by questions of the convergence of the series,
particularly by the presence of ``renormalon" terms which grow as $n!$,
reflecting the uncertainty in the analytic form of the QCD coupling
at low scales.  Virtually all QCD processes are complicated by the
presence of dynamical higher twist effects, including power-law suppressed
contributions due to multi-parton correlations, intrinsic transverse
momentum, and finite quark masses.  Many of these effects are inherently
nonperturbative in nature and require knowledge of hadron wavefunction
themselves.  The problem of interpreting perturbative QCD predictions is
further compounded by theoretical ambiguities due to the apparent
freedom in the choice of renormalization schemes, renormalizations
scales, and factorization procedures.

A central principle of renormalization theory is that
predictions which relate physical observables to each other cannot depend
on theoretical conventions.  For example, one can use any renormalization
scheme, such as the modified minimal subtraction dimensional
regularization scheme, and any choice of renormalization scale $\mu$ to
compute perturbative series observables $A$ and $B$.  However, all traces
of the choices of the renormalization scheme and scale must disappear
when we algebraically eliminate the $\alpha_{\overline MS}(\mu)$
and directly relate $A$ to $B$.  This is the
principle underlying ``commensurate scale relations"
(CSR)~ \cite{Brodsky:1995eh},
which are general leading-twist QCD predictions relating physical
observables to each other.  For example, the ``generalized Crewther
relation",  which is discussed in more detail below,  provides a
scheme-independent relation between the QCD corrections to the Bjorken
(or Gross Llewellyn-Smith) sum rule for deep inelastic lepton-nucleon
scattering,  at a given momentum transfer $Q$, to the radiative
corrections to the annihilation cross section
$\sigma_{e^+ e^- \to
\rm hadrons}(s)$, at a corresponding ``commensurate" energy scale $\sqrt s$
 \cite{Brodsky:1995eh,Brodsky:1996tb}.  The specific relation between the
physical scales $Q$
and $\sqrt s$ reflects the fact
that the radiative corrections to each process have
distinct quark mass thresholds.

Any perturbatively calculable physical quantity can be used to define an
effective charge \cite{Grunberg,DharGupta,GuptaShirkovTarasov} by incorporating
the entire radiative correction into its definition.
For example, the $e^+ e^- \gamma^* \to {\rm
hadrons}$ annihilation to muon pair cross section ratio can be written
\begin{equation}
R_{e^+ e^-}(s) \equiv R^0_{e^+ e^-}(s) [ 1 +
{\alpha_R(s)\over \pi}] ,
\end{equation}
where $R^0_{e^+ e^-}$ is the prediction at Born level.
Similarly, we can define the entire radiative correction to the Bjorken
sum rule as the effective charge $\alpha_{g_1}(Q^2)$ where
$Q$ is the corresponding momentum transfer:
\begin{eqnarray}
\int_0^1 d x \left[ g_1^{ep}(x,Q^2) - g_1^{en}(x,Q^2) \right]
   &\equiv& {1\over 6} \left|g_A \over g_V \right|
   C_{\rm Bj}(Q^2) \nonumber \\
 &=& {1\over 6} \left|g_A
\over g_V \right| \left[ 1- {3\over 4} C_F{\alpha_{g_1}(Q^2) \over
   \pi} \right]  .
\end{eqnarray}

By convention,
each effective charge is normalized to $\alpha_s$ in the weak coupling
limit.  One can define effective charges for virtually any quantity
calculable in perturbative QCD; \eg\ moments of structure functions,
ratios of form factors, jet observables, and the effective potential
between massive quarks.  In the case of decay constants of the $Z$ or the
$\tau$, the mass of the decaying system serves as the physical scale in
the effective charge.  In the case of multi-scale observables, such as the
two-jet fraction in $e^+ e^-$ annihilation, the multiple arguments of the
effective coupling $\alpha_{2 jet}(s,y)$ correspond to the overall
available energy $s$ variables such as $y = {\rm max}_{ij}
(p_i+p_j)^2/s$ representing the maximum jet mass fraction.

Commensurate scale relations take the general form
\begin{equation}
\alpha_A(Q_A) = C_{AB}[\alpha_B(Q_B)] \ .
\end{equation}
The function $C_{AB}(\alpha_B)$ relates the observables $A$ and $B$ in
the conformal limit; \ie, $C_{AB}$ gives the functional dependence
between the effective charges which would be obtained if the theory had
zero $\beta$ function.  The conformal coefficients can be distinguished
from the terms associated with the $\beta$ function at each order in
perturbation theory from their color and flavor dependence, or by an
expansion about a fixed point.

The ratio
of commensurate scales is determined by the requirement that all terms
involving the $\beta$ function are incorporated into the arguments of the
running couplings, as in the original BLM procedure.
Physically, the ratio of scales corresponds to the fact that the physical
observables have different quark threshold and distinct sensitivities to
fermion loops.  More generally, the differing scales are in effect
relations between mean values of the physical scales which appear in loop
integrations.  Commensurate scale relations are transitive; \ie, given the
relation
between effective charges for observables $A$ and $C$ and $C$ and $B$,
the resulting between $A$ and $B$ is independent of $C$.  In particular,
transitivity implies $\Lambda_{AB} = \Lambda_{AC} \times \Lambda_{CB}$.

One can consider QCD predictions as functions of analytic
variables of the number of colors $N_C$ and flavors $N_F$.  For example,
one can show at all orders of perturbation theory that PQCD predictions
reduce to those of an Abelian theory at $N_C
\to 0$ with ${\widehat \alpha} = C_F \alpha_s$ and ${\widehat N_F} = N_f/T
C_F$ held fixed.  In particular, CSRs obey the ``Abelian correspondence
principle" in that they give the correct Abelian relations at $N_c \to 0.$

Similarly, commensurate scale relations obey the ``conformal
correspondence principle":  the CSRs reduce to correct conformal
relations when $N_C$ and $N_F$ are tuned to produce zero
$\beta$ function.  Thus conformal symmetry provides a {\it template} for
QCD predictions, providing relations between observables
which are present even in theories which are not scale invariant.  All
effects of the nonzero beta function are encoded in the appropriate
choice of relative scales $\Lambda_{AB} = Q_A/Q_B$.

The scale $Q$ which enters a given effective charge corresponds to a
physical momentum scale.  The total logarithmic derivative of each
effective charge effective charge $\alpha_A(Q)$ with respect to its
physical scale is given by the Gell Mann-Low equation:
\begin{equation}
{d\alpha_A(Q,m) \over d \log Q} = \Psi_A (\alpha_A(Q,m), Q/m) ,
\end{equation}
where the functional dependence of $\Psi_A$ is specific to its own
effective charge.  Here $m$ refers to the quark's pole mass.  The pole
mass is universal in that it does not depend on the choice of effective
charge.  The Gell Mann-Low relation is reflexive in that $\psi_A$ depends
on only on the coupling $\alpha_A$ at the same scale.  It should be
emphasized that the Gell Mann-Low equation deals with physical quantities
and is independent of the renormalization procedure and choice of
renormalization scale.  A central feature of quantum chromodynamics is
asymptotic freedom; \ie, the monotonic decrease of the QCD coupling
$\alpha_A(\mu^2)$ at large spacelike scales.  The empirical test of
asymptotic freedom is the verification of the negative sign of the Gell
Mann-Low function at large momentum transfer, which must be true for any
effective charge.

In perturbation theory,
\begin{equation}
\Psi_A = - \psi_A^{\{0\}} {\alpha_A^2\over \pi}
  - \psi_A^{\{1\}}{\alpha_A^3\over \pi^2}
 - \psi_A^{\{2\}} {\alpha_A^4\over \pi^3} + \cdots
\end{equation}
At large scales $Q^2 \gg m^2$, the first two terms are universal and
identical to the first two terms of the $\beta$ function
$\psi_A^{\{0\}}=\beta_0 = {11 N_C \over 3} - {2 \over 3} N_F,
\psi_A^{\{1\}}=\beta_1,
$ whereas
$\psi_A^{\{n\}}$ for $n \ge 2$ is process dependent.  The quark mass
dependence of the $\psi$ function is analytic, and in the
case of $\alpha_V$ scheme is known to two loops.

The commensurate scale relation between $\alpha_A$ and $\alpha_B$ implies
an elegant relation between their conformal dependence $C_{AB}$ and their
respective Gell Mann Low functions:
\begin{equation}
\psi_B = {d C_{BA}\over d \alpha_A} \times \psi_A  .
\end{equation}
Thus given the result for $N_{F,V}(m/Q)$ in $\alpha_V$
scheme we can use the CSR to derive $N_{F,A}(m/Q)$ for any other effective
charge, at least to two loops.  The above relation also shows that if
one effective charge has a fixed point $\psi_A[\alpha_A(Q^{FP}_A)] = 0$,
then all effective charges $B$ have a corresponding fixed point
$\psi_B[\alpha_B(Q^{FP}_B)] = 0$ at the corresponding commensurate scale
and value of effective charge.

In quantum electrodynamics, the running coupling $\alpha_{QED}(Q^2)$,
defined from
the Coulomb scattering of two infinitely heavy test charges at the
momentum transfer
$t = -Q^2$, is
taken as the standard observable.  Is there a preferred effective charge
which we should use to characterize the
coupling strength in QCD?  In the case of QCD,  the heavy-quark potential
$V(Q^2)$ is defined via a Wilson loop from the interaction energy of
infinitely heavy quark and antiquark at momentum transfer $t = -Q^2.$ The
relation
$V(Q^2) = - 4 \pi C_F
\alpha_V(Q^2)/Q^2$ then defines
the effective charge $\alpha_V(Q).$
As in the corresponding case of Abelian QED, the scale $Q$ of the coupling
$\alpha_V(Q)$ is identified with the exchanged momentum.  Thus there is never
any ambiguity in the interpretation of the scale.  All vacuum polarization
corrections
due to fermion pairs are incorporated in $\alpha_V$ through the usual
vacuum polarization
kernels which depend on the physical mass thresholds.  Other
observables could be used to define the standard QCD coupling, such as the
effective
charge defined from heavy quark radiation  \cite{Uraltsev}.

Commensurate
scale relations between $\alpha_V$ and the QCD
radiative corrections to other observables have no scale or scheme
ambiguity, even in
multiple-scale problems such as multi-jet production.  As is the case in
QED, the
momentum scale which appears as the argument of
$\alpha_V$ reflect the mean virtuality of the exchanged gluons.
Furthermore, we can
write a commensurate scale relation between $\alpha_V$ and an analytic
extension of
the
$\alpha_{\overline {MS}}$ coupling, thus transferring all of the
unambiguous scale-fixing
and analytic properties of the physical $\alpha_V$ scheme to the
$\overline {MS}$
coupling.

An elegant example is the relation between the rate for semi-leptonic
$B$-decay and $\alpha_V$:
\begin{equation}
\Gamma(b \to X_u \ell \nu) =
          {G_F^2 {\vert V_{ub}\vert}^2 M_b^2\over 192 \pi^3}
           \left[1-2.41 {\alpha_V(0.16 M_b)\over \pi} -
               1.43 {\alpha_V(0.16 M_b)\over \pi}^2 \right]  ,
\end{equation}
where $M_b$ is the scheme independent $b-$quark pole mass.  The
coefficient of $\alpha^2_{\overline MS}(\mu)$ in the usual expansion
with $\mu = m_b$ is 26.8.

Some other examples of CSR's at NLO:
\begin{equation}
\alpha_R(\sqrt s) =
\alpha_{g_1}(0.5 \sqrt s) -
{\alpha^2_{g1}(0.5 \sqrt s)\over \pi} +
{\alpha^3_{g1}(0.5 \sqrt s)\over \pi^2}
\end{equation}
\begin{equation}
\alpha_R(\sqrt s) =
\alpha_V(1.8 \sqrt s) + 2.08
{\alpha^2_V(1.8 \sqrt s)\over \pi} - 7.16
{\alpha^3_V(1.8 \sqrt s)\over \pi^2}
\end{equation}
\begin{equation}
\alpha_{\tau}(\sqrt s) =
\alpha_V(0.8 \sqrt s) + 2.08
{\alpha^2_V(0-.8 \sqrt s)\over \pi} - 7.16
{\alpha^3_V(0.8 \sqrt s)\over \pi^2}
\end{equation}
\begin{equation}
\alpha_{g1}(\sqrt s) =
\alpha_V(0.8 Q) + 1.08
{\alpha^2_V(0.8 Q)\over \pi} - 10.3
{\alpha^3_V(0.8 Q)\over \pi^2} \ .
\end{equation}
For numerical purposes in each case we have used $N_F=5$ and $\alpha_V=
0.1$ to compute the NLO correction to the CSR scale.

Commensurate scale relations thus provide
fundamental and precise scheme-independent tests of QCD, predicting how
observables
track not only in relative normalization, but also in their commensurate scale
dependence.

\section{The Generalized Crewther Relation}

The generalized Crewther relation can be derived by calculating the
QCD radiative corrections to the deep inelastic sum rules and $R_{e^+ e^-}$ in a
convenient renormalization scheme such as the modified minimal subtraction
scheme $\overline{\rm MS}$.  One then algebraically eliminates
$\alpha_{\overline
{MS}}(\mu)$.
Finally, BLM scale-setting \cite{BLM} is used to eliminate the $\beta$-function
dependence of the coefficients.  The form of the resulting relation between the
observables thus matches the result which would have been obtained had QCD
been a
conformal theory with zero $\beta$ function.  The final result relating the
observables is independent of the choice of intermediate
$\overline{\rm MS}$ renormalization scheme.

More specifically, consider the Adler function \cite{Adler} for the $e^+
e^-$ annihilation cross section
\begin{equation} D(Q^2)=-12\pi^2 Q^2{d\over dQ^2}\Pi(Q^2),~
\Pi(Q^2) =-{Q^2\over 12\pi^2}\int_{4m_{\pi}^2}^{\infty}{R_{e^+ e^-}(s)ds\over
s(s+Q^2)}.
\end{equation}
The entire radiative correction to this function is defined as
the effective charge
$\alpha_D(Q^2)$:
\begin{eqnarray}
    D \left( Q^2/ \mu^2, \alpha_{\rm s}(\mu^2) \right) &=&
D \left (1, \alpha_{\rm s}(Q^2)\right) \label{3} \\
&\equiv&
    3 \sum_f Q_f^2 \left[ 1+ {3\over 4} C_F{\alpha_D(Q^2) \over \pi}
                   \right]
    +( \sum_f Q_f)^2C_{\rm L}(Q^2) \nonumber \\
&\equiv& 3 \sum_f Q_f^2 C_D(Q^2)+( \sum_f Q_f)^2C_{\rm L}(Q^2),
\nonumber
\end{eqnarray}
where $C_F={N_C^2-1\over 2 N_C}. $
The coefficient $C_{\rm L}(Q^2)$ appears at the third order in
perturbation theory and is related to the ``light-by-light scattering type"
diagrams.  (Hereafter $\alpha_{\rm s}$ will denote the ${\overline{\rm MS}}$
scheme strong coupling constant.)

It is straightforward to algebraically relate $\alpha_{g_1}(Q^2)$ to
$\alpha_D(Q^2)$ using the known expressions to three loops in the
$\overline{\rm MS}$ scheme.  If one chooses the renormalization scale to resum
all of the quark and gluon vacuum polarization corrections into
$\alpha_D(Q^2)$,  then
the final result turns out to be remarkably simple \cite{Brodsky:1996tb}
 $(\widehat\alpha = 3/4\, C_F\ \alpha/\pi):$
\begin{equation}
\widehat{\alpha}_{g_1}(Q)=\widehat{\alpha}_D( Q^*)-
\widehat{\alpha}_D^2( Q^*)+\widehat{\alpha}_D^3( Q^*) +
\cdots,
\end{equation}
where
\begin{eqnarray}
\ln \left({ {Q}^{*2} \over Q^2} \right) &=&
{7\over 2}-4\zeta(3)+\left(\frac{\alpha_D ( Q^*)}{4\pi}
\right)\Biggl[ \left(
            {11\over 12}+{56\over 3} \zeta(3)-16{\zeta^2(3)}
      \right) \beta_0\cr &&
      +{26\over 9}C_{\rm A}
      -{8\over 3}C_{\rm A}\zeta(3)
      -{145\over 18} C_{\rm F}
      -{184\over 3}C_{\rm F}\zeta(3)
      +80C_{\rm F}\zeta(5)
\Biggr]. \nonumber
\label{EqLogScaleRatio}
\end{eqnarray}
where in QCD, $C_{\rm A}=N_C = 3$ and $C_{\rm F}=4/3$.
This relation shows how
the coefficient functions for these two different processes are
related to each other at their respective commensurate scales.  We emphasize
that the $\overline{\rm MS}$ renormalization
scheme is used only for calculational convenience; it serves simply as an
intermediary between observables.  The renormalization group
ensures
that the forms of the CSR relations in perturbative QCD are independent
of the choice of an intermediate renormalization scheme.

The Crewther relation was originally derived assuming that the theory is
conformally
invariant; \ie, for zero $\beta$ function.  In the physical case, where the
QCD coupling runs,  all non-conformal effects are resummed into the
energy and momentum transfer scales of
the effective couplings $\alpha_R$ and $\alpha_{g1}$.  The general
relation between these two effective charges for non-conformal
theory thus takes the form of a geometric series
\begin{equation}
        1- \widehat \alpha_{g_1} =
\left[ 1+ \widehat \alpha_D( Q^*)\right]^{-1} \ .
\end{equation}
We have dropped the small light-by-light scattering contributions.
This is again a special advantage of relating observable to
observable.
The coefficients are independent of
color and are the same in Abelian, non-Abelian, and conformal gauge theory.  The
non-Abelian structure of the theory is reflected in the expression for the scale
${Q}^{*}$.

Is experiment consistent with the generalized Crewther relation?  Fits
 \cite{MattinglyStevenson} to the experimental measurements of the
$R$-ratio above the thresholds for the production of $c\overline{c}$ bound
states
provide the empirical constraint:
$\alpha_{R}({\sqrt s} =5.0~{\rm GeV})/\pi \simeq 0.08\pm 0.03.$
The prediction for the effective coupling for
the deep inelastic sum rules at the commensurate momentum transfer $Q$
is then
$\alpha_{g_1}(Q=12.33\pm 1.20~{\rm GeV})/\pi
\simeq \alpha_{\rm GLS}(Q=12.33\pm 1.20~{\rm GeV})/\pi
\simeq 0.074\pm 0.026.$
Measurements of the Gross-Llewellyn Smith sum rule have so far only been
carried out
at relatively small values of $Q^2$ \cite{CCFRL1,CCFRL2};
however, one can use the results of the theoretical
extrapolation \cite{KS} of the experimental data presented in \cite{CCFRQ}:
$ \alpha_{\rm GLS}^{\rm extrapol}(Q=12.25~{\rm GeV})/\pi\simeq 0.093\pm 0.042.$
This range overlaps with the prediction from the generalized
Crewther relation.  It is clearly important to have higher precision
measurements to
fully test this fundamental QCD prediction.

\section{General Form of Commensurate Scale Relations}

In general, commensurate scale
relations connecting the effective charges for observables $A$ and $B$ have the
form
\begin{equation}
\alpha_A(Q_A) =
\alpha_B(Q_B) \left(1 + r^{(1)}_{A/B} {\alpha_B(Q_B)\over \pi} + r^{(2)}_{A/B}
{\alpha_B(Q_B)\over
\pi}^2 + \cdots\right),
\label{eq:CSRg}
\end{equation}
where the coefficients $r^{{n}}_{A/B}$
are
identical to the coefficients obtained in a con\-formally invariant theory
with $\beta_B(\alpha_B) \equiv (d/d\ln Q^2) \alpha_B(Q^2) = 0$.
The ratio of the scales $Q_A/Q_B$ is thus fixed by the requirement that
the couplings sum all of the effects of the non-zero $\beta$ function.  In
practice the
NLO and NNLO coefficients and relative scales can be identified from the flavor
dependence of the perturbative series; \ie\ by shifting scales such that the
$N_F$-dependence associated with $\beta_0 = 11/3 C_A - 4/3 T_F N_F$ and
$\beta_1 =
-34/3 C_A^2 + {20\over 3} C_A T_F N_F + 4 C_F T_F N_F$
does not appear in
the coefficients.  Here $C_A=N_C$, $C_F=(N^2_C-1)/2N_C$ and $T_F=1/2$.
The shift in scales which gives conformal coefficients in effect pre-sums
the large and strongly divergent terms in the PQCD series which grow as
$n!  (\beta_0
\alpha_s)^n$, \ie, the infrared renormalons associated with coupling-constant
renormalization  \cite{tHooft,Mueller,LuOneDim,BenekeBraun}.

The
renormalization scales $Q^*$ in the BLM method are physical in the sense that
they reflect the mean virtuality of the gluon propagators.  This
scale-fixing procedure is consistent with scale fixing in QED, in agreement
with
in the Abelian limit, $N_C \to
0$  \cite{BLM,Brodsky:1997jk,LepageMackenzie,Neubert,BallBenekeBraun}.
The ratio of scales
$\lambda_{A/B} = Q_A/Q_B$ guarantees that the
observables $A$ and $B$ pass through new quark thresholds at the same physical
scale.  One can also show that the commensurate scales satisfy the
transitivity rule
$\lambda_{A/B} = \lambda_{A/C} \lambda_{C/B},$ which ensures that predictions
are independent of the choice of an intermediate renormalization scheme or
intermediate observable $C.$

In general, we can write the relation between any two effective charges at
arbitrary
scales
$\mu_A$ and
$\mu_B$ as a correction to the corresponding relation obtained in a
conformally invariant
theory:
\begin{equation}
\alpha_A(\mu_A) = C_{AB}[\alpha_B(\mu_B)] +
\beta_B[\alpha_B(\mu_B)] F_{AB}[\alpha_B(\mu_B)]
\label{eq:ak}
\end{equation}
where
\begin{equation}
 C_{AB}[\alpha_B] = \alpha_B + \sum_{n=1} C_{AB}^{(n)}\alpha^n_B
\label{eq:al}
\end{equation}
is the functional relation when $\beta_B[\alpha_B]=0$.  In fact, if $\alpha_B$
approaches a fixed point $\bar\alpha_B$ where
$\beta_B[\bar\alpha_B]=0$,
then $\alpha_A$ tends to a fixed point given by
\begin{equation}
\alpha_A \to \bar\alpha_A = C_{AB}[\bar\alpha_B].
\label{eq:am}
\end{equation}
The commensurate scale relation for observables $A$ and $B$ has a similar
form, but
in this case the relative scales are fixed such that the non-conformal term
$F_{AB}$ is
zero.
Thus the commensurate scale relation $\alpha_A(Q_A) = C_{AB}[\alpha_B(Q_B)]$
at general commensurate scales is also the relation connecting the values
of the fixed
points for any two effective charges or schemes.  Furthermore, as
$\beta\rightarrow 0$,
the ratio of commensurate scales $Q^2_A/Q^2_B$ becomes the ratio of
fixed point scales $\bar Q^2_A/\bar Q^2_B$ as one approaches
the fixed point regime.

\section{Implementation of $\alpha_V$ Scheme}
\unboldmath

The effective charge $\alpha_V(Q)$ provides a physically-based alternative to
the usual modified minimal
subtraction ($\overline{\mbox{MS}}$) scheme.  All vacuum polarization
corrections due to fermion pairs are incorporated in $\alpha_V$ through
the usual vacuum polarization kernels which depend on the physical mass
thresholds.  When continued to time-like momenta, the coupling has the
correct analytic dependence dictated by the production thresholds in the
crossed channel.  Since $\alpha_V$ incorporates quark mass effects
exactly, it avoids the problem of explicitly computing and resumming
quark mass corrections which are related to the running of the coupling.
Thus the effective number of flavors $N_F(Q/m)$ is an analytic function
of the scale $Q$ and the quark masses $m$.  The effects of finite quark
mass corrections on the running of the strong coupling were first
considered by De R{\' u}jula and Georgi~ \cite{derujula}
within the momentum subtraction schemes (MOM)
(see also Georgi and Politzer~ \cite{Georgi_Politzer},
Shirkov and collaborators~ \cite{shirkov}, and Ch{\'y}la~ \cite{chyla}).

One important advantage of the physical charge approach is
its inherent gauge invariance to all orders in perturbation theory.  This
feature is not manifest in massive $\beta$-functions defined in non-physical
schemes such as the MOM schemes.  A second, more practical,
advantage is the automatic decoupling of heavy quarks according to the
Appelquist-Carazzone theorem \cite{ac}.

By employing the commensurate scale relations
other physical observables
can be expressed in terms of the analytic
coupling $\alpha_V$ without scale or scheme ambiguity.
This way the quark mass threshold effects
in the running of the coupling are taken into account by
 utilizing the mass dependence of the physical
$\alpha_V$ scheme.  In effect, quark thresholds are treated analytically to all
orders in $m^2/Q^2$; {\it i.e.},  the evolution of the physical $\alpha_V$
coupling in the intermediate regions reflects the actual mass dependence of a
physical effective charge and the analytic properties of particle production.
 Furthermore, the definiteness of the
dependence in the quark masses automatically constrains the
scale $Q$ in the argument of the coupling.
There is thus no scale ambiguity in perturbative expansions in $\alpha_V$.

In the conventional $\overline{\mbox{MS}}$ scheme,
the coupling is independent of the
quark masses since the quarks are treated as either massless or
infinitely heavy with respect to the running of the coupling.  Thus one
formulates different effective theories depending on the effective number of
quarks which is governed by the scale $Q$; the massless $\beta$-function
is used to describe the running in between the flavor thresholds.
These different theories are then matched to each other by imposing matching
conditions at the scale where the effective number of flavors is changed
(normally the quark masses).  The dependence on the matching scale
 can be made arbitrarily small by calculating the matching conditions
to high enough order.  For physical observables one can then include the effects
of finite quark masses by making a higher-twist expansion in $m^2/Q^2$ and
$Q^2/m^2$ for light and heavy quarks, respectively.  These higher-twist
contributions have to be calculated for each observables separately, so
that in principle one requires an all-order resummation of the
mass corrections to the effective Lagrangian to give correct
results.

The specification of the coupling and renormalization scheme also depends
on the definition of the quark mass.  In contrast to QED where the
on-shell mass provides a natural definition of lepton masses,  an
on-shell definition for quark masses is complicated by the confinement
property of QCD.  In this paper we will use the pole mass $m(p^2=m^2)=m$
which has the advantage of being scheme and renormalization-scale invariant.
Furthermore, when combined with the $\alpha_V$ scheme,
the pole mass gives predictions which are
free of the leading renormalon ambiguity.

A technical complication of massive schemes is that one cannot easily obtain
analytic solutions of renormalization group equations to the massive $\beta$
function, and the Gell-Mann Low function is scheme-dependent even at lowest
order.

In a recent paper we have presented a two-loop analytic extension of the
$\alpha_V$-scheme based on the recent results of Ref.~ \cite{melles98}.
The mass effects are in principle treated
exactly to two-loop order and are only limited in practice by
the uncertainties from numerical integration.
The desired features of gauge invariance and decoupling are manifest in
the form of the two-loop Gell-Mann Low function, and we give a simple
fitting-function which interpolates smoothly the exact two-loop results
obtained by using
the adoptive Monte Carlo integrator VEGAS \cite{vegas}.  Strong
consistency checks of the results are performed by comparing the Abelian
limit to the well known QED results in the on-shell scheme.  In addition,
the massless as well as the decoupling limit are reproduced exactly, and
the two-loop Gell-Mann Low function is shown to be renormalization scale
independent.

\begin{figure}[htb]
\begin{center}
\leavevmode
\epsfxsize=5in
\epsfbox{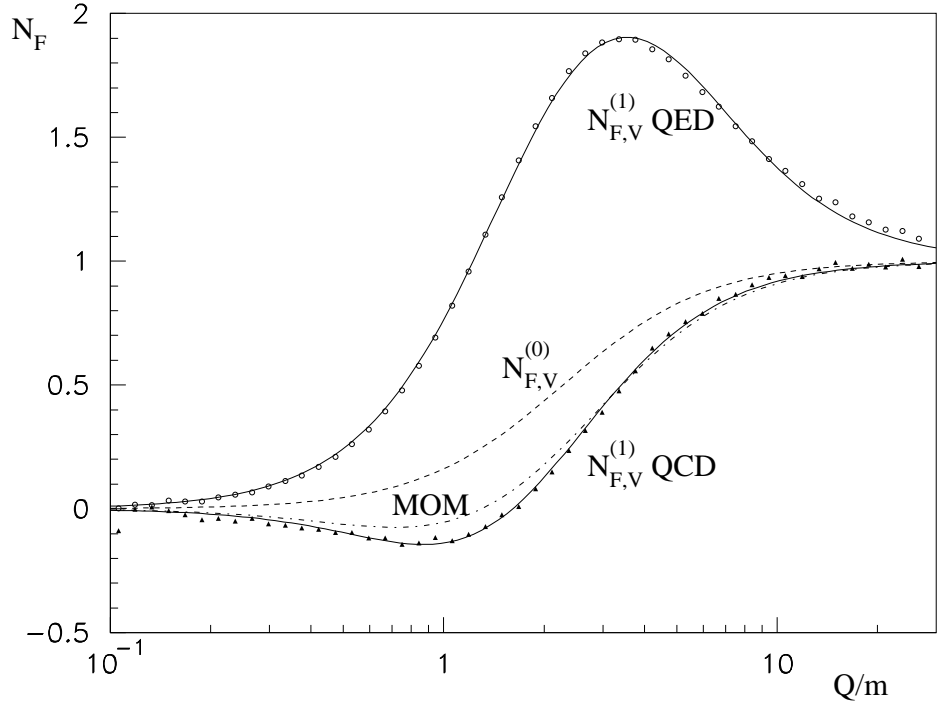}
\end{center}
\caption{The numerical results for the gauge-invariant $N_{F,V}^{(1)}$ in QED
(open circles) and QCD (triangles) with the best $\chi^2$ fits
superimposed respectively.  The dashed
line shows the one-loop $N_{F,V}^{(0)}$ function.
For comparison we
also show the gauge dependent two-loop result obtained in MOM schemes
(dash-dot) \protect \cite{yh,jt}.  At large $\frac{Q}{m}$ the theory becomes
effectively massless, and both schemes agree as expected.  The figure also
illustrates the decoupling of heavy quarks at small $\frac{Q}{m}$.}
\label{fig:nfV}
\end{figure}

The results of our numerical calculation of $N_{F,V}^{(1)}$ in the
$V$-scheme for QCD and QED are shown in Fig.~\ref{fig:nfV}.
The decoupling of heavy quarks becomes manifest at small $Q/m$, and
the massless limit is attained for large $Q/m$.  The QCD form actually
becomes negative at moderate values of $Q/m$, a novel feature of
the anti-screening non-Abelian contributions.  This property is
also present in the (gauge dependent) MOM results.
In contrast, in Abelian QED the two-loop contribution to
the effective number of flavors becomes larger than 1 at intermediate
values of $Q/m$.  We also display the one-loop
contribution $N^{(0)}_{F,V} \left( \frac{Q}{m} \right)$ which
monotonically interpolates between the decoupling and massless limits.
The solid curves displayed in
Fig.~\ref{fig:nfV} shows that the parameterizations
which we used for fitting the numerical results are quite accurate.

The relation of $\alpha_V(Q^2)$ to the conventional $\overline {MS}$
coupling is now known to NNLO  \cite{Peter},
but for clarity in this section only the NLO relation will be used.  The
commensurate scale relation is given by \cite{bgmr}
\begin{eqnarray}
\label{eq:csrmsovf}
\alpha_{\overline{\mbox{\tiny MS}}}(Q)
& = & \alpha_V(Q^{*}) + \frac{2}{3}N_C{\alpha_V^2(Q^{*}) \over \pi}
\nonumber \\
& = & \alpha_V(Q^{*}) + 2{\alpha_V^2(Q^{*}) \over \pi}\  ,
\end{eqnarray}
which is valid for $Q^2 \gg m^2$.  The coefficients in the
perturbation expansion have their conformal values, \ie, the same coefficients
would occur even if the theory had been conformally invariant with $\beta=0$.
The commensurate scale is given by
\begin{eqnarray}
Q^* & = & Q\exp\left[\frac{5}{6}\right] \ .
\end{eqnarray}
 The scale in the $\overline {MS}$ scheme is thus a factor
$\sim 0.4$ smaller than the physical scale.  The coefficient $2 N_C/3$ in
the NLO
coefficient is a feature of the non-Abelian couplings of QCD; the same
coefficient occurs even if the theory were conformally invariant with
$\beta_0=0.$

Using the above QCD results, we can transform any NLO prediction
given in $\overline{MS}$ scheme to a scale-fixed expansion in
$\alpha_V(Q)$.
We can also derive the connection between the $\overline{MS}$ and $\alpha_V$
schemes for Abelian perturbation theory using the limit $N_C \to 0$ with
$C_F\alpha_s$ and $N_F/C_F$ held fixed  \cite{Brodsky:1997jk}.

The use of $\alpha_V$ and related physically defined effective charges such as
$\alpha_p$ (to NLO the effective charge defined from the (1,1) plaquette,
$\alpha_p$ is the same as $\alpha_V$) as expansion parameters has been
found to be valuable in lattice
gauge theory, greatly increasing the convergence of perturbative expansions
relative to
those using the bare lattice coupling  \cite{LepageMackenzie}.  Recent lattice
calculations of the
$\Upsilon$- spectrum \cite{Davies} have been used with BLM
scale-fixing to determine a NLO normalization
of the static heavy quark potential:  $
\alpha_V^{(3)}(8.2 \GeV) = 0.196(3)$ where the effective number of light
flavors is
$n_f = 3$.  The
corresponding modified minimal subtraction coupling evolved to the
$Z$ mass and five flavors is $ \alpha_{\overline{MS}}^{(5)}(M_Z) =
0.1174(24)$.  Thus a high precision value for $\alpha_V(Q^2)$ at a specific
scale is
available from lattice gauge theory.  Predictions for other QCD observables
can be
directly referenced to this value without the scale or scheme ambiguities,
thus greatly
increasing the precision of QCD tests.

One can also use $\alpha_V$ to characterize the coupling which appears in
the hard
scattering contributions of exclusive process amplitudes at large momentum
transfer,
such as elastic hadronic form factors, the photon-to-pion transition form
factor at
large momentum transfer \cite{BLM,BJPR} and exclusive weak decays of heavy
hadrons  \cite{BHS}.  Each gluon
propagator with four-momentum $k^\mu$ in the hard-scattering quark-gluon
scattering amplitude $T_H$ can be associated with the coupling
$\alpha_V(k^2)$ since the
gluon exchange propagators closely resembles the interactions encoded in the
effective potential $V(Q^2)$.  [In Abelian theory this is exact.]
Commensurate scale
relations can then be
established which connect the hard-scattering subprocess amplitudes which
control exclusive processes to other QCD observables.

We can anticipate that eventually
nonperturbative
methods such as lattice gauge theory or discretized light-cone quantization will
provide a complete form for the heavy quark potential in $QCD$.  It
is reasonable to assume that $\alpha_V(Q)$ will not diverge at small space-like
momenta.  One possibility is that $\alpha_V$ stays relatively constant
$\alpha_V(Q) \simeq 0.4$ at low momenta, consistent with fixed-point behavior.
There is, in fact, empirical evidence for freezing of the $\alpha_V$ coupling
from the observed systematic dimensional scaling behavior of exclusive
reactions  \cite{BJPR}.  If this is in fact the case, then the range of QCD
predictions can be extended to quite low momentum scales, a regime normally
avoided because of the apparent singular structure of perturbative
extrapolations.

There are a number of other advantages of the $V$-scheme:
\begin{enumerate}
\item
Perturbative expansions in $\alpha_V$ with the scale set by the momentum
transfer cannot
have any
$\beta$-function dependence in their coefficients since all running
coupling effects are
already summed into the definition of the potential.  Since
coefficients involving
$\beta_0$ cannot occur in an expansions in $\alpha_V$,  the divergent infrared
renormalon series of the form $\alpha^n_V\beta_0^n n!$ cannot occur.  The
general convergence properties of the scale $Q^*$ as an expansion in $\alpha_V$
is not known  \cite{Mueller}.

\item
The effective coupling $\alpha_V(Q^2)$ incorporates vacuum polarization
contributions with finite fermion masses.  When continued to time-like
momenta, the coupling has the correct analytic dependence dictated by the
production thresholds in the $t$ channel.  Since $\alpha_V$ incorporates quark
mass effects exactly, it avoids the problem of explicitly computing and
resumming
quark mass corrections.

\item
The $\alpha_V$ coupling is the natural expansion parameter for
processes involving non-relativistic momenta, such as heavy quark production at
threshold where the Coulomb interactions, which are enhanced at low relative
velocity $v$ as $\pi \alpha_V/v$, need to be
re-summed  \cite{Voloshin,Hoang,Fadin}.
The effective Hamiltonian for nonrelativistic QCD is thus most naturally
written in
$\alpha_V$ scheme.
The threshold corrections
to heavy quark production in $e^+ e^-$ annihilation depend on
$\alpha_V$ at specific scales $Q^*$.  Two distinct ranges of scales arise
as arguments of
$\alpha_V$ near threshold: the relative momentum of the quarks governing the
soft gluon exchange responsible for the Coulomb potential, and a high momentum
scale, induced by
hard gluon exchange, approximately equal to twice the quark mass for the
corrections
 \cite{Hoang}.  One thus can use threshold production to obtain a direct
determination of $\alpha_V$ even at low scales.  The corresponding QED results
for $\tau$ pair production allow for a measurement of the magnetic moment of the
$\tau$ and could be tested at a future $\tau$-charm
factory  \cite{Voloshin,Hoang}.

\end{enumerate}

We also note that
computations in different sectors of the Standard Model have been
traditionally carried out using different renormalization schemes.
However, in a grand
unified theory, the forces between all of the particles in the fundamental
representation should become universal above the grand unification scale.
Thus it is
natural to use $\alpha_V$ as the effective charge for all sectors of a grand
unified theory,  rather than in
a convention-dependent coupling such as $\alpha_{\overline {MS}}$.

\section{ Extension of the $\bar{MS}$ Scheme}

The standard ${\overline {MS}}$ scheme
is not an analytic function of the renormalization scale at heavy quark
thresholds;
in the running of the coupling the quarks are taken as massless, and
at each quark threshold the value of $N_F$ which appears in the $\beta$
function is
incremented.  Thus Eq. (\ref{eq:csrmsovf}) is technically only valid far
above a heavy quark threshold.  However, we can use this commensurate scale
relation to define
an extended
$\overline {MS}$ scheme which is continuous and analytic at any scale.  The new
modified scheme inherits all of the good properties of the $\alpha_V$ scheme,
including its correct analytic properties as a function of the quark masses
and its
unambiguous scale fixing  \cite{bgmr}.
Thus we define
\begin{equation}
\widetilde {\alpha}_{\overline{\mbox{\tiny MS}}}(Q)
= \alpha_V(Q^*) + \frac{2N_C}{3} {\alpha_V^2(Q^{**})\over\pi} +
\cdots ,
\label{alpmsbar2}
\end {equation}
for all scales $Q$.  This equation not only provides an analytic
extension of the $\overline{MS}$ and similar schemes, but it also ties down the
renormalization scale to the physical masses of the quarks as they
enter into the vacuum polarization contributions to $\alpha_V$.

The modified scheme \amst\ provides an analytic interpolation of
conventional $\overline{MS}$ expressions by utilizing the mass dependence of the
physical \av\ scheme.  In effect, quark thresholds are treated
analytically to all orders in $m^2/Q^2$; \ie, the evolution of the analytically
extended coupling in the intermediate regions reflects the actual mass
dependence of a physical effective charge and the analytic properties of
particle production.
Just as in Abelian QED, the mass dependence of the effective potential
and the analytically extended scheme \amst\ reflects the analyticity of the
physical thresholds for particle production in the
crossed channel.  Furthermore, the definiteness of the dependence in the quark
masses automatically constrains the renormalization scale.  There
is thus no scale ambiguity in perturbative expansions in \av\ or \amst.

In leading order the effective number of flavors in the modified scheme
\amst\ is given to a very good approximation by the simple form \cite{bgmr}
\begin{equation}
\widetilde {N}_{F,\overline{\mbox{\tiny MS}}}^{(0)}\left(\frac{m^2}{Q^2}\right)
\cong \left(1 + {5m^2 \over {Q^2\exp({5\over 3})}} \right)^{-1}
\cong \left( 1 + {m^2 \over Q^2} \right)^{-1}.
\end{equation}
Thus the contribution from one flavor is $\simeq 0.5$ when
the scale $Q$ equals the quark mass $m_i$.  The standard procedure
of matching $\alpha_{\overline{\mbox{\tiny MS}}}(\mu)$ at the quark
masses serves as a zeroth-order approximation to the continuous $N_F$.

\begin{figure}[htb]
\begin{center}
\leavevmode
\epsfxsize=4in
\epsfbox{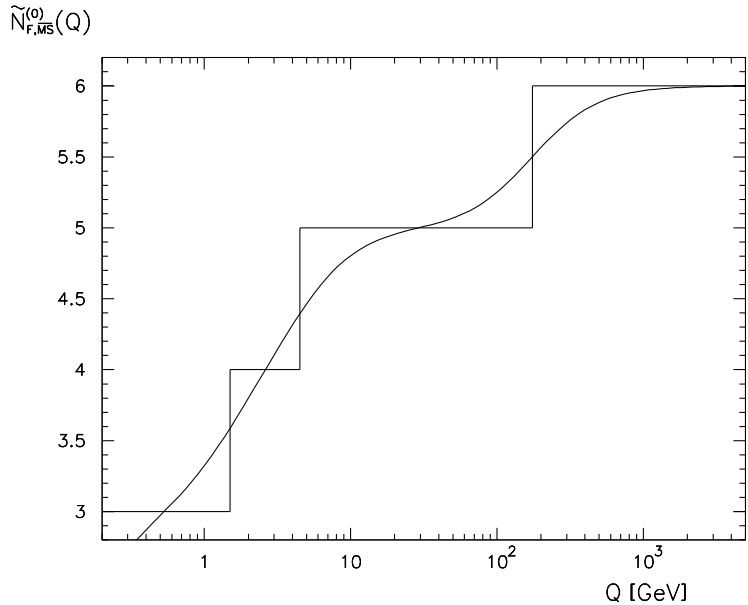}
\end{center}
\caption[*]{The continuous
$\widetilde {N}_{F,\overline{\mbox{\tiny MS}}}^{(0)}$ in the analytic
extension of the $\overline{\mbox{MS}}$ scheme as a
function of the physical scale $Q$.  (For reference the
continuous $N_F$ is also compared with
the conventional procedure of taking $N_F$ to be a step-function at the
quark-mass thresholds.)}
\label{fig:nfsum}
\end{figure}

Adding all flavors together gives the total
$\widetilde {N}_{F,\overline{\mbox{\tiny MS}}}^{(0)}(Q)$
which is shown in Fig.~\ref{fig:nfsum}.  For reference, the
continuous $N_F$ is also compared with
the conventional procedure of taking $N_F$ to be a step-function at the
quark-mass thresholds.
The figure shows clearly that there are hardly any plateaus at all
for the continuous
$\widetilde {N}_{F,\overline{\mbox{\tiny MS}}}^{(0)}(Q)$ in
between the quark masses.
Thus there is really no scale below 1 TeV where
$\widetilde {N}_{F,\overline{\mbox{\tiny MS}}}^{(0)}(Q)$
can be approximated by a constant; for all $Q$ below 1 TeV there is always
one quark
with mass $m_i$ such that $m_i^2 \ll Q^2$ or $Q^2 \gg m_i^2$ is not
true.
We also note that if one would use any other scale than the
BLM-scale for $\widetilde {N}_{F,\overline{\mbox{\tiny MS}}}^{(0)}(Q)$,
the result would be to increase the difference between the analytic
$N_F$ and the standard procedure of using the step-function at the
quark-mass thresholds.

\begin{figure}[htb]
\begin{center}
\leavevmode
\epsfxsize=4in
\epsfbox{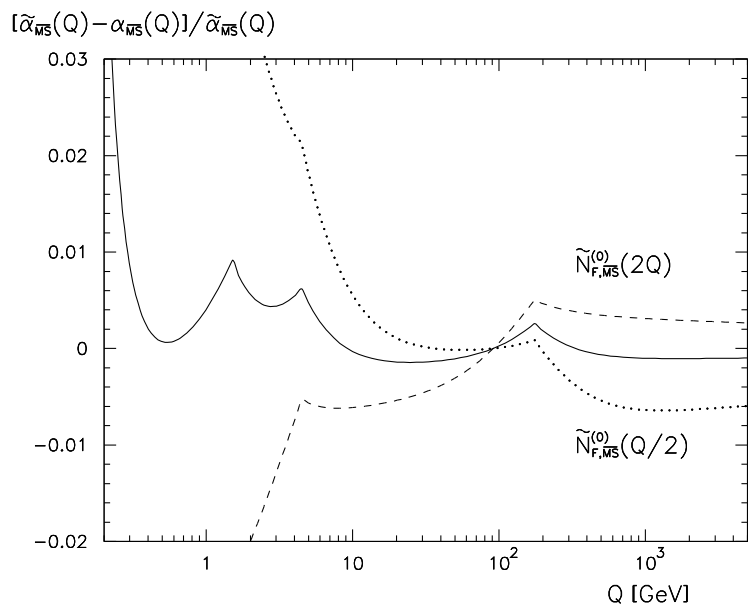}
\end{center}
\caption[*]{The solid curve shows the relative difference between the
solutions to
the 1-loop renormalization group equation using continuous $N_F$,
$\widetilde{\alpha}_{\overline{\mbox{\tiny MS}}}(Q)$, and conventional discrete
theta-function thresholds, $\alpha_{\overline{\mbox{\tiny MS}}}(Q)$.
The dashed (dotted) curves shows the same quantity but using the scale $2Q$
($Q/2$)
in $\widetilde {N}_{F,\overline{\mbox{\tiny MS}}}^{(0)}$.  The solutions
have been
obtained numerically starting from the world average \cite{Burrows}\
$\alpha_{\overline{\mbox{\tiny MS}}}(M_Z) = 0.118$.}
\label{fig:adiff}
\end{figure}

Figure~\ref{fig:adiff} shows the relative difference between the two
different solutions of the 1-loop renormalization group equation,
\ie\ $(\widetilde{\alpha}_{\overline{\mbox{\tiny MS}}}(Q)-
           {\alpha}_{\overline{\mbox{\tiny MS}}}(Q) )/
           \widetilde{\alpha}_{\overline{\mbox{\tiny MS}}}(Q)$.
The solutions have been obtained numerically starting from the
world average \cite{Burrows}
$\alpha_{\overline{\mbox{\tiny MS}}}(M_Z) = 0.118$.
The figure shows that
taking the quark masses into account in the running leads to
effects of the order of one percent which are most especially
pronounced near thresholds.

The extension
of the $\overline{\mbox{MS}}$-scheme proposed here provides a coupling
which is an analytic function of both the scale and the quark masses.
The new
modified coupling $\widetilde {\alpha}_{\overline{\mbox{\tiny MS}}}(Q)$
inherits most of the good properties of the $\alpha_V$ scheme, including its
correct analytic properties as a function of the quark masses and its
unambiguous scale fixing~ \cite{bgmr}.
However, the conformal coefficients in the commensurate scale
relation between the $\alpha_V$ and
$\overline{\mbox{MS}}$ schemes does not preserve one of the
defining criterion
of the potential expressed in the bare charge, namely the non-occurrence of
color factors corresponding to an iteration of the potential.  This
is probably an effect of the breaking of conformal invariance by
the $\overline{\mbox{MS}}$ scheme.  The breaking of conformal symmetry
has also been observed when dimensional regularization is used as a
factorization scheme in both
exclusive \cite{Brodsky:1986ve,Frishman,Muller} and
inclusive \cite{Blumlein} reactions.  Thus, it does not turn out to be
possible to extend the modified scheme
${\widetilde \alpha}_{\overline{\mbox{\tiny MS}}}$ beyond leading order
without running into an intrinsic contradiction with conformal
symmetry.

The observation of rapidly increasing structure functions in deep
inelastic scattering at small-$x_{bj}$ and the observation
of rapidly increasing diffractive processes such as $\gamma^* p \to \rho
p$ at high energies at HERA is in agreement with the expectations of the
BFKL \cite{BFKL} QCD high-energy limit.  The highest eigenvalue,
$\omega^{\rm max}$, of the leading order (LO) BFKL equation \cite{BFKL} is
related to the intercept of the Pomeron which in turn governs
the high-energy asymptotics of the cross sections: $\sigma \sim
s^{\alpha_{I \negthinspace P}-1} = s^{\omega^{\rm max}}$.
The BFKL Pomeron intercept in LO turns out to be rather large:
$\alpha_{I \negthinspace P} - 1 =\omega_L^{\rm max} =
12 \, \ln2 \, ( \alpha_S/\pi ) \simeq 0.55 $ for
$\alpha_S=0.2$; hence, it is very important to know the next-to-leading order
(NLO) corrections.

Recently the NLO corrections to the BFKL resummation
of energy logarithms were calculated \cite{FL,CC98} by employing the
modified minimal subtraction scheme ($\overline{\mbox{MS}}$)
 \cite{Bar78} to regulate the ultraviolet divergences with arbitrary
scale setting.  The NLO corrections \cite{FL,CC98}
to the highest eigenvalue of the BFKL equation turn out to be negative
and even larger than the LO contribution for $\alpha_s > 0.157$.
It is thus important to analyze the NLO BFKL resummation of
energy logarithms
 \cite{FL,CC98} in physical renormalization schemes and apply the BLM-CSR
method.  In fact, as shown in a recent paper \cite{BBFKL}, the
reliability of QCD predictions for the intercept of the BFKL Pomeron at
NLO when evaluated using BLM scale setting
 \cite{BLM} within non-Abelian
physical schemes, such as the momentum space
subtraction (MOM) scheme \cite{Cel79,Pas80}
or the $\Upsilon$-scheme based on $\Upsilon \rightarrow ggg$ decay,
is significantly improved compared to the $\overline{\mbox{MS}}$-scheme.

The renormalization scale ambiguity problem can be resolved if one can
optimize the choice of scales and renormalization schemes according to some
sensible criteria.  In the BLM optimal scale
setting \cite{BLM}, the renormalization scales are chosen such that
all vacuum polarization effects from the QCD $\beta$-function are resummed
into the running couplings.  The coefficients of the perturbative series are
thus identical to the perturbative coefficients of the corresponding
conformally invariant theory with $\beta=0$.

In the present case one can show that within the
V-scheme (or the $\overline{\mbox{MS}}$-scheme)
the BLM procedure does not change significantly the value of
the NLO coefficient $r(\nu)$.
This can be understood since the V-scheme, as well as
$\overline{\mbox{MS}}$-scheme, are adjusted primarily to
the case when in the LO there are dominant QED (Abelian) type
contributions, whereas in the BFKL case there are important LO
gluon-gluon (non-Abelian) interactions.
Thus one can choose for the BFKL case the MOM-scheme
 \cite{Cel79,Pas80} or the $\Upsilon$-scheme based on
$\Upsilon \rightarrow ggg$ decay.

Adopting BLM scale setting, the NLO BFKL eigenvalue
in the MOM-scheme is
\begin{equation}
\omega_{BLM}^{MOM}(Q^{2},\nu) =
N_C \chi_{L} (\nu) \frac{\alpha_{MOM}(Q^{MOM \, 2}_{BLM})}{\pi}
\Biggl[1 +
r_{BLM}^{MOM} (\nu) \frac{\alpha_{MOM}(Q^{MOM \, 2}_{BLM})}{\pi} \Biggr] ,
\end{equation}
$$
r_{BLM}^{MOM} (\nu) = r_{MOM}^{conf} (\nu) \, .
$$

The $\beta$-dependent part of the $r_{MOM}(\nu)$ defines the
corresponding BLM optimal scale
\begin{equation}
Q^{MOM \, 2}_{BLM} (\nu) = Q^2 \exp
\Biggl[ - \frac{4 r_{MOM}^{\beta}(\nu)}{\beta_0} \Biggr]
= Q^2 \exp \Biggl[ \frac
{1}{2}\chi_L (\nu) - \frac{5}{3} + 2 \biggl(1+\frac{2}{3} I \biggr) \Biggr].
 \nonumber
\label{qblm}
\end{equation}
At $\nu=0$ we
have $Q^{MOM \, 2}_{BLM} (0) = Q^2 \bigl( 4 \exp [2(1+2 I /3)-5/3] \bigr) \simeq
Q^2  \, 127$.  Note that $Q^{MOM \, 2}_{BLM}(\nu)$ contains a large factor,
$\exp [- 4 T_{MOM}^{\beta}/\beta_0 ] = \exp [2(1+2 I /3)] \simeq 168$, which
reflects a large kinematic difference between MOM- and
$\overline{\mbox{MS}}$- schemes \cite{Cel83,BLM}.

One of the striking features of this analysis is that the NLO value for
the intercept of the BFKL Pomeron, improved by the BLM procedure, has a
very weak dependence on the gluon virtuality $Q^2$.
This agrees with the conventional Regge-theory where
one expects an universal intercept of the Pomeron without any $Q^2$-dependence.
The minor $Q^2$-dependence obtained, on one side, provides near insensitivity
of the results to the precise value of $\Lambda$, and, on the other side, leads
to approximate scale and conformal invariance.  Thus one may use conformal
symmetry \cite{Lipatov97,Lipatov86} for the continuation of the present results
to the case $t \neq 0$.

The NLO corrections to the BFKL
equation for the QCD Pomeron thus become controllable
and meaningful provided one uses physical renormalization scales and
schemes relevant to non-Abelian gauge theory.  BLM optimal scale setting
automatically sets the appropriate physical renormalization scale by
absorbing the non-conformal $\beta$-dependent coefficients.  The
strong renormalization scheme dependence of the NLO corrections to BFKL
resummation then largely disappears.  This is in contrast to the unstable
NLO results obtained in the conventional $\overline{\mbox{MS}}$-scheme with
arbitrary choice of renormalization scale.
A striking feature of the NLO BFKL Pomeron intercept in
the BLM approach is its very weak $Q^2$-dependence, which provides
approximate conformal invariance.

The new results presented here open new windows for applications
of NLO BFKL resummation to high-energy phenomenology.

Recently the $L3$ collaboration at LEP{L3} has presented new results for
the virtual photon cross section $\sigma(\gamma^*(Q_A) \gamma^*(Q_b) \to
{\rm hadrons}$ using double tagged $e^+ e^- \to e^+ e^- {\rm hadrons}.$
This process provides a remarkably clean possible test of the perturbative
QCD pomeron since there are no initial hadrons  \cite{bhsprl}.
The calculation of $\sigma (\gamma^* \gamma^*)$ and is discussed in
detail in references~ \cite{bhsprl}.  We note here some important features:

i) for large virtualities,  $\sigma (\gamma^* \gamma^*)$ the
longitudinal cross section $\sigma_{LL}$ dominates and scales like
$1/Q^2$, where $Q^2 \sim {\mbox {\rm max}} \{ Q_A^2, Q_B^2\} $.  This is
characteristic of the perturbative QCD prediction.  Models based on
Regge factorization (which work well in the soft-interaction regime
dominating $\gamma \, \gamma$ scattering near the mass shell) would
predict a higher power in $1/Q$.

ii) $\sigma (\gamma^* \gamma^*)$ is affected by logarithmic
corrections in the energy $s$ to all orders in $\alpha_s$.  As a result
of the BFKL summation of these contributions, the cross section rises
like a power in $s$, $\sigma \propto s^\lambda$.  The Born
approximation to this result (that is, the ${\cal O} (\alpha_s^2) $
contribution, corresponding to single gluon exchange
gives a constant cross section, $\sigma_{\rm Born} \propto s^0$.
A fit to photon-photon
sub-energy dependence measured by L3 at
$\sqrt s_{e^+e^-} = 91~{\rm GeV}$ and $\VEV{Q_A^2}=\VEV{Q_A^2}
= 3.5~{\rm GeV}^2$
gives
$\alpha_P-1 = 0.28 \pm 0.05$.  The L3 data at $\sqrt s_{e^+e^-} = 183~{\rm
GeV}$ and
$\VEV{Q_A^2}=\VEV{Q_A^2} = 14~{\rm GeV}^2,$ gives
$\alpha_P-1 = 0.40 \pm 0.07$
which shows a rise of the virtual photon cross section much stronger
than single gluon or soft pomeron exchange, but it is compatible with the
expectations from the NLO scale- and scheme-fixed BFKL predictions.  It
will be crucial to measure the $Q_A^2$ and $Q_B^2$ scaling and
polarization dependence and compare with the detailed predictions
of PQCD \cite{bhsprl}.

Commensurate scale relations have a number of attractive properties:
\begin{enumerate}
\item
The ratio of physical scales $Q_A/Q_B$ which appears in commensurate scale
relations
reflects the relative position of physical thresholds, \ie\ quark
anti-quark pair
production.
\item
The functional dependence and perturbative expansion of the CSR are identical to
those of a conformal scale-invariant theory where $\beta_A(\alpha_A)=0$
and $\beta_B(\alpha_B)=0$.
\item
In the case of theories approaching fixed-point behavior
$\beta_A(\bar\alpha_A)=0$ and
$\beta_B(\bar\alpha_B)=0$, the commensurate scale relation relates both
the ratio of
fixed point couplings $\bar\alpha_A/\bar\alpha_B$, and the ratio of
scales as the fixed point is approached.
\item
Commensurate scale relations satisfy the Abelian correspondence principle
 \cite{Brodsky:1997jk};
\ the non-Abelian gauge theory prediction reduces to Abelian theory for
$N_C \to 0$ at
fixed $ C_F\alpha_s$ and fixed $N_F/C_F$.
\item
The perturbative expansion of a commensurate scale relation has the same
form as a
conformal theory, and thus has no
$n!$ renormalon growth arising from the $\beta$-function  \cite{Gardi}.
It is an interesting conjecture whether the perturbative expansion relating
observables to observable are in fact free of all $n!$ growth.  The
generalized Crewther relation, where the commensurate relation's perturbative
expansion forms a geometric series to all orders, has convergent behavior.
\end{enumerate}

Virtually any perturbative QCD prediction can be written in the form of a
commensurate
scale relation, thus eliminating any uncertainty due to renormalization
scheme or scale
dependence.  Recently it has been shown \cite{Brodsky:1998ua} how the
commensurate scale relation between the radiative corrections to
$\tau$-lepton decay and
$R_{e^+e^-}(s)$
can be generalized and empirically tested for arbitrary $\tau$ mass and nearly
arbitrarily functional dependence of the $\tau$ weak decay matrix element.

An essential feature of the \av(Q) scheme is the absence of any
renormalization scale ambiguity, since $Q^2$ is,  by definition, the square of
the physical momentum transfer.  The \av\ scheme naturally takes into
account quark mass
thresholds,  which is of particular phenomenological importance to QCD
applications in the mass region close to threshold.
As we have seen, commensurate scale relations provide
an analytic extension of the conventional \ms\ scheme in which many of
the advantages of the \av\ scheme are inherited by the \amst\ scheme,
but only minimal changes have to be made.
Given the commensurate scale relation connecting \amst\ to \av\, expansions in
\amst\ are effectively expansions in \av\ to the given order in perturbation
theory at a corresponding commensurate scale.

The calculation of $\psi_V^{(1)}$, the two-loop
term in the Gell-Mann Low function for the $\alpha_V$ scheme,
with massive quarks gives for the first time a gauge invariant
and renormalization scheme independent two-loop result for the
effects of quarks masses in the running of the coupling.  Renormalization
scheme independence is achieved by using the pole mass definition for
the ``light" quarks which contribute to the scale dependence of
the static heavy quark potential.  Thus the pole mass and the $V$-scheme
are closely connected and have to be used in conjunction to give
reasonable results.

It is interesting that the effective number of flavors in the two-loop
coefficient of the Gell-Mann Low function in the $\alpha_V$ scheme,
$N_{F,V}^{(1)}$, becomes negative for intermediate values of
$Q/m$.  This feature can be
understood as anti-screening from the non-Abelian contributions and
should be contrasted with the QED case where the effective number of
flavors becomes larger than $1$ for intermediate $Q/m$.  For small
$Q/m$ the heavy quarks decouple explicitly as expected in a physical
scheme, and for large $Q/m$ the massless result is retained.

The analyticity of the $\alpha_V$ coupling can be utilized to
obtain predictions for any perturbatively calculable observables
including the
finite quark mass effects associated with the running of the
coupling.  By employing the commensurate scale relation method,
observables which have been calculated in the
$\overline{\mbox{MS}}$ scheme can be related to the analytic V-scheme
without any scale ambiguity.  The commensurate scale relations
provides the relation between the physical scales of two effective charges
where they pass through a common flavor threshold.  We also note the
utility of the \av\ effective charge in
supersymmetric and grand unified theories, particularly since the
unification of couplings and masses would be expected to occur in terms
of physical quantities rather than parameters defined by theoretical
convention.

As an example we have showed in Ref.  \cite{bgmr} how to calculate the
finite quark mass
corrections connected with the running of the coupling for
the non-singlet hadronic width of the Z-boson compared
with the standard treatment in the $\overline{\mbox{MS}}$ scheme.
The analytic treatment in the V-scheme gives a simple and straightforward
way of incorporating these effects for any observable.  This should
be contrasted with the $\overline{\mbox{MS}}$ scheme where higher
twist corrections due to finite quark mass threshold effects have to be
calculated separately for each observable.
The V-scheme is especially
suitable for problems where the quark masses are important such as for
threshold production of heavy quarks and the hadronic width of the
$\tau$ lepton.

It has also been shown that the NLO corrections to
the BFKL equation for the QCD Pomeron become controllable
and meaningful provided one uses physical renormalization scales and
schemes relevant to non-Abelian gauge theory.  BLM optimal scale setting
automatically sets the appropriate physical renormalization scale by
absorbing the non-conformal $\beta$-dependent coefficients.  The
strong renormalization scheme dependence of the NLO corrections to BFKL
resummation then largely disappears.  This is in contrast to the unstable
NLO results obtained in the conventional $\overline{\mbox{MS}}$-scheme with
arbitrary choice of renormalization scale.
A striking feature of the NLO BFKL Pomeron intercept in
the BLM/CSR approach is its very weak $Q^2$-dependence, which provides
approximate conformal invariance.
The new results presented here open new windows for applications
of NLO BFKL resummation to high-energy phenomenology, particularly virtual
photon-photon scattering.

\section*{Outlook}

The traditional focus of theoretical work in QCD has been on hard
inclusive processes and jet physics where perturbative methods and
leading-twist factorization provide predictions up to next-to-next-to
leading order.  Most of these predictions appear to be validated by
experiment with good precision.  More recently, the
domain of reliable perturbative QCD predictions has been extended to
much more complex phenomena, such as the BFKL approach to the hard QCD pomeron
in deep inelastic scattering at small $x_{bj}$,
 \cite{Balitsky:1978ic} virtual photon scattering  \cite{Brodsky:1997sd},
and the energy dependence of hard virtual photon diffractive
processes, such as $\gamma^* p \to
\rho^0 p$  \cite{BGMFS}.

Exclusive hard-scattering reactions and hard diffractive reactions are now
providing
an invaluable window into the structure and
dynamics of hadronic amplitudes.  Recent measurements of the
photon-to-pion transition form factor at CLEO  \cite{Gronberg:1998fj}, the
diffractive dissociation of pions into jets at Fermilab  \cite{E791},
diffractive vector meson leptoproduction at Fermilab and HERA, and the new
program
of experiments on exclusive proton and deuteron processes at Jefferson
Laboratory
are now yielding fundamental information on hadronic wavefunctions,
particularly the
distribution amplitude of mesons.  There is now strong evidence for color
transparency from such processes.  Such information is also critical for
interpreting exclusive heavy hadron decays and the matrix elements and
amplitudes
entering $CP$-violating processes at the $B$ factories.

It many ways the study of quantum chromodynamics is just beginning.  The
most important features of the theory remain to be solved, such as the
problem of
confinement in QCD, the behavior of the QCD coupling in the infrared,
the phase
and vacuum structure/zero mode structure of QCD, the fundamental
understanding of hadronization and parton coalescence at the amplitude
level, and
the nonperturbative structure of hadron wavefunctions.
There are also
still many outstanding phenomenological puzzles in QCD.  The precise
interpretation of $CP$ violation and the weak interaction parameters in
exclusive
$B$ decays will require a full
understanding of the QCD physics of hadrons.

Light-cone quantization methods
appear to be especially well suited for progress in understanding the
relevant nonperturbative structure of the theory.  Since the Hamiltonian
approach
is formulated in Minkowski space, predictions for the hadronic phases needed
for CP violation studies can be obtained.  In these lectures I have
discussed how light-cone Fock-state
wavefunctions can be used to encode the properties of a hadron in terms of
its fundamental quark and gluon degrees of freedom.  Given the proton's
light-cone wavefunctions, one can compute not only the quark and gluon
distributions measured in deep inelastic lepton-proton scattering, but
also the multi-parton correlations which control 
helicity correlations in polarized leptoproduction \cite{Mulders:1999nt}, 
the distribution of
particles in the proton fragmentation region and dynamical higher twist
effects.  Light-cone wavefunctions also provide a systematic framework for
evaluating exclusive hadronic matrix elements, including timelike heavy
hadron decay amplitudes and form factors.

Commensurate scale relations promise a
new level of precision in perturbative QCD predictions which are devoid of
renormalization scale and renormalon ambiguities.
However, progress in QCD is driven by
experiment, and we are fortunate that there are
new experimental facilities such as Jefferson laboratory, the upcoming QCD
studies of exclusive processes $e^+ e^-$ and $\gamma \gamma$ processes at
the high
luminosity
$B$ factories, as well as the new accelerators and colliders now being
planned to
further advance the study of QCD phenomena.

\section*{Acknowledgments}
 Work supported by the Department of Energy, contract
DE--AC03--76SF00\-515.
Many of the results presented here are based on collaborations with a
number of colleagues, including
Steven Bass,
Victor Fadin,
Gregory Gaba\-dadze,
Mandeep Gill,
John Hiller,
Paul Hoyer,
Markus Diehl.
Dae Sung Hwang,
Chueng Ji,
Andrei Kataev,
Victor Kim,
Peter Lepage,
Lev Lipatov,
Hung Jung Lu,
Gary McCartor,
Michael Melles,
Chris Pauli,
Stephane Peigne,
Grigorii B. Pivovarov,
Johan Rathsman,
Ivan Schmidt,
and Prem Srivastava.
I thank
S. Dalley,
Yitzhak Frishman,
Einan Gardi,
Georges Grunberg,
Paul Hoyer,
Marek Karliner,
Carlos Merino,
Al Mueller,
and
Jose Pelaez
for helpful
conversations.  Parts of these lectures were also presented at the 12th
Nuclear Physics Summer School and Symposium (NUSS '99).
The section on commensurate scale relations is based on a review written in
collaboration with Johan Rathsman
 \cite{Brodsky:1999gm}.
I especially thank Piet Mulders and
his colleagues for their outstanding hospitality in Nijmegen.

\bigskip
\centerline
{APPENDIX I}
\centerline{LIGHT CONE QUANTIZATION AND PERTURBATION THEORY}
\bigskip

In this Appendix,  the canonical quantization of
QCD in the ghost free $A^+ =0$ light-cone gauge is given.   The discussion
follows that given in Refs. \cite{LepBrod,Brodsky:1989pv,BHLM}. The
light-cone quantization of QCD in Feynman gauge is given in Ref.
\cite{Srivastava:1999gi}
The quantization proceeds in several steps.  First one
identifies the independent dynamical degrees of freedom in the
Lagrangian.  The theory is quantized by defining commutation
relations for these dynamical fields at a given light-cone time
$\tau = t+z$ (we choose $\tau =0$).  These commutation relations
lead immediately to the definition of the Fock state basis.
Expressing dependent fields in terms of the independent fields,
we then derive a light-cone Hamiltonian, which determines the
evolution of the state space with changing $\tau$.
Finally the rules for $\tau$-ordered perturbation theory or given.
The major purpose of this exercise is to illustrate the origins
and nature of the Fock state expansion, and of light-cone
perturbation theory.  Subtleties due to the
large scale structure of non-Abelian gauge fields (\eg\
`instantons'), chiral symmetry breaking, and the like are ignored.
Although these have a profound effect on the structure of the
vacuum, the theory can still be described with a Fock state
basis and some sort of effective Hamiltonian.  Furthermore,
the short distance interactions of the theory are unaffected
by this structure, or at least this is the central ansatz of
perturbative QCD.

\medskip
\noindent
{\bf Quantization}

The Lagrangian (density) for QCD can be written
\begin{equation}
{\cal L} = -{1\over 2} \ {\rm Tr} \left( F^{\mu\nu} \,
F_{\mu\nu} \right) + {\overline \psi} \left (i \not\D -
m \right) \psi
\end{equation}
where $F^{\mu\nu} = \partial^\mu A^\nu - \partial^\nu A^\mu
+ ig [A^\mu, A^\nu]$ and $iD^\mu = i\partial^\mu - gA^\mu$.
Here the gauge field $A^\mu$ is a traceless $3\times 3$ color
matrix ($A^\mu \equiv \sum_a \, A^{a\mu} T^a$, Tr$(T^aT^b)
= 1/2\delta^{ab}$, $[T^a,T^b] = ic^{abc} T^c, \ldots$), and
the quark field $\psi$ is a color triplet spinor (for simplicity,
we include only one flavor).  At a given light-cone time, say
$\tau =0$, the independent dynamical fields are $\psi_\pm \equiv
\Lambda_\pm \psi$ and $A_\perp^i$ with conjugate fields
$i\psi_+^{\dag}$ and $\partial^+ A_\perp^i$, where $\Lambda_\pm
= \gamma^o\gamma^\pm /2$ are projection operators $(\Lambda_+
\Lambda_- = 0, \ \Lambda_\pm^2 = \Lambda_\pm, \ \Lambda_+ +
\Lambda_- = 1)$ and $\partial^\pm = \partial^0 \pm \partial^3$.
Using the equations of motion, the remaining fields in ${\cal L}$
can be expressed in terms of $\psi_+, \ A_\perp^i$:
\begin{eqnarray}
\psi_- &\equiv& \Lambda_- \psi = {1\over i\partial^+}
\left[ i{\vec D}_\perp \cdot {\vec \alpha}_\perp + \beta m \right]
\psi_+ \nonumber \\
&=& \widetilde \psi_- - {1\over i\partial^+} \ g{\vec A}_\perp
\cdot {\vec \alpha}_\perp \, \psi_+ \ , \nonumber \\
A^+ &=& 0 \ , \nonumber \\
A^- &=&{2\over i\partial^+} \, i{\vec \partial}_\perp \cdot
{\vec A}_\perp + {2g\over (i\partial^+)^2} \left\{ \left[
i\partial^+ A_\perp^i, A_\perp^i \right] + 2\psi_+^{\dag}
\, T^a \, \psi_+ \, T^a \right\} \nonumber \\
&\equiv& \wA^- + {2g\over (i\partial^+)^2}
\left\{ \left[ i\partial^+ A_\perp^i, A_\perp^i \right]
+ 2\psi_+^{\dag} \, T^a \, \psi_+ \, T^a \right\} \ ,
 \end{eqnarray}
with $\beta = \gamma^o$ and ${\vec \alpha}_\perp = \gamma^o
{\vec \gamma}$.

To quantize, we expand the fields at $\tau =0$ in terms of
creation and annihilation operators,
\begin{eqnarray}
\psi_+ (x)
&=& \int_{k^+ >0} {dk^+ \, d^2 k_\perp \over k^+ \, 16\pi^3} \sum_\lambda
\left\{ b({\underline k}, \lambda) \ u_+({\underline k}, \lambda) \
e^{-ik\cdot x}
\right.\nonumber \\
&&+ \left. d^{\dag} ({\underline k}, \lambda) \ v_+ ({\underline k},
\lambda) \ e^{ik\cdot x} \right\} \ , \quad \tau = x^+ = 0 \nonumber \\
A_\perp^i (x)
&=& \int_{k^+ >0} {dk^+ \, d^2 k_\perp \over k^+ \, 16\pi^3} \sum_\lambda
\left\{ a({\underline k}, \lambda) \ \epsilon_\perp^i (\lambda) \
e^{-ik\cdot x} +
c^. c^. \right\} \ , \nonumber \\
 \tau &=& x^+ = 0 \ ,
\end{eqnarray}
with commutation relations $({\underline k} = (k^+, {\vec k}_\perp))$:
\begin{eqnarray}
\left\{ b({\underline k}, \lambda), \ b^{\dag}
({\underline p}, \lambda) \right\} &=& \left\{ d({\underline k},
\lambda), \ d^{\dag} ({\underline p}, \lambda^\prime) \right\} \nonumber \\
&=& \left[ a({\underline k}, \lambda), \ a^{\dag} ({\underline p},
\lambda^\prime) \right] \nonumber \\
&=& 16\pi^3 \, k^+ \, \delta^3 ({\underline k} - {\underline p})
\, \delta_{\lambda\lambda^\prime} \ , \nonumber \\
\left\{ b,b \right\} = \left\{ d,d \right\} &=& \ldots = 0 \ ,
 \end{eqnarray}
where $\lambda$ is the quark or gluon helicity.  These definitions
imply canonical commutation relations for the fields with their
conjugates $(\tau = x^+ = y^+ = 0, ^
{\underline x} = (x^-, x_\perp),\ldots)$:
\begin{eqnarray}
 \left\{ \psi_+ ({\underline x}),
\ \psi_+^{\dag} ({\underline y}) \right\} &=&
\Lambda_+ \, \delta^3 ({\underline x} -
{\underline y}) \ , \nonumber \\
\left[ A^i ({\underline x}), \ \partial^+ A_\perp^j ({\underline y})
\right] &=& i\delta^{ij} \, \delta^3 ({\underline x}
- {\underline y}) \ .
\end{eqnarray}

The creation and annihilation
operators define the Fock state basis for the theory at $\tau =0$,
with a vacuum $\ket{0}$ defined such that $b\ket{0} = d\ket{0} =
a\ket{0} = 0$.  The evolution of these states with $\tau$ is
governed by the light-cone Hamiltonian, $H_{LC} = P^-$, conjugate
to $\tau$.  The Hamiltonian can be
readily expressed in terms of $\psi_+$ and $A_\perp^i$:
\begin{equation}H_{LC} = H_0 + V \ , \end{equation}
where
\begin{eqnarray}
H_0 &=&\int d^3x \left\{ {\rm Tr} \left(
\partial_\perp^i A_\perp^j \partial_\perp^i A_\perp^j \right)\right.\nonumber
 \\ &+& \left. \psi_+^{\dag} \left( i\partial_\perp \cdot \alpha_\perp +
\beta m\right) {1\over i\partial^+} \left( i\partial_\perp
\cdot \alpha_\perp + \beta m \right) \psi_+ \right\} \nonumber \\
&=&\sum_{\lambda \atop {\rm colors}} \int {dk^+ \, d^2k_\perp
\over 16\pi^3 \, k^+} \left\{ a^{\dag} ({\underline k}, \lambda)
\, a({\underline k}, \lambda) {k_\perp^2 \over k^+} + b^{\dag}
({\underline k}, \lambda) \, b({\underline k}, \lambda) \right.\nonumber \\
&\times& \left. {k_\perp^2 + m^2 \over k^+} +
d^{\dag} ({\underline k}, \lambda) \, b({\underline k}, \lambda)
\, {k_\perp^2 + m^2 \over k^+} \right\} + {\rm constant}
\end{eqnarray}
is the free Hamiltonian and $V$ the interaction:
\begin{eqnarray}
 V &=&  \int d^3x \ \left\{ 2g \ {\rm Tr} \left(
i\partial^\mu \wA^\nu \left[ \wA_\mu, \wA_\nu \right] \right)
- {g^2\over 2} \, {\rm Tr} \left( \left[ \wA^\mu, \wA^\nu \right]
\left[ \wA_\mu, \wA_\nu \right] \right) \right. \nonumber \\
&+& g\bwpsi \, \not\A \, \wpsi + g^2 \, {\rm Tr} \left( \left[
i\partial^+ \wA^\mu, \wA_\mu \right] {1\over (i\partial^+)^2}
\, \left[ i\partial^+ \wA^\nu, \wA_\nu \right] \right) \nonumber \\
&+& g^2 \bwpsi \, \not\A \, {\gamma^+ \over 2i\partial^+}
\, \not\A \, \wpsi - g^2 \bwpsi \gamma^+
\left( {1\over (i\partial^+)^2} \, \left[ i\partial^+
\wA^\nu, \wA_\nu \right] \right) \wpsi \nonumber \\
&+& \left. {g^2\over 2} \ \bpsi \gamma^+ \ T^a \psi
\, {1\over (i\partial^+)^2} \, \bpsi \gamma^+
\ T^a \psi \right\} \ ,
\end{eqnarray}
with $\wpsi = \wpsi_- + \psi_+$ ($\rightarrow \psi$ as $g \rightarrow 0$)
and $\wA^\mu = (0,\wA^-,A_\perp^i)$ ($\rightarrow A^\mu$ as
$g \rightarrow 0$).  The Fock states are obviously eigenstates of
$H_0$ with
\begin{equation}
H_0 \ket{n:k_i^+, k_{\perp i}} = \sum_i \left( {k_\perp^2 + m^2
\over k^+} \right)_i \ket{n:k_i^+, k_{\perp i}} \ .
\end{equation}
It is equally obvious that they are not eigenstates of $V$,
though any matrix element of $V$ between Fock states is
trivially evaluated.

\vspace{.5cm}
\begin{figure}[htb]
\begin{center}
\leavevmode
{\epsfbox{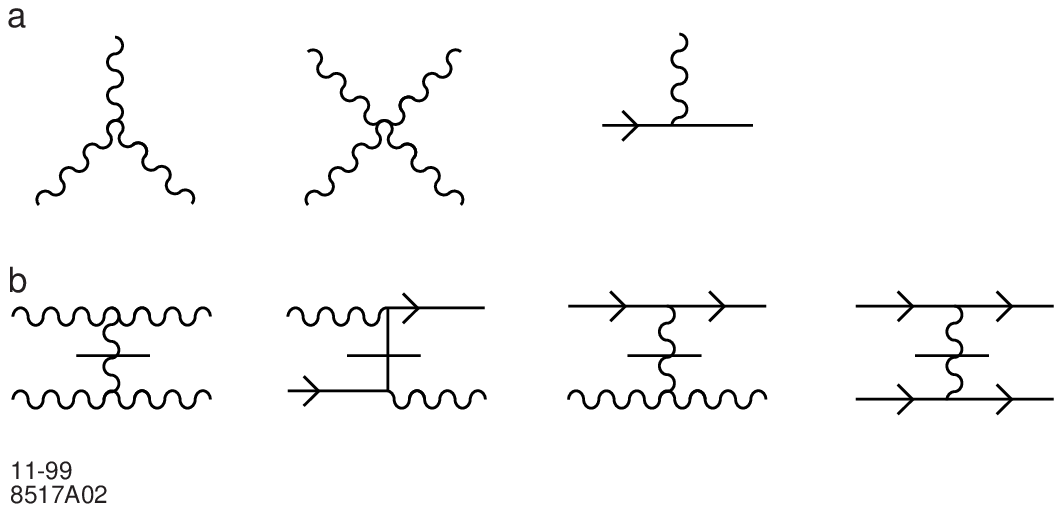}}
\end{center}
\caption[*]{Diagrams which appear in the interaction Hamiltonian
for QCD on the light cone. The propagators with horizontal
bars represent ``instantaneous" gluon and quark exchange which
arise from reduction of the dependent fields in $A^+=0$ gauge.
(a) Basic interaction vertices in QCD.
(b) ``Instantaneous" contributions.
}
\label{figa}
\end{figure}

The first three terms in $V$ correspond
to the familiar three and four gluon vertices, and the
gluon-quark vertex [ Fig. \ref{figa} (a)].  The remaining terms
represent new four-quanta
interactions containing instantaneous fermion and gluon
propagators [Fig. \ref{figa} (b)].  All terms conserve total
three-momentum ${\underline k} = (k^+, {\vec k}_\perp)$,
because of the integral over ${\underline x}$ in $V$.
Furthermore, all Fock states other than the vacuum have
total $k^+ > 0$, since each individual bare quantum has
$k^+ > 0.$   Consequently the Fock state vacuum
must be an eigenstate of $V$ and therefore an eigenstate
of the full light-cone Hamiltonian.

\medskip
\noindent
{\bf Light-Cone Perturbation Theory}

We define light-cone Green's functions to be the probability
amplitudes that a state starting in Fock state $\ket{i}$ ends
up in Fock state $\ket{f}$ a (light-cone) time $\tau$ later
\begin{eqnarray}
 \langle f|i \rangle \ G(f,i;\tau) &\equiv&
\langle f | e^{-iH_{LC}\tau/2} | i \rangle \nonumber \\
&=& i \int {d\epsilon \over 2\pi} \ e^{-i\epsilon\tau/2}
\ G(f,i;\epsilon) \, \langle f|i \rangle \ ,
 \end{eqnarray}
where Fourier transform $G(f,i;\epsilon)$ can be written
\begin{eqnarray}
\langle f|i \rangle \ G(f,i;\epsilon) & = &
\Bigl\langle f \left| {1\over \epsilon - H_{LC} + i0_+}
\right| i \Bigr\rangle \nonumber \\
&=& \Bigl\langle f \left| {1\over \epsilon - H_{LC} + i0_+}
+ {1\over \epsilon - H_0 + i0_+} \, V \,
{1\over \epsilon - H_0 + i0_+} \right. \nonumber \\
&+&  {1\over \epsilon - H_0 + i0_+} \, V \,
{1\over \epsilon - H_0 + i0_+} \, V \,
{1\over \epsilon - H_0 + i0_+} \nonumber \\
&+& \left.\ldots \right| i \Bigr\rangle \ .
 \end{eqnarray}
The rules for $\tau$-ordered perturbation theory follow
immediately  when $(\epsilon
- H_0)^{-1}$ is replaced by its spectral decomposition.
\begin{equation}
{1\over \epsilon - H_0 + i0_+} = \sum_{n,\lambda_i}
\int \prod^\sim \ {dk_i^+ \, d^2k_{\perp i} \over 16\pi^3
\, k_i^+} \ {\ket{n:{\underline k}_i, \lambda_i} \,
\bra{n:{\underline k}_i, \lambda_i} \over \epsilon
- \sum\limits_i (k^2 + m^2)_i / k_i^+ + i0_+}
\end{equation}
The sum becomes a sum over all states $n$
intermediate between two interactions.

To calculate
$G(f,i;\epsilon)$ perturbatively then, all $\tau$-ordered diagrams
must be considered, the contribution from each graph computed
according to the following rules:\cite{LepBrod}

\begin{enumerate}
\item
Assign a momentum $k^\mu$ to each line such that the total
$k^+, k_\perp$ are conserved at each vertex, and such that
$k^2 = m^2$, \ie\  $k^- = (k^2 + m^2)/k^+$.  With fermions
associate an on-shell spinor.
\begin{equation}
u({\underline k}, \lambda) = {1\over \sqrt{k^+}} \left(
k^+ + \beta m + {\vec \alpha}_\perp \cdot {\vec k}_\perp
\right) \ \left\{ \matrix{ \chi(\uparrow) &\lambda=\uparrow \cr
\chi(\downarrow) &\lambda=\downarrow \cr} \right.
\end{equation}
or
\begin{equation}
v({\underline k}, \lambda) = {1\over \sqrt{k^+}} \left(
k^+ - \beta m + {\vec \alpha}_\perp \cdot {\vec k}_\perp
\right) \ \left\{ \matrix{ \chi(\downarrow) &\lambda=\uparrow
\cr \chi(\uparrow) &\lambda=\downarrow \cr} \right.
\end{equation}
where $\chi(\uparrow) = 1/\sqrt{2} \, (1,0,1,0)$ and
$\chi(\downarrow) = 1/\sqrt{2} \, (0,1,0,-1)^T$.  For
gluon lines, assign a polarization vector $\epsilon^\mu =
(0, \ 2{\vec \epsilon}_\perp \cdot {\vec k}_\perp/k^+,
\ {\vec \epsilon}_\perp)$ where ${\vec \epsilon}_\perp
(\uparrow) = -1/\sqrt{2} \, (1,i)$ and ${\vec \epsilon}_\perp
(\downarrow) = 1/\sqrt{2} \, (1,-i)$.
\item
Include a factor $\theta(k^+)/k^+$ for each internal line.
\item
For each vertex include

\vspace{.5cm}
\begin{figure}[htb]
\begin{center}
\leavevmode
{\epsfbox{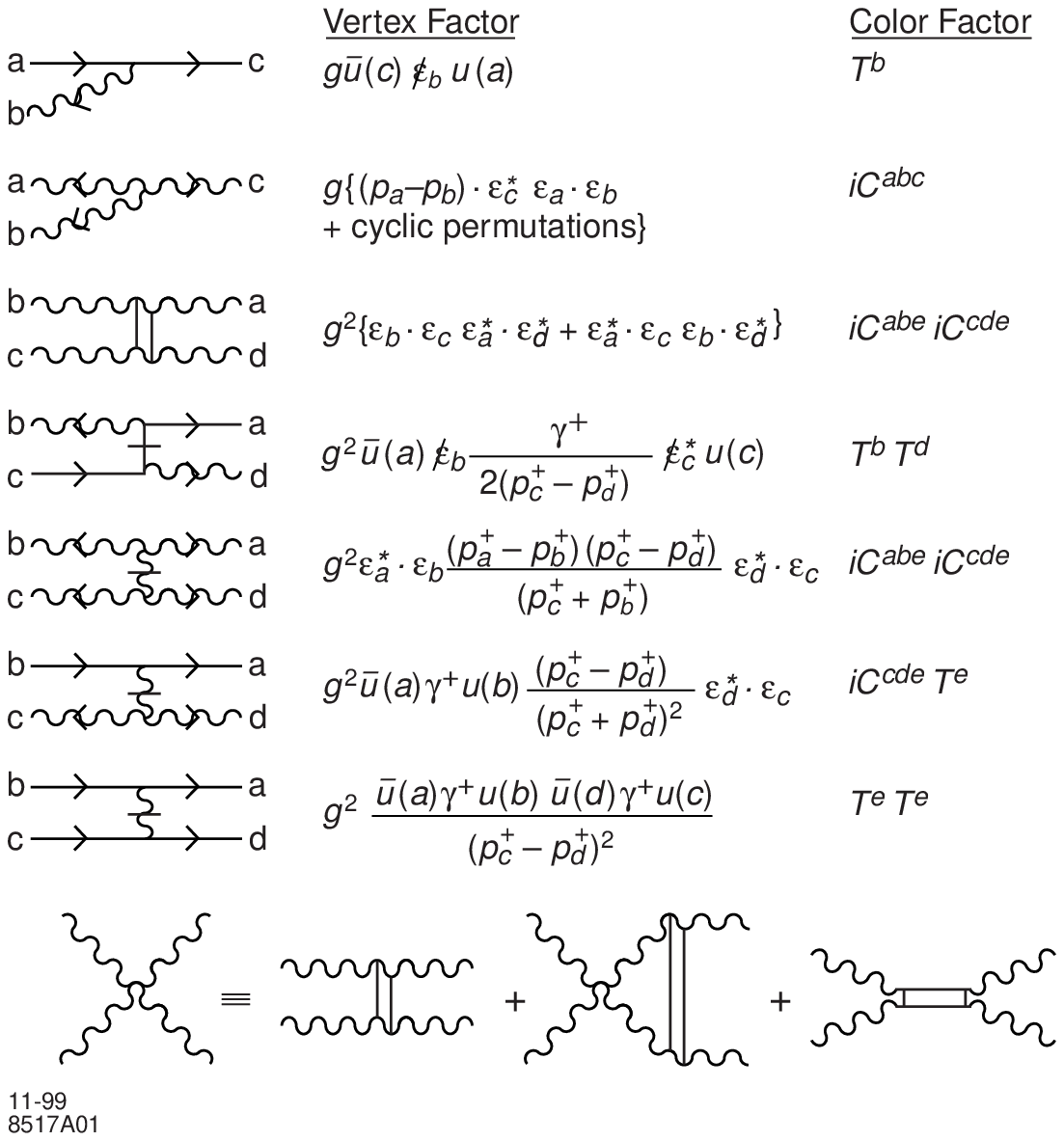}}
\end{center}
\caption[*]{Graphical rules for QCD in light-cone
perturbation theory.
}
\label{figb}
\end{figure}
factors as illustrated in
Fig. \ref{figb}.  To convert incoming into outgoing lines or vice
versa replace
\begin{equation}
u \leftrightarrow v \ , \qquad {\overline u} \leftrightarrow
-{\overline v} \ , \qquad \epsilon \leftrightarrow \epsilon^\ast
\end{equation}
in any of these vertices.

\item
For each intermediate state there is a factor
\begin{equation}
{1\over \epsilon - \sum\limits_{\rm interm} \, k^- + i0_+}
\end{equation}
where $\epsilon$ is the incident $P^-$, and the sum is
over all particles in the intermediate state.
\item
Integrate $\int dk^+ d^2k_\perp /16\pi^3$ over each
independent $k$, and sum over internal helicities and colors.
\item
Include a factor $-1$ for each closed fermion loop, for
each fermion line that both begins and ends in the initial
state (\ie \  ${\overline v} \ldots u$), and for each diagram
in which fermion lines are interchanged in either of the
initial or final states.

As an illustration, the second diagram in Fig. \ref{figb}\ contributes
\begin{eqnarray}
& &{1\over \epsilon - \sum\limits_{i=b,d} \, \left(
{k_\perp^2 + m^2 \over k^+} \right)_i} \cdot {\theta
(k_a^+ - k_b^+) \over k_a^+ - k_b^+} \nonumber \\[2ex]
&\times &{g^2 \sum\limits_\lambda \, {\overline u} (b) \,
\epsilon^\ast ({\underline k}_{\,a} - {\underline k}_{\,b},
\lambda) \, u(a) \, {\overline u} (d) \not\ep
({\underline k}_{\,a} - {\underline k}_{\,b}, \lambda) \,
u(c) \over \epsilon - \sum\limits_{i=b,c} \, \left(
{k_\perp^2 + m^2 \over k^+} \right)_i - {(k_{\perp a} -
k_{\perp b})^2 \over k_a^+ - k_b^+}} \nonumber \\[2ex]
&\cdot&
{1\over \epsilon - \sum\limits_{i=a,c} \, \left(
{k_\perp^2 + m^2 \over k^+} \right)_i}
\end{eqnarray}
(times a color factor) to the $q{\overline q} \rightarrow
q{\overline q}$ Green's function.  (The vertices for quarks
and gluons of definite helicity have very simple expressions
in terms of the momenta of the particles.)
The same rules apply for scattering amplitudes,
but with propagators omitted for external lines, and with
$\epsilon = P^-$ of the initial (and final) states.
 \end{enumerate}

Finally, notice that this quantization procedure and perturbation
theory (graph by graph) are manifestly invariant under a large
class of Lorentz transformations:
\begin{enumerate}
\item
boosts along the 3-direction---\ie \ $p^+ \rightarrow Kp^+,
\ p^- \rightarrow K^{-1}p^-, \ p_\perp \rightarrow p_\perp$ for
each momentum;
\item
transverse boosts---\ie \  $p^+ \rightarrow p^+, \ p^- \rightarrow
p^- + 2p_\perp \cdot Q_\perp + p^+ Q_\perp^2, \ p_\perp
\rightarrow p_\perp + p^+Q_\perp$ for each momentum ($Q_\perp$
like $K$ is dimensionless);
\item
rotations about the 3-direction.
\end{enumerate}

It is these invariances which lead to the frame independence
of the Fock state wave functions.

\bigskip
\begin{center}
 APPENDIX II  \\
LIGHT CONE FOCK REPRESENTATION OF\\
 ELECTROWEAK CURRENTS
\end{center}
\bigskip

The light-cone Fock representation provides an explicit form for the
matrix elements of electroweak currents
$\VEV{A|J^\mu|B}$  between hadrons $B$ and $A$. The discussion in this
appendix follows that of Ref. \cite{Brodsky:1998hn}  The underlying
formalism is the light-cone Hamiltonian Fock expansion in which  hadron
wavefunctions are decomposed on the free Fock basis of QCD. In this
formalism, the full Heisenberg current
$J^\mu$ can be equated to the current $j^\mu$ of the non-interacting theory
which in turn has simple matrix elements on the free Fock basis.

Elastic form factors at
space-like momentum transfer
$q^2 = - Q^2 < 0$ are most simply  evaluated from matrix elements of the
``good"
current $j^+ = j^0 + j^z$ in the preferred Lorentz frame
where
$q^+ = q^0+ q^z = 0$ \cite{DY,West,BD}.   The  $j^+$ current has the
advantage that it does not have large matrix elements to pair
fluctuations, so that only diagonal, parton-number-conserving transitions
need to be considered. The use of the  $j^+$ current and the $q^+ =
0$ frame brings out  striking advantage of the light-cone quantization
formalism: only diagonal, parton-number-conserving Fock state matrix elements
are required. However, in the case of the time-like form factors which occur in
semileptonic heavy hadron decays, we need to choose a frame with
$q^+ > 0$, where
$q^\mu$ is the four-momentum of the lepton pair.  Furthermore, in order to sort
out the contributions to the various weak decay form factors, we need to
evaluate the ``bad'' $-$ current $j^- = j^0 - j^z$ as well as the ``good''
current $j^+$.  In such cases
we will also require off-diagonal Fock state transitions; \ie\ the convolution
of Fock state wavefunctions differing by two quanta, a $q {\bar{q'}}$ pair.
The entire electroweak current matrix element is then in general given by the
sum of the diagonal $n \to n$ and off-diagonal
$n+1 \to n-1$ transitions. As we shall see, an important feature of a general
analysis is the emergence of
singular
$\delta(x)$ ``zero-mode" contributions from the off-diagonal matrix elements  if
the choice of frame dictates $q^+ = 0.$ The formulas \cite{Brodsky:1998hn}
are in principle exact, given the light-cone wavefunctions of hadrons.

\vspace{.5cm}
\begin{figure}[htb]
\begin{center}
\leavevmode
\epsfbox{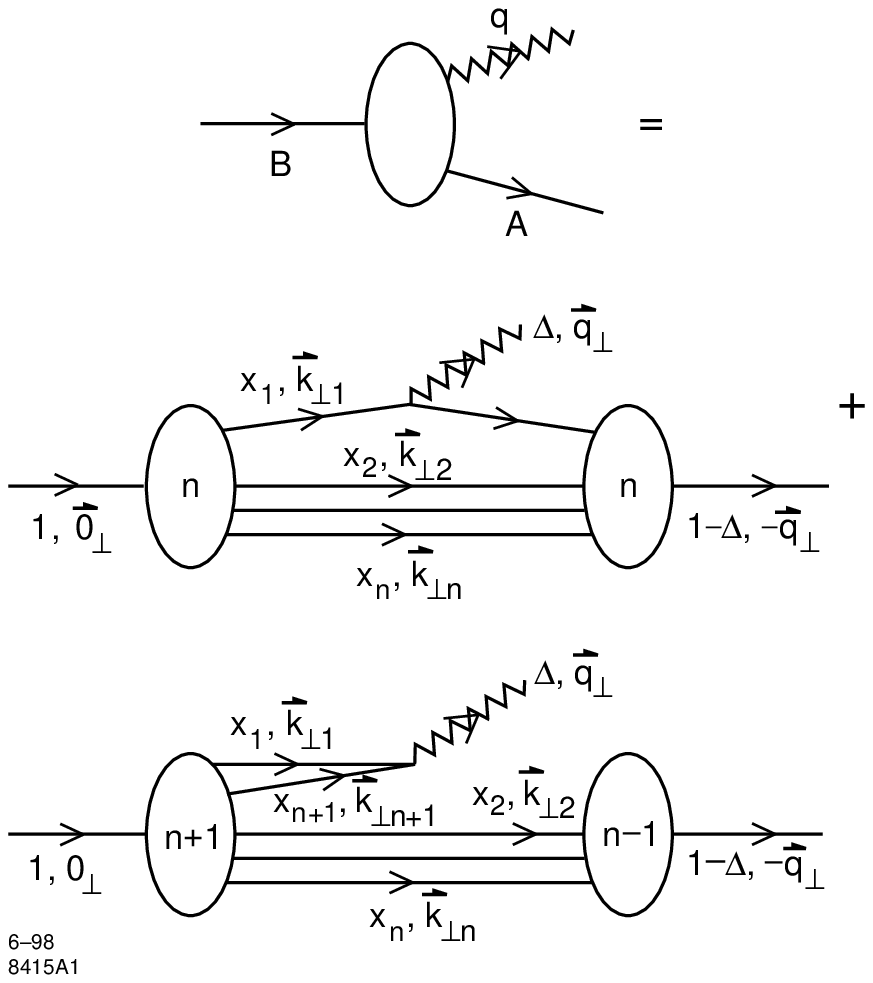}
\end{center}
\caption[*]{Exact representation of electroweak decays and time-like form
factors in the
light-cone Fock representation.
}
\label{figc}
\end{figure}

The evaluation of the timelike semileptonic decay amplitude $B \to A \ell
{\bar{\nu}}$ requires the matrix element  of the weak current between
hadron states
$\VEV{A \vert j^\mu(0) \vert B}$.
Here $x ={k^+\over  P^+} = {k^0 + k^3 \over P^0 + P^3}$ and we use the
metric convention
$a\cdot b={1\over 2}(a^+b^-+a^-b^+)-{\vec{a}}_{\perp}\cdot {\vec{b}}_{\perp}$.
 (See Fig. \ref{figc}.)
The interaction current then has simple matrix
elements of the free Fock amplitudes, with the provisal that all $x_i > 0.$  We
shall adopt the choice of a  Lorentz general frame where the outgoing leptonic
current carries
$q^\mu = \left(q^+, q_\perp, q^- \right)= \left(\Delta P^+, q_\perp,
{q^2+q^2_\perp\over \Delta P^+}\right)$.
The value of $\Delta= q^+/ P^+$ is determined from four-momentum conservation:
\begin{equation}
{q^2 + q_\perp^2\over \Delta } + {m_A^2+q_\perp^2\over 1-\Delta} = m_B^2.
\label{ff1}
\end{equation}
In the limit $\Delta \to 0$, the matrix element for the $+$ vector current
should coincide
with the Drell-Yan West formula \cite{DY,West,BD}.

For the $n \to n$ diagonal term ($\Delta n = 0$),
the final-state hadron wavefunction has arguments
$x_1-\Delta \over 1-\Delta$,
${\vec{k}}_{\perp 1} - {1-x_1\over 1-\Delta} {\vec{q}}_\perp$ for
the struck quark
and $x_i\over 1-\Delta$,
${\vec{k}}_{\perp i} + {x_i\over 1-\Delta} {\vec{q}}_\perp$
for the $n-1$ spectators.
We thus have a formula for the diagonal (parton-number-conserving) matrix
element of the form:
\begin{eqnarray}
{\VEV{A \vert J^\mu \vert B}}_{\Delta n = 0} &=&
\sum_{n, ~ \lambda}
\prod_{i=1}^{n} \int^1_{\Delta} dx_1
\int^1_0 dx_{i(i\ne 1)} \int {d^2{\vec{k}}_{\perp i} \over 2 (2\pi)^3 }
\nonumber \\[2ex]
&\times&  \delta\left(1-\sum_{j=1}^n x_j\right) ~ \delta^{(2)}
\left(\sum_{j=1}^n {\vec{k}}_{\perp j}\right)  \nonumber\\[1ex]
&&\times\
\psi^\dagger_{A (n)}(x^\prime_i, {\vec{k}}^\prime_{\perp i},\lambda_i) ~ j ^\mu
~ \psi_{B (n)}(x_i, {\vec{k}}_{\perp i},\lambda_i),
\label{t1}
\end{eqnarray}
where
\begin{equation}
\left\{ \begin{array}{lll}
x^\prime_1 = {x_1-\Delta \over 1-\Delta}\, ,\
&{\vec{k}}^\prime_{\perp 1} ={\vec{k}}_{\perp 1}
- {1-x_1\over 1-\Delta} {\vec{q}}_\perp
&\mbox{for the struck quark}\\[1ex]
x^\prime_i = {x_i\over 1-\Delta}\, ,\
&{\vec{k}}^\prime_{\perp i} ={\vec{k}}_{\perp i}
+ {x_i\over 1-\Delta} {\vec{q}}_\perp
&\mbox{for the $ (n-1)$ spectators.}
\end{array}\right.
\label{t2}
\end{equation}
A sum over all possible helicities $\lambda_i$ is understood.
If quark masses are neglected the vector and axial currents conserve helicity.
 We also can check that $\sum_i^n x^\prime_i = 1$,
$\sum_i^n {\vec{k}}^\prime_{\perp i} = {\vec{0}}_\perp$.

For the $n+1 \to n-1$ off-diagonal term ($\Delta n = -2$),
let us consider the case where
partons $1$ and
$n+1$ of the initial wavefunction annihilate into the leptonic current leaving
$n-1$ spectators.
Then $x_{n+1} = \Delta - x_{1}$,
${\vec{k}}_{\perp n+1} = {\vec{q}}_\perp-{\vec{k}}_{\perp 1}$.
The remaining $n-1$ partons have total momentum
$((1-\Delta)P^+, -{\vec{q}}_{\perp})$.
The final wavefunction then has arguments
$x^\prime_i = {x_i \over (1- \Delta)}$ and
${\vec{k}}^\prime_{\perp i}=
{\vec{k}}_{\perp i} + {x_i\over 1-\Delta} {\vec{q}}_\perp$.
We thus obtain the formula for the off-diagonal matrix element:
\begin{eqnarray}
{\VEV{A\vert J^\mu \vert B}}_{\Delta n = -2} &=&
\sum_{n ~ \lambda}
\int^{\Delta}_0 dx_1 \int^1_0 dx_{n+1}
\int {d^2{\vec{k}}_{\perp 1} \over 2 (2\pi)^3 }
\int {d^2{\vec{k}}_{\perp n+1} \over 2 (2\pi)^3 }
\nonumber \\[2ex]
 &&\times \ \prod_{i=2}^{n}
\int^1_0 dx_{i} \int {d^2{\vec{k}}_{\perp i} \over 2 (2\pi)^3 }
\delta\left(1-\sum_{j=1}^{n+1} x_j\right) ~
\delta^{(2)}\left(\sum_{j=1}^{n+1} {\vec{k}}_{\perp j}\right)
\nonumber \\[2ex]
&&\times \
\psi^\dagger_{A (n-1)}(x^\prime_i,{\vec{k}}^\prime_{\perp i},\lambda_i)
~ j^\mu
~ \psi_{B (n+1)}
\nonumber \\[1ex]
&& \times \ (\{x_1, x_i, x_{n+1} = \Delta - x_{1}\},
\nonumber\\[2ex]
&&\times \
 \{ {\vec{k}}_{\perp 1},
{\vec{k}}_{\perp i},
{\vec{k}}_{\perp n+1} = {\vec{q}}_\perp-{\vec{k}}_{\perp 1}\},
\nonumber\\[2ex]
&&\times \
\{\lambda_1,\lambda_{i},\lambda_{n+1} = - \lambda_{1}\}).
\label{t3}
\end{eqnarray}
Here $i=2,3,\cdots ,n$ with
\begin{equation}
x^\prime_i = {x_i\over 1-\Delta}\, ,\qquad
{\vec{k}}^\prime_{\perp i} ={\vec{k}}_{\perp i}
+ {x_i\over 1-\Delta} {\vec{q}}_\perp
\label{t3a}
\end{equation}
label the $n-1$ spectator
partons which appear in the final-state hadron wavefunction.
We can again check that the arguments of the final-state wavefunction
satisfy
$\sum_{i=2}^n x^\prime_i = 1$,
$\sum_{i=2}^n {\vec{k}}^\prime_{\perp i} = {\vec{0}}_\perp$.
Similarly, in gauge theory with spin-half charged constituents,
matrix elements of the
``bad'' currents $J^\perp$ and $J^-$ receive $\Delta n = \pm 1$ and
$\Delta n =  - 3$  contributions
from the induced instantaneous fermion exchange currents
$q \rightarrow \gamma^* qg$, $gq \rightarrow \gamma^*q$,
and $gq \bar q  \rightarrow \gamma^*$.  In the
case of scalars, these contributions arise
from the 4 point ``seagull'' interactions.  Note that these terms do not
occur for matrix elements of $J^+$.

The free current matrix elements $j^\mu$  in the light-cone
representation are easily constructed.
For example, the vector current of
quarks
takes the form
\begin{equation}
j^\mu =
{\bar u(x^\prime,k^\prime_\perp,\lambda^\prime)
\gamma^\mu
 u(x,k_\perp,\lambda) \over {\sqrt{k^+}} {\sqrt{k^{+ \prime}}}}
\end{equation}
and
\begin{equation}
j^+ = 2 \delta_{\lambda , \lambda^\prime }\ .
\end{equation}
The other light-cone spinor matrix elements  of
$j^\mu$ can be obtained from the
tables in Ref. \cite{LB}.
In the case of spin zero partons
\begin{equation}
j^+ = {x + x^\prime \over \sqrt{ x  x^\prime}}
\end{equation}
and
\begin{equation}
j^- =  {k^- + k^{\prime- }\over\sqrt { x x^\prime} P^+}.
\end{equation}
However, instead of evaluating each $k^-$  in the $j^-$ current from the
on-shell
condition
$k^- k^+ = m^2$, one must instead  evaluate the $k^-$ of the struck
partons from
energy conservation
$k^- = p^-_{\rm initial} - p^-_{\rm spectator}$.  This effect is seen explicitly
when one integrates the covariant current over the denominator poles in the
$k^-$ variable.  It can also be understood as due to the implicit inclusion of
local instantaneous exchange contributions obtained in  light-cone
quantization \cite{CRY,BRS}.  The mass
$m_{\rm spectator}^2$ which is needed for the evaluation of
$j^-$ current in the diagonal case is the mass of the entire spectator system.
Thus
$m^2_{\perp {\rm spectator}} = m^2_{\rm spectator} +{\vec{k}}^2_{\perp {\rm
spectator}}$, where ${\vec{k}}_{\perp {\rm spectator}}= \sum_j {\vec{k}}_{\perp
j}$ and $m^2_{\perp {\rm spectator}}/x_{\rm spectator} = \sum_j m^2_j/x_j$,
summed
over the $j$ spectators.
This is an important simplification for  phenomenology, since we can change
variables to $m_{\rm spectator}^2$ and $d^2{\vec{k}}_{\perp {\rm
spectator}}$ and
replace all of the spectators by a spectral integral over the cluster mass
$m_{\rm spectator}^2$.

The proper treatment of the $J^-$ current implies
consistency conditions which must be obeyed by the light-cone
wavefunctions. For example, current conservation for the form factors of
spin zero hadrons requires
\begin{equation}
(2p-q)^\mu F(q^2) = \VEV{p-q\,|\,J^\mu(0)\, |\, p}
\label{vir1}
\end{equation}
and thus
\begin{equation}
\VEV{p-q\,|\,J^+\,|\,p} =
\frac{(2p-q)^+}{(2p-q)^-}\
\VEV{p-q\,|\,J^-\,|\,p} \ .
\label{vir2}
\end{equation}
We have explicitly verified this new type of virial theorem in a
simple scalar composite model in Ref. \cite{Brodsky:1998hn} .

The off-diagonal $n+1 \rightarrow n-1$ contributions provide a new
perspective on the
physics of $B$-decays.  A semileptonic decay involves not only matrix
element where a
quark changes flavor, but also a contribution where the leptonic pair is
created from the
annihilation of a $q {\bar{q'}}$ pair within the Fock states of the initial $B$
wavefunction.  The semileptonic decay thus can occur from the annihilation of a
nonvalence quark-antiquark pair in the initial hadron.  This feature will carry
over to exclusive hadronic
$B$-decays, such as
$B^0
\rightarrow
\pi^-D^+$.  In this case the pion can be produced from the coalescence of a
$d\bar u$ pair emerging from the initial higher particle number Fock
wavefunction of the $B$.  The $D$ meson is then formed from the remaining quarks
after the internal exchange of a $W$ boson.

A remarkable advantage of the light-cone formalism that all
matrix elements of local operators can be written down exactly in terms of
simple
convolutions of light-cone Fock wavefunctions.
The light-cone wavefunctions depend
only on the hadron itself; they are process-independent. The formalism is
relativistic and
frame-independent---the incident four-vectors can be chosen in any frame.
Note that the  matrix element of a current in the covariant Bethe-Salpeter
formalism requires the construction of the current from insertions into an
infinite number of irreducible kernels.  The Bethe-Salpeter formalism becomes
even more intractable for bound-states of more than two particles.

\bigskip

\begin{center}
APPENDIX III \\
BARYON FORM FACTORS AND EVOLUTION EQUATIONS
\end{center}
\bigskip

The baryon form factor is a
prototype for the calculation of the QCD hard scattering contribution
for the whole range of exclusive processes at large momentum transfer.
Away from possible special points in the $x_i$ integrations
a general hadronic amplitude can be written to leading order in
$1/Q^2$ as a convolution of a connected hard-scattering amplitude
$T_H$ convoluted with the meson and baryon distribution amplitudes:
\begin{equation}
\phi_M(x,Q)= \int^{\mid{\cal E}\mid <Q^2}{d^2k_\bot\over 16\pi^2}
\psi^Q_{q\bar q}(x,\vec k_\bot)\quad ,
\end{equation}
and
\begin{equation}
\phi_B(x_i,Q)= \int^{\mid{\cal E}\mid <Q^2} [d^2k_\bot]
\psi_{qqq}(x_i,\vec k_{\bot i}) \ .
\end{equation}
Here ${\cal E} ={\cal M}^2_{qqq} - M_B^2$ is the invariant off-shellness
of the  three-quark baryon light-cone wavefunction.

The hard scattering amplitude $T_H$ is computed by replacing each
external hadron line by massless valence quarks each collinear
with the hadron's momentum $p^\mu_i\cong x_ip^\mu_H$.
For example the baryon form factor at large $Q^2$ has the form
\cite{LepBrod,helicity}
\begin{equation}
G_M(Q^2)=\int[dx][dy]\phi^\star(y_i,\bar Q) T_H(x,y;Q^2)\phi(x,\bar Q)
\end{equation}
where $T_H$ is the $3q+\gamma\rightarrow 3q^\prime$ amplitude.
For the proton and neutron we have to leading order
[$C_B = 2/3$]
\begin{eqnarray}
T_p&=&{128\pi^2C^2_B\over (Q^2+M^2_0)^2}\ T_1  \nonumber \\
T_n&=&{128\pi^2C^2_B\over 3(Q^2+M^2_0)^2}\left[T_1-T_2\right]
\end{eqnarray}
where
\begin{eqnarray}
T_1 =&-& {\alpha_s(x_3y_3Q^2)\ \alpha_s(1-x_1)(1-y_1)Q^2)\over
x_3(1-x_1)^2\ y_3(1-y_1)^2} \nonumber \\
 &+& {\alpha_s(x_2y_2Q^2)\ \alpha_s\left((1-x_1)(1-y_1)Q^2\right) \over
x_2(1-x_1)^2\ y_2(1-y_1)^2}\nonumber \\
&-& {\alpha_s(x_2y_2Q^2)\ \alpha_s(x_3y_3Q^2) \over
x_2x_3(1-x_3)\ y_2y_3(1-y_1)}\quad ,
\end{eqnarray}
and
\begin{equation}
T_2 = - {\alpha_s(x_1y_1Q^2)\ \alpha_s(x_3y_3Q^2) \over
x_1x_3(1-x_1)\ y_1y_3(1-y_3)} \quad .
\end{equation}
$T_1$ corresponds to the amplitude where the photon interacts
with the quarks (1) and (2) which have helicity parallel to the
nucleon helicity, and $T_2$ corresponds to the amplitude where
the quark with opposite helicity is struck.
The running coupling constants have arguments $\hat Q^2$
corresponding to the gluon momentum transfer of each diagram.
Only the large $Q^2$ behavior is predicted by the theory; we
utilize the parameter $M_0$ to represent the effect of power-law
suppressed terms from mass insertions, higher Fock states, etc.

The $Q^2$-evolution of the baryon distribution amplitude can be
derived from the operator product expansion of three quark fields
or from the gluon exchange kernel.
The baryon evolution equation to leading order in $\alpha_s$ is
\cite{helicity}
\begin{equation}
x_1x_2x_3\left\{{\partial\over\partial\zeta}\tilde\phi(x_i,Q) +
{3\over 2}{C_F\over \beta_0}\tilde\phi(x_i,Q)\right\} =
{C_B\over \beta_0}\int^1_0[dy]V(x_i,y_i)\tilde\phi(y_i,Q).
\end{equation}
Here $\phi=x_1x_2x_3\tilde\phi, \zeta={\rm log}({\rm log}Q^2/
\Lambda^2)$,
$C_F = (n_c^2 -1)/2n_c = 4/3, \ C_B = (n_c +1)/2n_c =
2/3, \ \beta = 11 - (2/3)n_f$, and $V(x_i,y_i)$ is computed
to leading order in $\alpha_s$ from the single-gluon-exchange
kernel:
\begin{eqnarray}
V(x_i,y_i)&=&2x_ix_2x_3\sum_{i\neq j}\theta(y_i-x_i)\delta
(x_k-y_k){y_j\over x_j}
\left({\delta_{h_i\bar h_j}\over x_i+x_j} + {\Delta\over
y_i-x_i}\right) \nonumber \\
&=& V(y_i,x_i)\quad .
\end{eqnarray}
The infrared singularity at $x_i=y_i$ is cancelled because the
baryon is a color singlet.

The baryon evolution equation automatically sums to leading order
in $\alpha_s(Q^2)$ all of the contributions from multiple
gluon exchange which determine the tail of the valence
wavefunction and thus the $Q^2$-dependence of the
distribution amplitude.
The general solution of this equation is
\begin{equation}
\phi (x_i,Q) = x_1 x_2 x_3 \, \sum_{n=0}^\infty \, a_n
\left( \ell n \, {Q^2 \over \Lambda^2} \right)^{-\gamma_n}
\, {^ \phi}_n (x_i) \quad ,
\end{equation}
where the anomalous dimensions $\gamma_n$ and the eigenfunctions
${\widetilde \phi}_n (x_i)$ satisfy the characteristic
equation:
\begin{equation}
x_1 x_2 x_3 \left( -\gamma_n + {3C_F \over 2\beta} \right)
\, {\widetilde \phi}_n (x_i) = {C_B \over \beta} \int_0^1
[dy] \ V(x_i, y_i) \, {\widetilde \phi}_n (y_i) \quad .
\end{equation}

A  useful technique for obtaining the solution
to  evolution equations
is to construct completely antisymmetric representations as
a polynomial orthonormal basis for the distribution
amplitude of multiquark bound states.  In this way one
obtain a distinctive classification of nucleon $(N)$ and
delta $(\Delta)$ wave functions and the corresponding $Q^2$
dependence which discriminates $N$ and $\Delta$ form factors.
This technique is developed
in detail in Ref. \cite{JiBaryon}.  The conformal representation of
baryon distribution amplitudes is given in Ref. \cite{Braun:1999te}.

Taking into account the evolution of the baryon distribution
amplitude, the nucleon magnetic form factors at large $Q^2$,
has the form \cite{LepBrod,helicity}\
\begin{equation}
G_M(Q^2)\rightarrow{\alpha^2_s(Q^2)\over Q^4}\sum_{n,m} b_{nm}
\left({\rm log}{Q^2\over \Lambda^2}\right)^{\gamma^B_n-\gamma^B_n}
\left[1+{\cal O}\left(\alpha_s(Q^2),{m^2\over Q^2}\right)\right]\quad .
\end{equation}
where the $\gamma_n$ are computable anomalous dimensions of the baryon
three-quark wave function at short distance
and the $b_{mn}$ are determined
from the value of the distribution amplitude $\phi_B(x,Q^2_0)$
at a given point $Q_0^2$ and the normalization of $T_H$.
Asymptotically, the dominant term has the minimum anomalous dimension.
The dominant part of the form factor comes from the region
of the $x_i$ integration where each
quark has a finite fraction of the light cone momentum.
The integrations over $x_i$ and $y_i$
have potential endpoint singularities.
However, it is easily seen that any anomalous contribution [\eg\
from the region $x_2,x_3\sim{\cal O}(m/Q), x_1\sim 1-{\cal O}
(m/Q)$] is asymptotically suppressed at large $Q^2$ by a
Sudakov form factor arising from the virtual correction to the
$\bar q\gamma q$ vertex when the quark legs are near-on-shell
[$p^2\sim{\cal O}(mQ)$] \cite{helicity,MuellerDuncan}.
This Sudakov suppression of the endpoint region requires an all
orders resummation of perturbative contributions,
and thus the derivation of the baryon form factors is not
as rigorous as for the meson form factor, which has no
such endpoint singularity \cite{MuellerDuncan}.

One can also use PQCD to predict ratios of
various baryon and isobar form factors assuming isospin or
$SU(3)$-flavor symmetry for the basic wave function structure.
Results for the neutral weak and charged weak form factors assuming
standard $SU(2)\times U(1)$ symmetry are given in Ref. \cite{Zaidi}.

\bigskip

\begin{center}
Comparison Between Time-Ordered and $\tau$-Ordered Perturbation Theory
\end{center}

\begin{tabular}{ll}
\hline
\hfil Equal $t$ &\hfil Equal $\tau = t+z$ \\
\hline
$k^o = \sqrt{{\vec k}^2 + m^2}$ \ (particle mass shell)
&$k^- = {  k^2_\perp + m^2 \over   k^+}$
\ (particle mass shell) \\[2ex]
$\sum \, {\vec k}$ conserved &$\sum \, {\vec k}_\perp, \ k^+$
conserved \\[2ex]
${\cal M}_{ab} = V_{ab} + \sum\limits_c V_{ac} \, {{  1}\over
\sum_a \, {  k^o} - \sum_c \, {  k^o + i\epsilon}} \, V_{ac}$
&${\cal M}_{ab} = V_{ab} + \sum\limits_c V_{ac} \, {{  1}\over
\sum_a \, {  k^-} - \sum_c \, {  k^- + i\epsilon}} \,
V_{cb}$ \\[2ex]
$n!$ time-ordered contributions &$k^+ > 0$ only \\[2ex]
Fock states $\psi_n ({\vec k}_i)$
&Fock states $\psi_n ({\vec k}_{\perp i}, x_i)$ \\[2ex]
$\sum\limits_{i=1}^n \, {\vec k}_i = {\vec P} = 0$
&$x = {  k^+ \over   P^+} \ , \ \sum\limits_{i=1}^n
\, x_i =1 \ , \ \sum\limits_{i=1}^n \, {\vec k}_{\perp i} = 0$ \\[1ex]
{ } &\qquad \qquad \ $(0 < x_i < 1)$ \\[2ex]
${\cal E} = P^o - \sum\limits_{i=1}^n \, k_i^o$
&${\cal E} = P^+ \left( P^- - \sum\limits_{i=1}^n \,
k_i^- \right)$ \\[1ex]
\quad $= M - \sum\limits_{i=1}^n \, \sqrt{k_i^2 + m_i^2}$
&\quad $= M^2 - \sum\limits_{i=1}^n \left( {  k_\perp^2
+ m^2 \over   x} \right)_i$ \\
\hline \end{tabular}

\end{document}